\documentclass{aa}
\usepackage[varg]{txfonts}
\usepackage{graphicx,aas_macros,natbib}
\usepackage{subfigure,aas_macros}

\newcommand\msun{\rm{M_{\odot}}}

\def\stacksymbols #1#2#3#4{\def\theguybelow{#2}
        \def\verticalposition{\lower#3pt}
        \def\spacingwithinsymbol{\baselineskip0pt\lineskip#4pt}
        \mathrel{\mathpalette\intermediary#1}}
\def\intermediary #1#2{\verticalposition\vbox{\spacingwithinsymbol
        \everycr={}\tabskip0pt
        \halign{$\mathsurround0pt#1\hfil##\hfil$\crcr#2\crcr
                \theguybelow\crcr}}}

\def\lta{\stacksymbols{<}{\sim}{2.5}{.2}}
\def\gta{\stacksymbols{>}{\sim}{2.5}{.2}}

\begin{document}

\title{Chaotic cold accretion on to black holes in rotating atmospheres}
\author{M. Gaspari$^1$\thanks{E-mail: mgaspari@mpa-garching.mpg.de} \and F. Brighenti$^2$ \and P. Temi$^3$}
\institute{$^1$Max Planck Institute for Astrophysics, Karl-Schwarzschild-Strasse 1, 85741 Garching, Germany\\ 
$^2$Astronomy Department, University of Bologna, Via Ranzani 1, 40127 Bologna, Italy\\
$^3$Astrophysics Branch, NASA/Ames Research Center, MS 245-6, Moffett Field, CA 94035, USA}

\abstract{
The fueling of black holes is one key problem in the evolution of baryons in the universe. 
Chaotic cold accretion (CCA) profoundly differs from classic accretion models, as Bondi and thin disc theories. Using 3D high-resolution hydrodynamic simulations, we now probe the impact of {\it rotation} on the hot and cold accretion flow in a typical massive galaxy. In the hot mode, with or without turbulence,  
the pressure-dominated flow forms a geometrically thick rotational barrier, suppressing the black hole accretion rate to $\sim$1/3 of the spherical case value. 
When radiative cooling is dominant, the gas loses pressure support
and quickly circularizes in a cold thin disk; the accretion rate is decoupled from the cooling rate, although it is  
higher than that of the hot mode.
In the more common state of a turbulent and heated atmosphere,
{\it chaotic cold accretion} drives the dynamics 
if the gas velocity dispersion exceeds the rotational velocity, i.e., turbulent Taylor number ${\rm Ta_t}<1$. Extended multiphase filaments condense out of the hot phase via thermal instability (TI) and rain toward the black hole, boosting the accretion rate up to 100 times the Bondi rate ($\dot M_\bullet \sim \dot M_{\rm cool}$). Initially, turbulence broadens the angular momentum distribution of the hot gas, allowing the cold phase to condense with prograde or retrograde motion. Subsequent chaotic collisions between the cold filaments, clouds, and a clumpy variable torus promote the cancellation of angular momentum, leading to high accretion rates. As turbulence weakens (${\rm Ta_t} > 1$), the broadening of the distribution and the efficiency of collisions diminish, damping the accretion rate $\propto{\rm Ta_t}^{-1}$, until the cold disk drives the dynamics. This is exacerbated by the increased difficulty to grow TI in a rotating halo.
The simulated sub-Eddington accretion rates cover the range inferred from AGN cavity observations. 
CCA predicts inner flat X-ray temperature and $r^{-1}$ density profiles, as recently discovered in M 87 and NGC 3115.  
The synthetic H$\alpha$ images reproduce the main features of cold gas observations in massive ellipticals, as the line fluxes and the filaments versus disk morphology. Such dichotomy is key for the long-term AGN feedback cycle.
As gas cools, filamentary CCA develops and boosts AGN heating; the cold mode is thus reduced and the rotating disk remains the sole cold structure. Its consumption leaves the atmosphere in hot mode with suppressed accretion and feedback, reloading the cycle.
}


\authorrunning{Gaspari et al.}
\titlerunning{Raining on to black holes -- Accretion driven by TI in rotating halos}
\maketitle

\section{Introduction and observations}\label{s:intro}
\noindent
In current astrophysics, a marked disconnection exists between theoretical works focusing on the vicinity of supermassive black holes (SMBHs) and those studying the large-scale properties of the host (a massive elliptical, galaxy group or cluster).
The former often employ idealized, constant boundary conditions, while the latter are forced to rely on 
semianalytic subgrid models. The intermediate zone, between a few 10 kpc of the host galaxy and the 
sub-pc core, is however a crucial region that often determines the fueling and feeding 
of the black hole.

\citeauthor{Gaspari:2013_cca} (2013; hereafter G13) aimed to fill this gap, showing that realistic turbulence, cooling, and heating affecting the hot gaseous halo, can dramatically change the accretion flow on to black holes, departing from the idealized picture of Bondi (\citeyear{Bondi:1952}) formula. As the cooling time becomes relatively low compared with the dynamical time ($t_{\rm cool}\lta\,10\, t_{\rm ff}$), cold clouds and filaments condense out of the hot phase via nonlinear thermal instability (TI). Chaotic collisions promote the funneling of the cold phase toward the BH, leading to episodic spikes in the accretion rate up to $\sim100\times$ the Bondi rate.
For the more poetic minds, chaotic cold accretion (CCA) can be viewed as `raining on to black holes'.

Because of the simplicity of the Bondi formula, it is tempting to exploit it
in theoretical and observational studies (e.g., \citealt{Reynolds:1996, Loewenstein:2001, Churazov:2002, Baganoff:2003, DiMatteo:2003, DiMatteo:2005, Springel:2005, Allen:2006, Croton:2006,Hopkins:2006, Rafferty:2006, Cattaneo:2007, Sijacki:2007, Hardcastle:2007, Booth:2009, Cattaneo:2009, Barai:2014,Nemmen:2015}). 
Knowing the black hole mass $M_\bullet$ and the gas entropy at large radii ($K\propto T/\rho^{\gamma-1}$, where $\gamma$ is the adiabatic index, $\rho$ the gas density, and $T$ the gas temperature) allows us to immediately retrieve the accretion rate via
\begin{equation}\label{e:MdotB}
\dot M_{\rm B} = \lambda\, 4\pi (GM_\bullet)^2\frac{\rho_{\infty}}{c^3_{{\rm s}, \infty}} \propto K_\infty^{-3/2},
\end{equation}
where $\lambda$ is a factor of order unity (varying as a function of $\gamma$ and the accretor radius; cf.~G13, Sec.~1).
However, Eq.~\ref{e:MdotB} can only be used if the hydrodynamic flow is adiabatic (no heating or cooling), unperturbed, spherically symmetric, with steady boundary conditions at infinity.
If one of these conditions is violated, then Bondi derivation and formula cannot be applied.
Bondi himself
warns in the first line of his 1952 abstract that
he is investigating a special accretion problem.
In the last decade, observations and simulations of gas in galaxies, groups, and clusters have proven that atmospheres are turbulent (e.g., \citealt{Norman:1999,Schuecker:2004, Dolag:2005, Kim:2005, Nagai:2007,Lau:2009, Vazza:2009, Borgani:2011,DePlaa:2012,Sanders:2013,Gaspari:2013_coma,Banerjee:2014,Gaspari:2014_coma2}; see also interstellar medium studies, \citealt{Elmegreen:2004}), while continuously shaped by the competition of cooling and heating processes (e.g., \citealt{Vikhlinin:2006, McNamara:2007,McNamara:2012,Diehl:2008b, Rasmussen:2009, Sun:2009a,Gaspari:2014_scalings}).

Tracking the realistic accretion rate and dominant mode of SMBH fueling is fundamental to understand and model the impact of AGN (`active galactic nucleus') feedback. The energy released by a SMBH can reach $10^{62}$ erg, which is able to affect the host galaxy and the surrounding group or cluster gaseous halo (e.g., \citealt{Gaspari:2011a,Gaspari:2011b,Gaspari:2012a,Gaspari:2012b}).
The AGN feedback can address many astrophysical problems, such as heating cooling flows, quenching star formation, forming buoyant bubbles and shocks, or ejecting metals and low-entropy gas at large radii (\citealt{Gaspari:2013_rev}, for a brief review). Overall, SMBHs can be seen as the thermostats regulating the baryonic structures throughout the cosmic evolution.

In recent times, it has become clear that most, if not all, massive galaxies retain a hot, X-ray emitting atmosphere 
down to galaxies with stellar masses $M_\ast\sim10^{11}\ \msun$
(\citealt{Anderson:2015}; see also \citealt{Planck:2013_YM}, Fig.~4), 
including spirals (e.g., \citealt{Anderson:2011_NGC1961,Dai:2012}). Therefore, cold gas condensation and accretion is expected to play central role in the evolution of SMBHs and their host galaxies, as supported by other works
(e.g., \citealt{Pizzolato:2005,Soker:2006,Soker:2009,Barai:2012}).
Although we focus in this study on a massive early-type galaxy (with inefficient star formation), 
CCA is also expected to be relevant in high-redshift and disk galaxies.
These systems host large reservoir of cold gas and thus do not necessitate TI condensation from a hot halo
to ignite CCA. On top of that, cold accretion can be augmented via cosmic-web inflows, and the related large-scale disk instabilities or tidal torques (e.g., \citealt{Dekel:2009,Hopkins:2010}).
Chaotic collisions via minor mergers can also boost cold accretion (e.g., \citealt{King:2006}). 

Observations of multiphase gas ($T<10^6$ K) 
embedded in the hot, X-ray emitting plasma of massive halos
have exponentially grown over the last decade. These observations detect
extended ionized gas in optical
H$\alpha$+[NII] (\citealt{Heckman:1989,Macchetto:1996,Crawford:1999,McDonald:2009,McDonald:2010,McDonald:2011a,McDonald:2012_Ha,Werner:2014}; \S\ref{s:comp}),
which is typically cospatial with infrared H$_2$ (e.g., \citealt{Jaffe:2005,Hatch:2005,Wilman:2009,Wilman:2011,Oonk:2010}), molecular gas traced by CO (\citealt{Lim:2000,Edge:2001,Salome:2003,Combes:2007,Salome:2008,Hamer:2014}),
far-infrared [CII], [OI] (\citealt{Mittal:2012,Werner:2013}), 
and far-ultraviolet CIV (\citealt{Sparks:2012}).
These data strongly favor the scenario of in-situ condensation via TI, as opposed to 
gas stripping from infalling galaxies.
Recently, the unprecedented resolution and sensitivity of ALMA has further proven 
the central role of condensed cold gas in the form of clouds, turbulent disks, and outflows
(\citealt{Combes:2013,Combes:2014,David:2014,McNamara:2014, Russell:2014}).
Remarkably, most cool-core systems with $t_{\rm cool}<1$ Gyr contain filamentary multiphase gas 
(\citealt{Cavagnolo:2008,Rafferty:2008}) and central radio sources (\citealt{Mittal:2009}), indicating that 
the cooling gas is the main driver of AGN feedback. 
The increase of radio-loud AGN with more massive halos (\citealt{Best:2007}), which have
higher cooling rates (\citealt{Shabala:2008}) and slower rotation, further supports the fueling via CCA.
It shall be noted that molecular gas does not need to be strictly correlated with AGN bubbles, jets, or outflows (e.g., \citealt{Werner:2014}), since CCA quickly consumes the infalling cold gas. Powerful AGN feedback can also drag it out of the galaxy (e.g., \citealt{Canning:2013}).

After the initial study by G13, 
many features of the newly proposed CCA remain to be tackled,
in this and future investigations.
One important open question concerns the role of rotation in being able to suppress accretion, 
in combination with cooling, heating, and turbulence (see also \citealt{Proga:2003,Krumholz:2005,Krumholz:2006,Pizzolato:2010,Hobbs:2011,Narayan:2011,Li_Ostriker:2013}).
Current surveys using integral-field spectroscopy \citep{Emsellem:2007,Emsellem:2011}  
show that early-type galaxies display slow or fast rotating stellar kinematics,
likely reflecting separate formation and evolution histories. 
While low-luminosity galaxies are typically fast rotators,
massive galaxies (as brightest cluster or group galaxies), the focus of the present work, belong to the `slow' rotator
family (e.g., \citealt{Jimmy:2013}; \citealt{Kormendy:2009} for a review). 
Recurrent gas-poor (dry) mergers (e.g., \citealt{Bois:2011}) or AGN outflows 
(e.g., \citealt{Gaspari:2012b}) can both contribute in reducing angular momentum.
Slow rotators or massive ellipticals have angular momentum parameter $\lambda_{R_{\rm e}}\sim0.1$\,-\,$0.3$ 
(\citealt{Emsellem:2007}), corresponding to stellar rotational velocities $v_{\ast, {\rm rot}} \sim 0.1$\,-\,$0.3\, \sigma_\ast$ 
(\citealt{Binney:1990,Caon:2000,Pinkney:2003,Jimmy:2013}).
A notable example with significant rotation is NGC 4649.

Compared with the stellar kinematics, the rotation of the gas in observed massive ellipticals is more uncertain.
Roughly 70\% of the hot gas within the effective radius likely comes
from stellar mass loss (e.g., \citealt{Brighenti:1999a}), thereby gas rotation is expected to share
similar specific angular momentum as the local stars, 
$v_{\rm rot}\sim 0.1$\,-\,$0.3\, \sigma_\ast$ (e.g., \citealt{Caon:2000}). 
Evidence for gas rotation in the inner part of 
the galaxy is given by the X-ray ellipticity, typically $\lta0.2$, which steadily declines below that of the stars
at $r\gta 1$ kpc (\citealt{Diehl:2007,Brighenti:2009}). The negative slope of the X-ray ellipticity is typically moderate (\citealt{Diehl:2007}), suggesting that transport processes,
such as turbulence, are required to circularize the isophotes and to prevent the rapid spin-up of the gas due to cooling flows (e.g., \citealt{Brighenti:1996,Brighenti:2000_rot}).
Nonzero gas angular momentum can also be associated with subsonic sloshing motions and cold fronts due to infalling substructures (\citealt{Markevitch:2007,ZuHone:2013}) or with galaxy peculiar velocity 
(e.g., \citealt{Zabludoff:1993}).

In this study, we perform astrophysical experiments, 
dissecting each physics in a methodical way to disentangle its impact on the hot and cold
accretion flow affected by rotation in a massive early-type galaxy.
The key objective is to further unveil and understand CCA, 
rather than covering any possible accretion scenario.
In \S\ref{s:init}, we review the physical and numerical ingredients of the simulations. 
In \S\ref{s:adi}, we study the rotating flow in the adiabatic galactic atmosphere.
In \S\ref{s:cool}, we analyze pure cooling and the related cold thin disk.
Section \ref{s:stir} shows the impact of realistic turbulence in the rotating hot flow. 
In \S\ref{s:stir_cool}, we combine cooling and stirring, while in \S\ref{s:heat} we present the complete CCA
evolution in a rotating halo, including heating and varying levels of turbulence. 
In \S\ref{s:comp}, we compute synthetic H$\alpha$ images and compare them with the most recent observations 
obtained with the SOAR telescope (\citealt{Werner:2014}).
In \S\ref{s:disk}, we summarize and discuss our findings in the context of the long-term AGN feedback loop. 
Remarkably, CCA is unhindered as long as a key dimensionless quantity, which we define as `turbulent Taylor'\footnote{This quantity shares similarities with the classic Taylor number, which characterizes the importance of centrifugal forces relative to viscous forces, i.e., ${\rm Ta} = \omega^2 R^4/\nu^2$, where $\omega$ is the angular velocity, $R$ is the cylindrical radius, and $\nu$ is the kinematic viscosity. Notice that ${\rm Ta}\propto {\rm Ta_t}^2$.} number, remains below unity, i.e.,
\begin{equation}\label{e:Tt}
{\rm Ta_t} \equiv \frac{v_{\rm rot}}{\sigma_v} < 1,
\end{equation}   
where $v_{\rm rot}$ and $\sigma_v$ are the rotational velocity and turbulent velocity dispersion of the gas, respectively.

\section[]{Physics and numerics} \label{s:init}
\noindent
The implemented physics and numerics are described in depth in G13 (Sec.~2). 
Here we summarize the essential features and new ingredients, such as rotation (\S\ref{s:init1}).

\subsection{Initial conditions and rotation} \label{s:init1}
\noindent
We study the accretion flow in a typical massive elliptical galaxy (NGC 5044) embedded in the gaseous intragroup medium.
The initial temperature and gravitational potential profiles are unchanged compared with our previous work. The $T$ profile is directly derived from {\it Chandra} and {\it XMM} observations (\citealt{Gaspari:2011b}). 
Since we focus on a virialized galaxy, we use a
static potential $\phi$, which is given by the central SMBH ($M_\bullet=3\times10^9$ $\msun$), the galactic stellar component ($M_{\ast}\simeq 3.4\times10^{11}$ $\msun$, with effective radius $\simeq10$ kpc), and a 
dark matter Navarro-Frenk-White (\citeyear{Navarro:1996}) halo 
with virial mass $M_{\rm vir} \simeq 3.6\times10^{13}$ $\msun$ and concentration $\simeq 9.5$.
The Schwarzschild and Bondi radii are $R_{\rm S}\equiv 2GM_\bullet/c^2 \simeq 3\times10^{-4}$ pc and $r_{\rm B}\equiv GM_\bullet/c_{\rm s, \infty}^2 \simeq 85$ pc, respectively (the sound speed is $c_{\rm s, \infty}\simeq390$ km s$^{-1}$ near 1 kpc). We integrate the system for a long-term evolution, $\sim200\,t_{\rm B}$, where
$t_{\rm B}\equiv r_{\rm B}/c_{\rm s, \infty}\simeq 210$ kyr.

The density profile is retrieved from hydrostatic equilibrium (neglecting the black hole).
We test the gas rotation, which induces a centrifugal force, effectively lowering the gravitational acceleration $g$ along $R\equiv (x^2 + y^2)^{1/2}$. This partially changes the stratification of the hot gaseous halo. 
The new hydrostatic equilibrium can be better visualized and set up separating the two main directions, $R$ and $z$,
as follows:
\begin{equation}\label{e:HSER}
\frac{\partial}{\partial R}\ln\rho = - \left(\frac{\partial\phi}{\partial R} - \frac{v^2_{\rm rot}}{R} \right)\,c^{-2}_{\rm s,i} - \frac{\partial}{\partial R}{\ln T},
\end{equation}
\begin{equation}\label{e:HSEz}
\frac{\partial}{\partial z}\ln\rho = - \frac{\partial\phi}{\partial z}\,c^{-2}_{\rm s,i} - \frac{\partial}{\partial z}{\ln T},
\end{equation}
where $c_{\rm s,i}^2=k_{\rm b} T/\mu m_{\rm p}$ is the isothermal sound speed (the mean particle weight is $\mu\simeq0.62$).
As is customary (e.g., \citealt{Strickland:2000}), the rotational velocity of the gas is parametrized as a fraction of the circular velocity,
\begin{equation}\label{e:vrot}
v_{\rm rot}(R) \equiv e_{\rm rot}\;v_{\rm circ} (R,0) = e_{\rm rot}\,\left(R\,\frac{\partial\phi(R,0)}{\partial R}\right)^{1/2},
\end{equation}
where $e_{\rm rot}$ is a free parameter ranging between 0 and 1, the latter corresponding to full rotational support.
We first integrate the hydrostatic equilibrium along $R$ (Eq.~\ref{e:HSER}), then along the $z$ direction (Eq.~\ref{e:HSEz}), linearly interpolating the retrieved 2D matrix in the discretized 3D domain. 
The central density normalization is the same as in G13. The initial profiles are shown in Figure \ref{f:pure_prof}.
As suggested by observations, the resultant rotational velocity is fairly constant with $R$ (outside the Keplerian influence region of the SMBH);
at 26 kpc $v_{\rm rot}$ is just $\sim$30\% higher than at 1 kpc. Using constant specific angular momentum at large radii ($v_{\rm rot}\propto R^{-1}$) should be thus avoided as an initial condition.

We discuss now the reference $e_{\rm rot}$.
The Jeans equation for a spherically symmetric, isotropic stellar system in equilibrium can be written as 
\begin{equation}\label{e:jeans}
v^2_{\ast,\rm rot}-\sigma_{\ast,r}^2 \left(\frac{\partial\ln\rho_\ast}{\partial\ln r} + \frac{\partial\ln\sigma^2_{\ast, r}}{\partial \ln r}\right) = v^2_{\rm circ}.
\end{equation}
Neglecting rotation, the radial stellar velocity dispersion for our simulated galaxy peaks at $\sigma_{\ast, r}\simeq 230$ km s$^{-1}$ (the 3D velocity dispersion is $\sigma_\ast=\sqrt{3}\,\sigma_{\ast, r}$). 
For weak rotation, the stellar velocity dispersion is a good proxy for the circular velocity.
Massive elliptical galaxies are known to have irregular velocity profiles,
at best with mild coherent rotation, in particular for the gas component (e.g., \citealt{Caon:2000}).
As discussed in \S\ref{s:intro}, gas rotation can typically reach $v_{\rm rot}\lta0.3\, \sigma_\ast$.
To maximize the impact of rotation, we thus adopt $e_{\rm rot} = 0.3$ as reference, leading to an average gas rotational velocity $v_{\rm rot}\approx100$ km s$^{-1}$ 
($r$\,$\sim$\,1\,-\,13 kpc).
Adopting slower rotation has negligible impact on the CCA dynamics, resembling G13 models, since $v_{\rm rot} \ll \sigma_v$. As faster rotation flattens the density profile and isophotes too much (\S\ref{s:intro}), 
it is better to compare models with fixed $e_{\rm rot}$ and varying $\sigma_v$ (the cooling rate also remains the same),
albeit ${\rm Ta_t}$ defines the self-similar dynamics in both cases (\S\ref{s:heat}).

\subsection{Hydrodynamics and source terms} \label{s:init2}
\noindent
We use a modified version of the adaptive-mesh-refinement code FLASH4 (\citealt{Fryxell:2000}) to integrate the well-known equations of hydrodynamics (e.g., \citealt{Gaspari:2011b}), 
adopting a very large dynamical range. The maximum resolution is $\simeq\,$0.8 pc, with radially concentric fixed meshes in cartesian coordinates (G13, Sec.~2.1).
The box width reaches 52 kpc ($\sim\,$$600\ r_{\rm B}$), an extension of almost a factor $10^5$ (see G13, for further details on resolution and convergence). Notice that using $e_{\rm rot}=0.3$ implies that the circularization radius is $r_{\rm circ} = (v_{\rm rot}/v_{\rm circ})\,r_{\rm init} = 0.3\,r_{\rm init}$. Even considering the steepening of $v_{\rm circ}$ due to the SMBH potential ($\sim$\,4000 km s$^{-1}$ at 1 pc), we can resolve circular motions down to a few 10 pc. Thermal instabilities form up to several kpc, hence we can clearly assess whether the cold gas is accreted or circularizes (see the thin disk evolution in \S\ref{s:cool}). 

In addition to hydrodynamics, we add the source terms related to the black hole sink, turbulence driving, radiative cooling, and distributed heating, testing step by step the contribution of each physics to the accretion process. Using the pseudo-relativistic \citet{Paczynski:1980} BH potential, the sonic point (for adiabatic index $\gamma=5/3$) is not located at $r=0$, but at a finite distance near the pc scale. This justifies the use of a central gas sink or void region ($\rho\simeq10^{-35}$ g\,cm$^{-3}$ and $v=0$) with $\approx$\,3 cells radius (which avoids artificial overpressure bounces), since the internal region is causally disconnected (G13, Sec.~2.2). 
Self-gravity is here not included; the cold clouds are not massive enough to overcome the external potential (see G13, end of Sec.~7.3 for an extended discussion).

Turbulence is implemented via a spectral forcing scheme, based on an Ornstein-Uhlenbeck random process, which drives a time-correlated and zero-mean acceleration field, reproducing experimental high-order structure functions (\citealt{Fisher:2008}).
The source of turbulence can be galaxy motions, substructure mergers, supernovae, or AGN feedback (e.g., \citealt{Norman:1999,Lau:2009,Vazza:2009,Gaspari:2012b}). We keep the subsonic turbulent velocity used in G13 as reference, $\sigma_v\sim150$ km s$^{-1}$ (3D Mach number ${\rm Ma}$\,$\sim$\,0.35), stirring the gas at injection scales $L\gta 4$ kpc, and allowing the gas to naturally develop a Kolmogorov-like cascade.
Long-term AGN feedback simulations typically retrieve such turbulence characteristics (e.g., \citealt{Gaspari:2012b,Vazza:2012}). Mergers inject higher kinetic energy at 100s kpc; however, velocities decay through the Kolmogorov cascade and, below 10 kpc, $\sigma_v$ reaches again a few 100 km\,s$^{-1}$ (\citealt{Gaspari:2014_coma2}).
The reference turbulence model has Taylor number ${\rm Ta_t}\sim0.7$, as we use $e_{\rm rot}=0.3$.
In \S\ref{s:stir}-\ref{s:heat},
we test weaker turbulence, i.e., ${\rm Ta_t}=1.5$ and 3.
Turbulent heating (with timescale $\propto {\rm Ma}^{-2}$) is subdominant during our entire evolution.

The hot plasma cools via X-ray radiation mainly because of Bremsstrahlung at $T \gta 10^7$ K and line emission at lower temperature. The radiative emissivity is $\mathcal{L}=n_{\rm e}n_{\rm i}\,\Lambda$, where $n_{\rm e}$ and $n_{\rm i}$ are the electron and ion number density, respectively.
The cooling function $\Lambda(T)$ is modeled following \citet{Sutherland:1993}, adopting solar metallicity.
As is customary, the cold phase has temperature floor at $10^4$ K where the abrupt recombination of hydrogen and the steepening of $\Lambda$ (with slope $\gg 2$) induces a thermally stable region (\citealt{Field:1965}).
The initial ratio of the cooling time, $t_{\rm cool}\equiv1.5\,n k_{\rm B}T/ n_{\rm e} n_{\rm i} \Lambda$, and free-fall time, $t_{\rm ff}\equiv(2r/g)^{1/2}$
has a minimum $\sim\,$4\,-\,5 near 250 pc.
As $t_{\rm cool}/t_{\rm ff}<10$ thermal instability can grow nonlinearly (\citealt{Gaspari:2012a, McCourt:2012, Sharma:2012}; for classic studies on TI see \citealt{Field:1965, Krolik:1983, Balbus:1989,Pizzolato:2005}). 

AGN feedback, in conjunction with stellar heating and mergers (e.g., \citealt{Brighenti:2003,Gaspari:2011a,Gaspari:2011b,Barai:2014}), acts to maintain the cool core of galaxies, groups, and clusters in global thermal quasiequilibrium ($\lta$\,10 percent), preserving core temperatures at $\sim$\,1/3 of the virial temperature (e.g., \citealt{Vikhlinin:2006, Diehl:2008b, Rasmussen:2009, Sun:2009a}).
We do not model AGN outflows or jets as in \citet{Gaspari:2012a,Gaspari:2012b}, since this is computationally unfeasible. 
In the final models (\S\ref{s:heat}), we inject distributed heating, mimicking a post-outburst phase, when the feedback heating has been properly deposited (\citealt{Gaspari:2012a}, Fig.~9). Computationally, we set the heating rate (per unit volume) to be equal to the average radiative emissivity in finite radial shells at each timestep, $\mathcal{H}\approx\langle\mathcal{L}\rangle$.

\section[]{Adiabatic accretion} \label{s:adi}

\begin{figure} 
    \centering
     \subfigure{\includegraphics[scale=0.31]{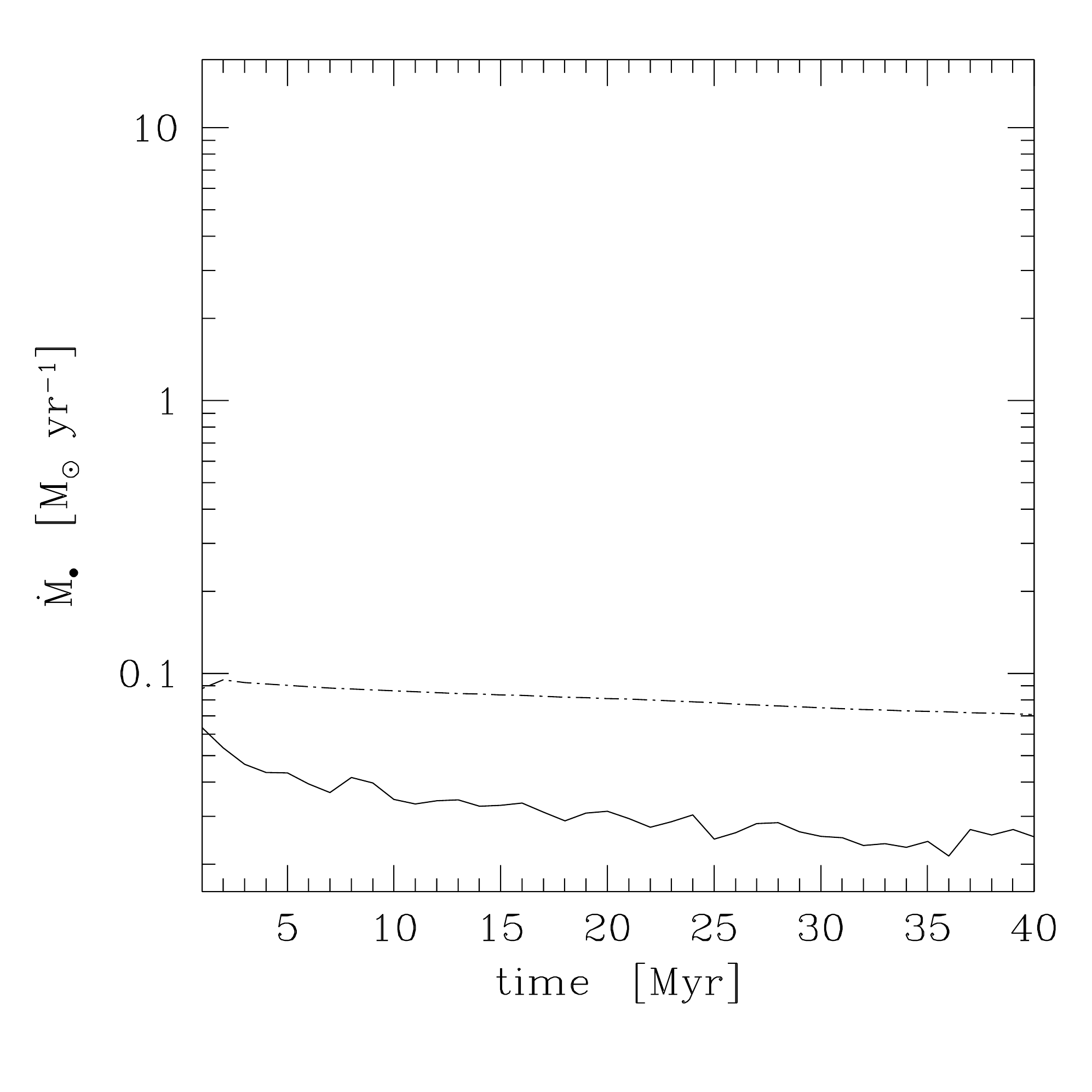}}
     \subfigure{\includegraphics[scale=0.31]{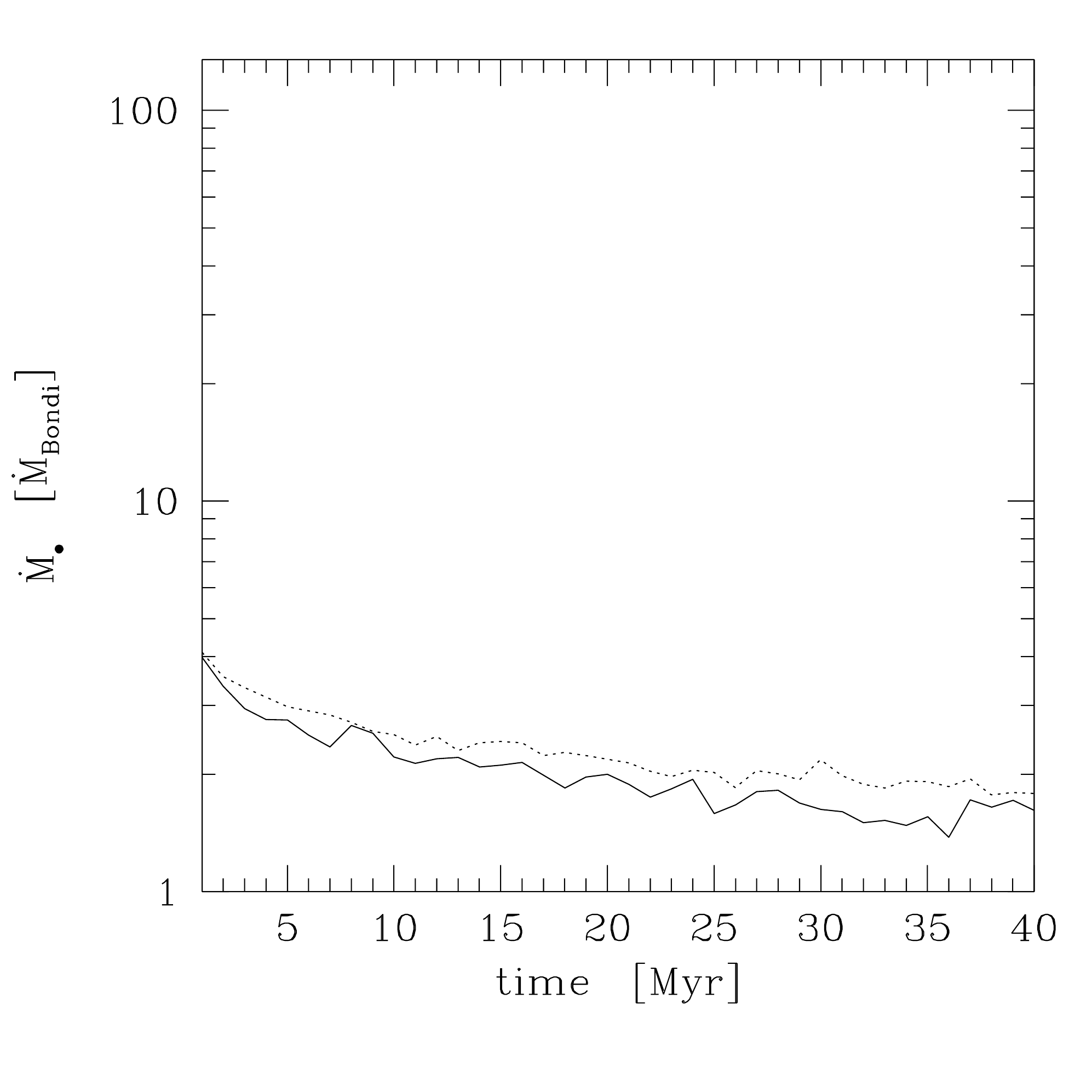}}
     \caption{Adiabatic rotating accretion: evolution of the accretion rate (1 Myr average).
      Top: Accretion rate in $\msun$ yr$^{-1}$ units, for the rotating ($e_{\rm rot} = 0.3$; solid) and nonrotating atmosphere (dot-dashed; see G13). Bottom: Accretion rate normalized to the runtime Bondi rate
      (averaged over $r\approx\,$1-2 kpc as in large-scale simulations) for the reference rotation $e_{\rm rot} = 0.3$ (solid) and $e_{\rm rot}=0.15$ (dotted).
      The central rotational barrier forming within $r_{\rm B}$ 
      suppresses accretion by a factor $\sim$\,3
      compared with the spherically symmetric accretion flow.} 
         \label{f:pure_mdot}
\end{figure}

\noindent
Following the procedure presented in G13, we test step by step the impact of each physical process.
We begin analyzing the model purely based on classic hydrodynamics.
No cooling, heating, or turbulence is affecting the accretion flow. The adiabatic flow shares 
tight connection with \citet{Bondi:1952} accretion, albeit the spherical symmetry is broken by the initial gas rotation
($e_{\rm rot} = 0.3$, i.e., $v_{\rm rot}\approx100$ km s$^{-1}$; see \S\ref{s:init1}).

\subsection{Accretion rate} \label{s:adi_mdot}
\noindent
In Fig.~\ref{f:pure_mdot}, we show the main diagnostic plot: the temporal evolution of the accretion rate during 40 Myr ($\sim$\,200$\,t_{\rm B}$).
The impact of rotation (solid line; top) is to suppress the accretion rate by a factor $\sim$\,3 compared with the nonrotating atmosphere (dashed line; Sec.~3.1 in G13), reaching $\dot M_\bullet\simeq0.025$ $\msun$ yr$^{-1}$.
The average rate is slightly decreasing due to the presence of the galactic gradients, progressively altering the `boundary' conditions at a few $r_{\rm B}$ ($\dot M_{\rm B}\propto K_\infty^{-3/2}$; the Bondi boundary conditions at infinity have no meaning in stratified atmospheres). 
Beside such minor trend, the accretion rate is solely stifled by the central rotationally-supported barrier,
reaching statistical steady state in a few 10 $t_{\rm B}$.

In the bottom panel (Fig.~\ref{f:pure_mdot}), the accretion rate is normalized to the Bondi rate (Eq.~\ref{e:MdotB}), averaged over $r\approx\,$1$\,$-$\,$2 kpc, as customarily employed in large-scale or cosmological simulations. 
Comparing the reference run with $e_{\rm rot}=0.3$ (solid) and the model with $e_{\rm rot}=0.15$ (dotted) indicates that
the final accretion rate is weakly lowered with increasing rotation.
Interestingly, adopting the Bondi formula at large radii predicts a fairly realistic accretion rate, 
within a factor $\lta$\,2.
Therefore, boosting the Bondi accretion rate by a large factor ($\sim\,$100; e.g., \citealt{DiMatteo:2005,Booth:2009}) is not required, even if $r_{\rm B}$ is under-resolved, at least in the regime of a hot and rotating atmosphere (a similar conclusion applies in the presence of turbulence, but not if cooling is dominant).

\subsection{Dynamics} \label{s:adi_dyn}


\begin{figure} 
    \begin{center}
     \subfigure{\includegraphics[scale=0.45]{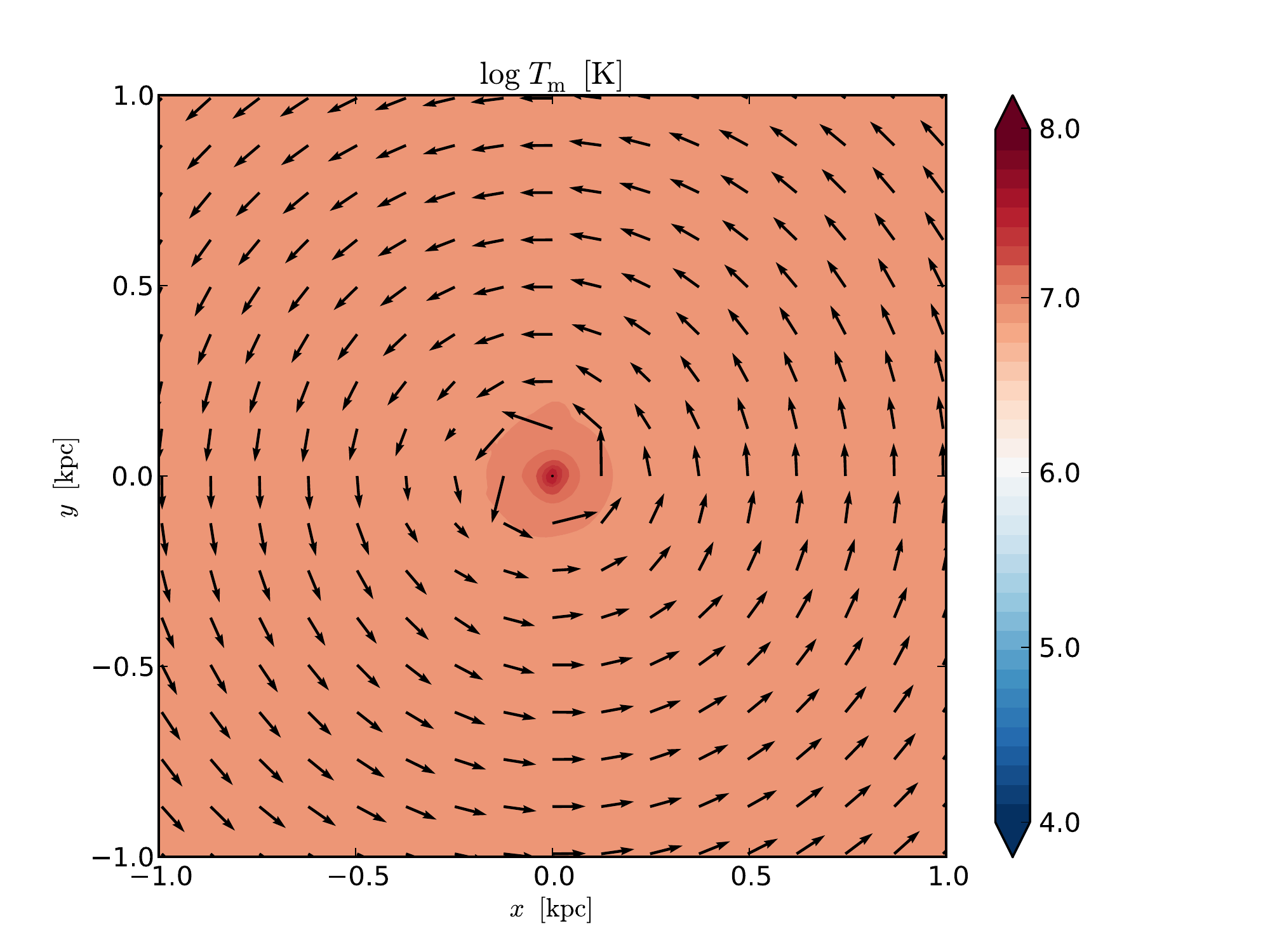}}
    \caption{Adiabatic accretion with $e_{\rm rot}=0.3$: $x$-$y$ rotational plane cross-section (2 kpc$^2$) of the mass-weighted temperature at final time; the normalization of the velocity field is $2000$ km s$^{-1}$ (such unit arrow is 1/8 of the image width). The gas progressively circularizes toward the equatorial plane, stifling accretion. Within the Bondi radius, the pressure-supported toroidal structure is variable, experiencing recurrent mild expansions and contractions, thus forming the pinwheel configuration. 
    } 
    \label{f:pure_T}
     \end{center}
\end{figure}


\begin{figure} 
    \begin{center}
       \subfigure{\includegraphics[scale=0.28]{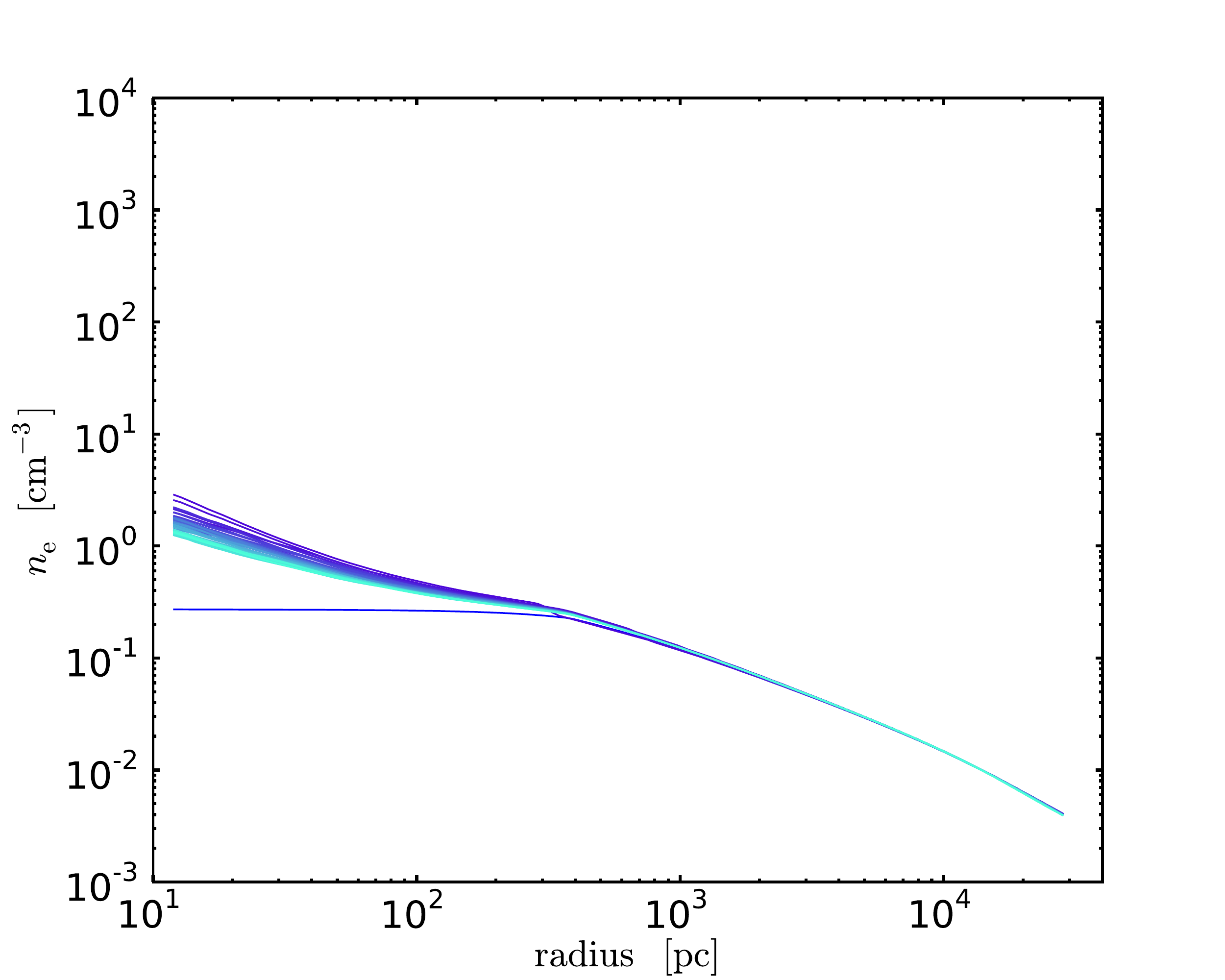}}
       \subfigure{\includegraphics[scale=0.28]{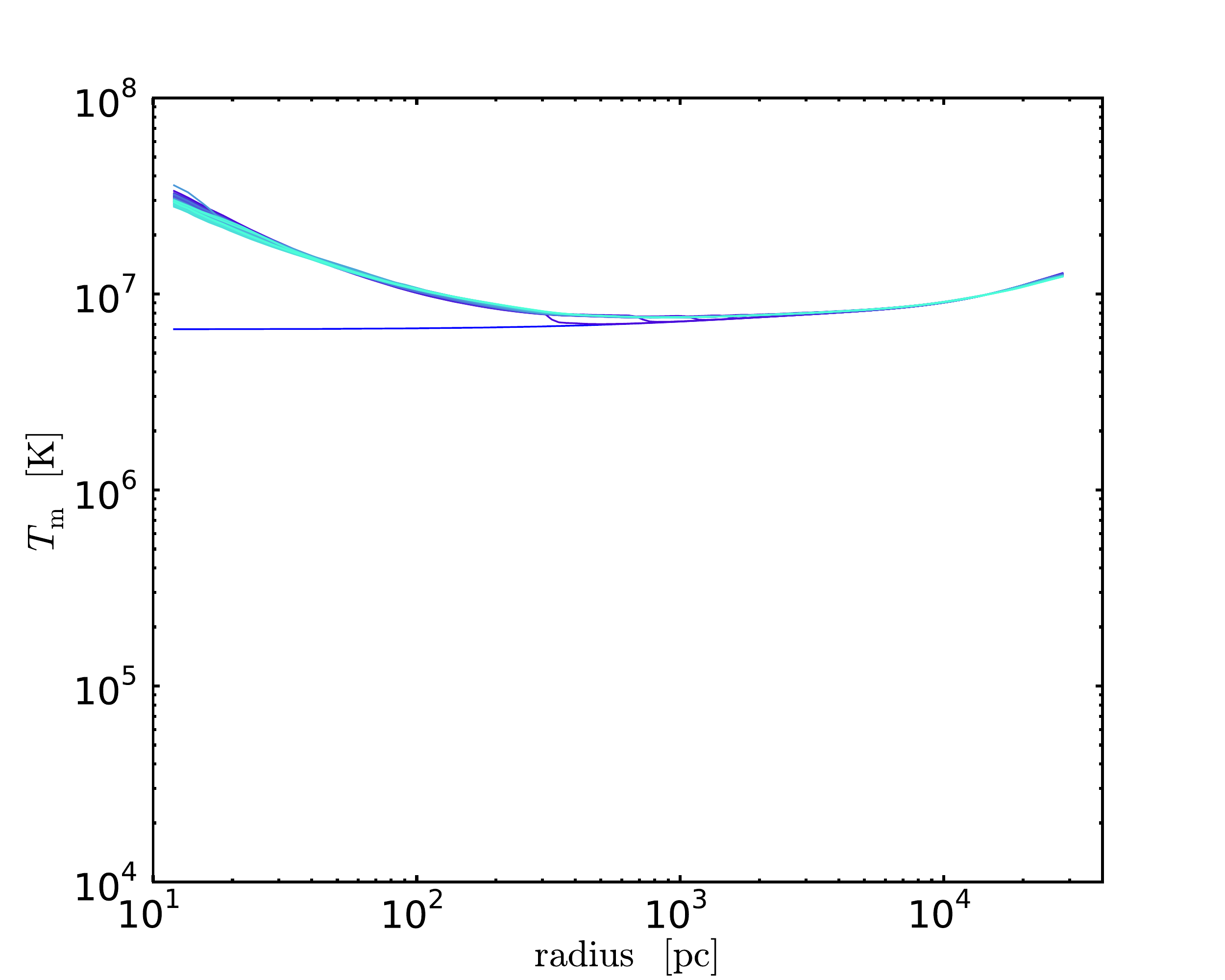}}
       \caption{Adiabatic accretion with $e_{\rm rot}=0.3$: evolution of the mass-weighted electron density (top; $n_e\simeq \rho/1.93\times10^{-24}$ cm$^{-3}$) and temperature (bottom) radial profiles, sampled every 1 Myr (from dark blue to cyan). The cuspy profiles are similar to the classic Bondi solution (G13), smoothly joining the galactic gradients at large $r$.}
       \label{f:pure_prof}
     \end{center}
\end{figure}  


\begin{figure} 
      \centering
      \subfigure{\includegraphics[scale=0.4]{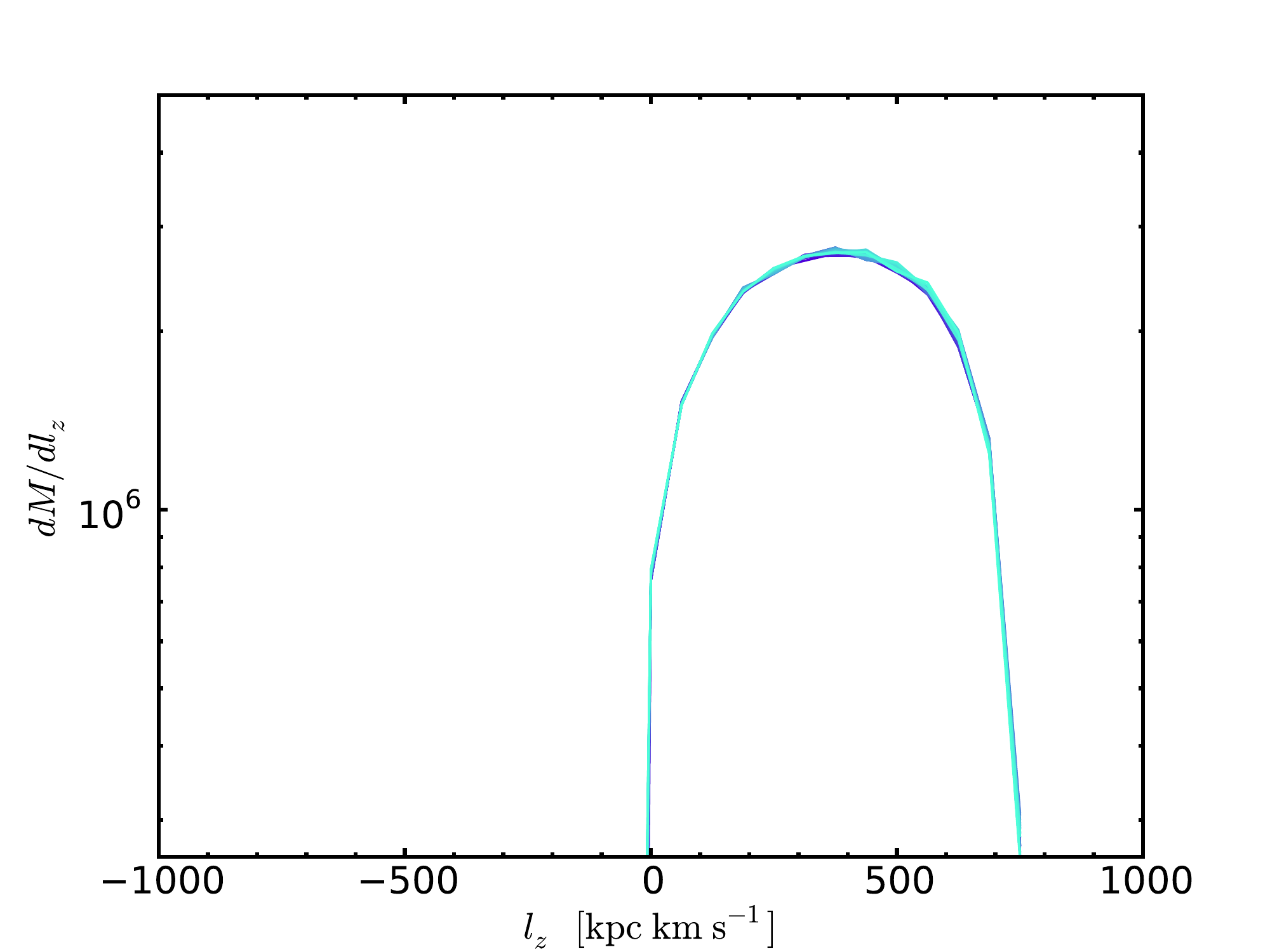}}
      \caption{Adiabatic accretion with $e_{\rm rot}=0.3$: mass `probability' distribution function (PDF) per bin of specific angular momentum along the rotation axis $z$ ($M_\odot$ per kpc km s$^{-1}$), for the gas within $r<8$ kpc. The color coding is the same as for the radial profiles (Fig.~\ref{f:pure_prof}), covering 40 Myr. With no additional physics, the system overall preserves the initial angular momentum distribution (coherent counter-clockwise rotation).
                 }
      \label{f:pure_lz}    
\end{figure}
 
\noindent
The suppression of accretion is related to the formation of a central toroidal structure, where gas with relatively high angular momentum circularizes, further blocking part of the inflow solid angle.
The hot gas can only accrete along the polar funnel perpendicular to the equatorial ($x$-$y$) plane.
The circulation within the toroidal region is variable, as reflected by the mild oscillations in $\dot M_\bullet$.
The configuration is partially unstable: the polar flow is inflowing, while the equatorial region is partially outflowing or circulating, leading to recurrent expansions and contractions (Fig.~\ref{f:pure_T}).

The `pinwheel' and toroidal configuration, as well as the suppression of the accretion rate, 
are consistent with the results of \citet{Proga:2003} and \citet{Krumholz:2005}, 
although our flow is embedded in galactic gradients. For low or moderate vorticity as in our typical system (hot plasma in most galaxies has subsonic and sub-Keplerian velocities\footnote{If the gas velocities are super Keplerian near $r_{\rm B}$, accretion is driven by the action of shock dissipation allowing the gas to become bound; in this regime, the higher the vorticity, the lower $\dot M_\bullet$, typically with a very small and stable torus
(cf.~\citealt{Krumholz:2005}).}), 
the accretion rate is weakly dependent on the initial angular momentum (Fig.~\ref{f:pure_mdot}, bottom), since the suppression in the final stage is related to the geometrically thick toroidal structure, which eventually builds up with similar shape.
Its typical radius is the BH influence radius $\sim r_{\rm B}$ with height $H\sim r_{B}$ -- for a Keplerian disk $H=c_{\rm s}/\omega=(r^3/r_{\rm B})^{1/2}$.
The polar funnel can be thus approximated as a cone with half-opening angle $\theta\sim \pi/4$, allowing gas accretion within a solid angle $\Omega=2\pi(1-\cos\theta)\simeq1.84$. Considering the two cones, the funnel has $\Omega\sim1/3$ of the spherical solid angle, broadly consistent with the simulated $\dot M_\bullet$ suppression (Fig.~\ref{f:pure_mdot}, top).

\subsection{Radial profiles and $l_z$ distribution} \label{s:adi_prof_pdf}
\noindent
In Fig.~\ref{f:pure_prof}, we show the mass-weighted radial profiles of electron gas density $n_{\rm e}$ (top) and temperature $T_{\rm m}$ (bottom), sampled every 1 Myr (dark blue to cyan color).
The cuspy profiles are similar to the classic Bondi solution (e.g., $T\propto r^{-1}$; G13), smoothly joining the galactic gradients at large radii. The slight central $n_{\rm e}$ decrease partly occurs
due to the variable boundary conditions near $r_{\rm B}$, and partly due to rotation.
Since the hot flow is mainly dominated by pressure, the toroidal region has large height ($H=c_{\rm s}/\omega$)
and is smooth, without a net demarcation line as in the radiative run forming a cold thin disk (\S\ref{s:cool}).      
The X-ray emission-weighted profiles (not shown) are not dissimilar because of the absence of the cold phase. Therefore, X-ray observations would see peaked gas temperature in the nucleus of the galaxy, if the hot mode is the currently dominant regime of accretion (typically occurring after AGN feedback has overheated the system; see \S\ref{s:disk}).

In Fig.~\ref{f:pure_lz}, we present another important diagnostic tool: the mass distribution function (PDF) of specific angular momentum along the rotation axis $l_z$, during the 40 Myr evolution. Since the model has no heating, cooling, or turbulence, and the accretion is substantially inhibited, the system overall conserves the angular momentum distribution. Notice that the PDF of $l_z$ has only positive values, the mark of coherent counter-clockwise rotation (the right tail is decreasing since we consider the gas within $r<8$ kpc).
Reshaping the angular momentum distribution via other physical processes is crucial to trigger boosted accretion.

\section[]{Accretion with cooling}  \label{s:cool}


\begin{figure}
      \begin{center}
      \subfigure{\includegraphics[scale=0.31]{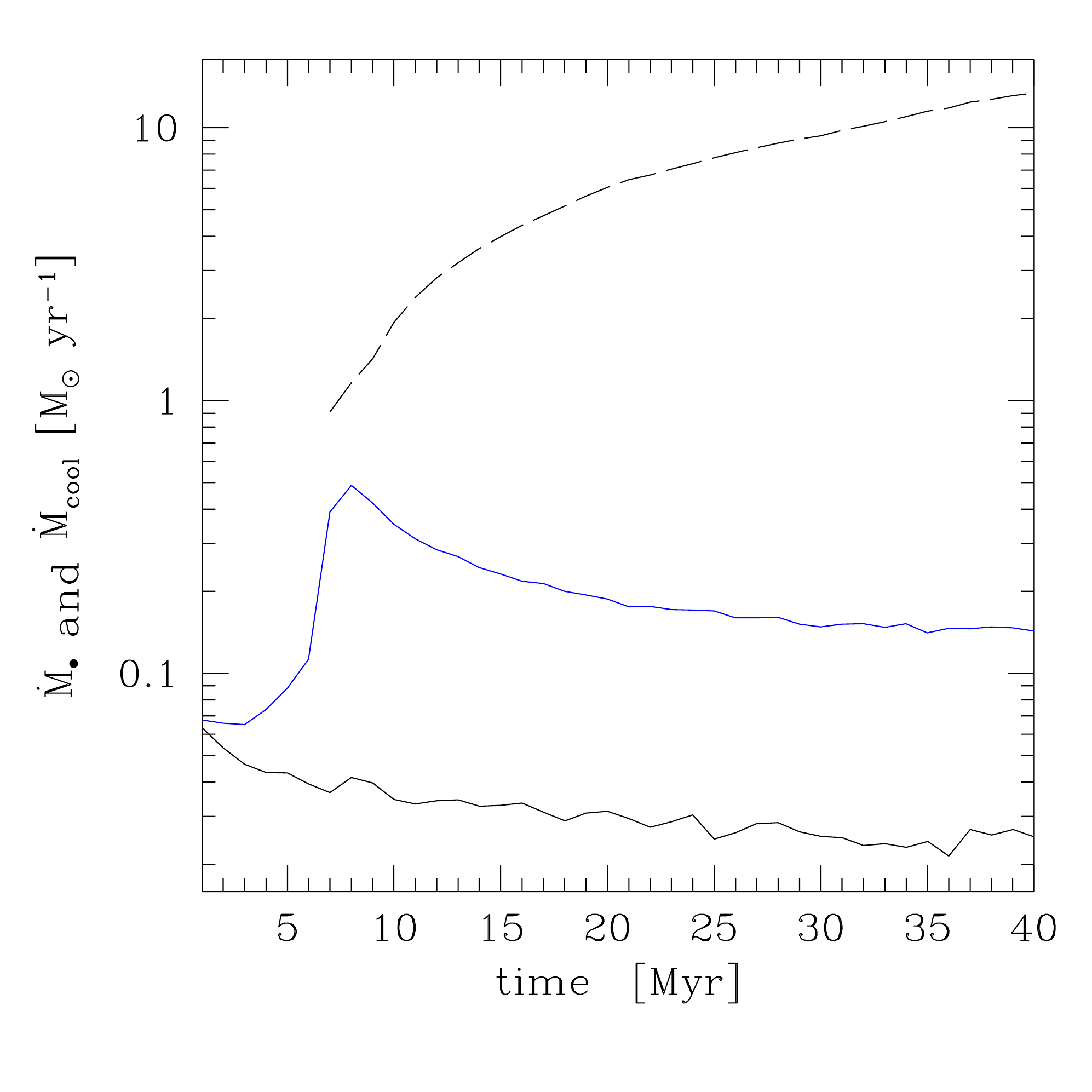}}
      \subfigure{\includegraphics[scale=0.31]{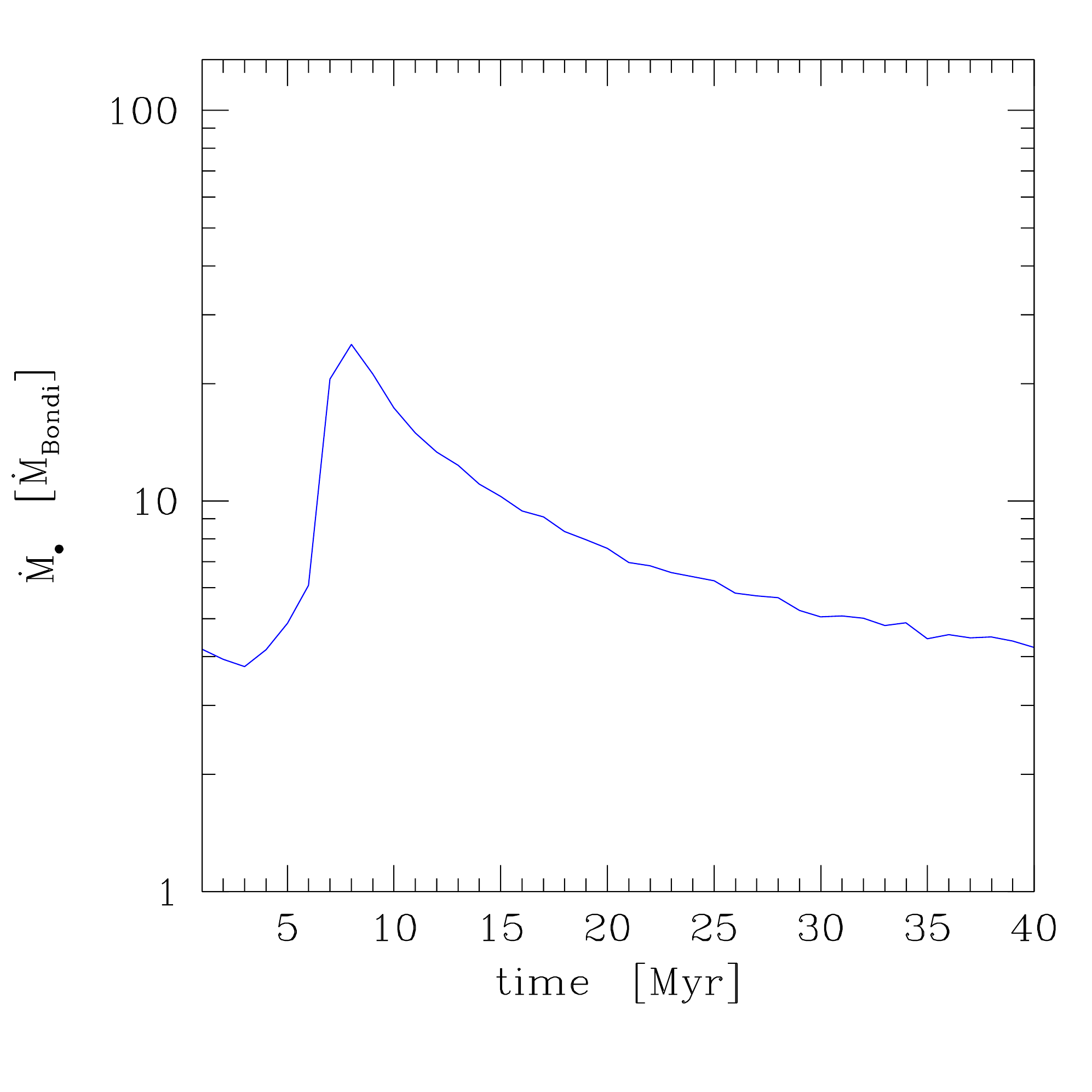}}
      \end{center}
      \caption{Accretion with cooling and $e_{\rm rot} = 0.3$ (blue): evolution of the accretion rate (physical and normalized to the kpc-scale, runtime Bondi rate -- top and bottom panel, respectively). 
      The dashed line is the average cooling rate (related to the gas with $T<10^5$ K). Solid black line is the adiabatic rotating model (\S\ref{s:adi}).
      Despite the substantial cooling rates, $\dot M_{\rm cool}\sim 15\ \msun$ yr$^{-1}$, the final accretion rate is two orders of magnitude lower because of the formation of a rotationally-supported thin disk. }
      \label{f:cool_mdot}
\end{figure}  


\begin{figure} 
     \subfigure{\includegraphics[scale=0.432]{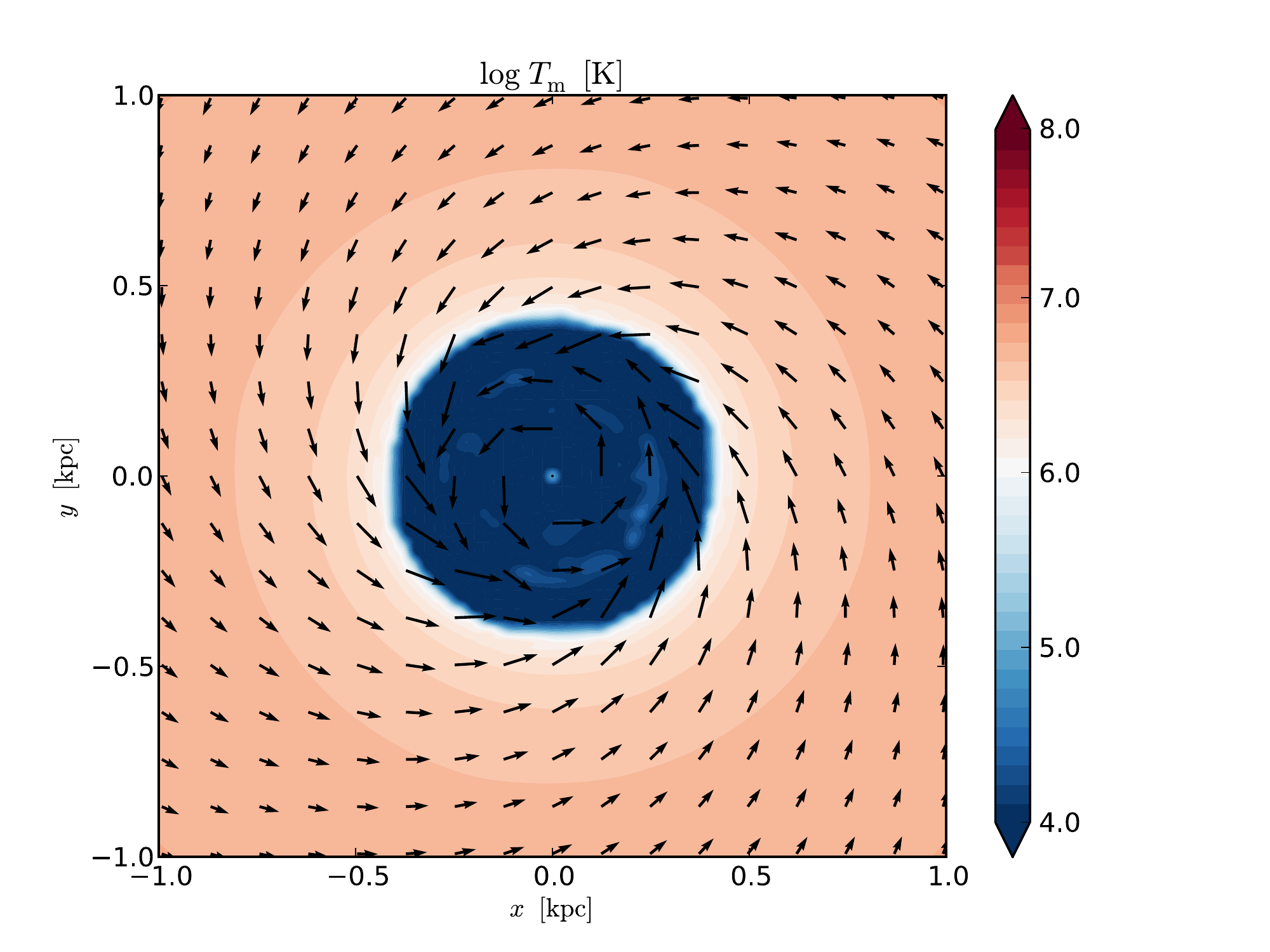}}
     \subfigure{\includegraphics[scale=0.432]{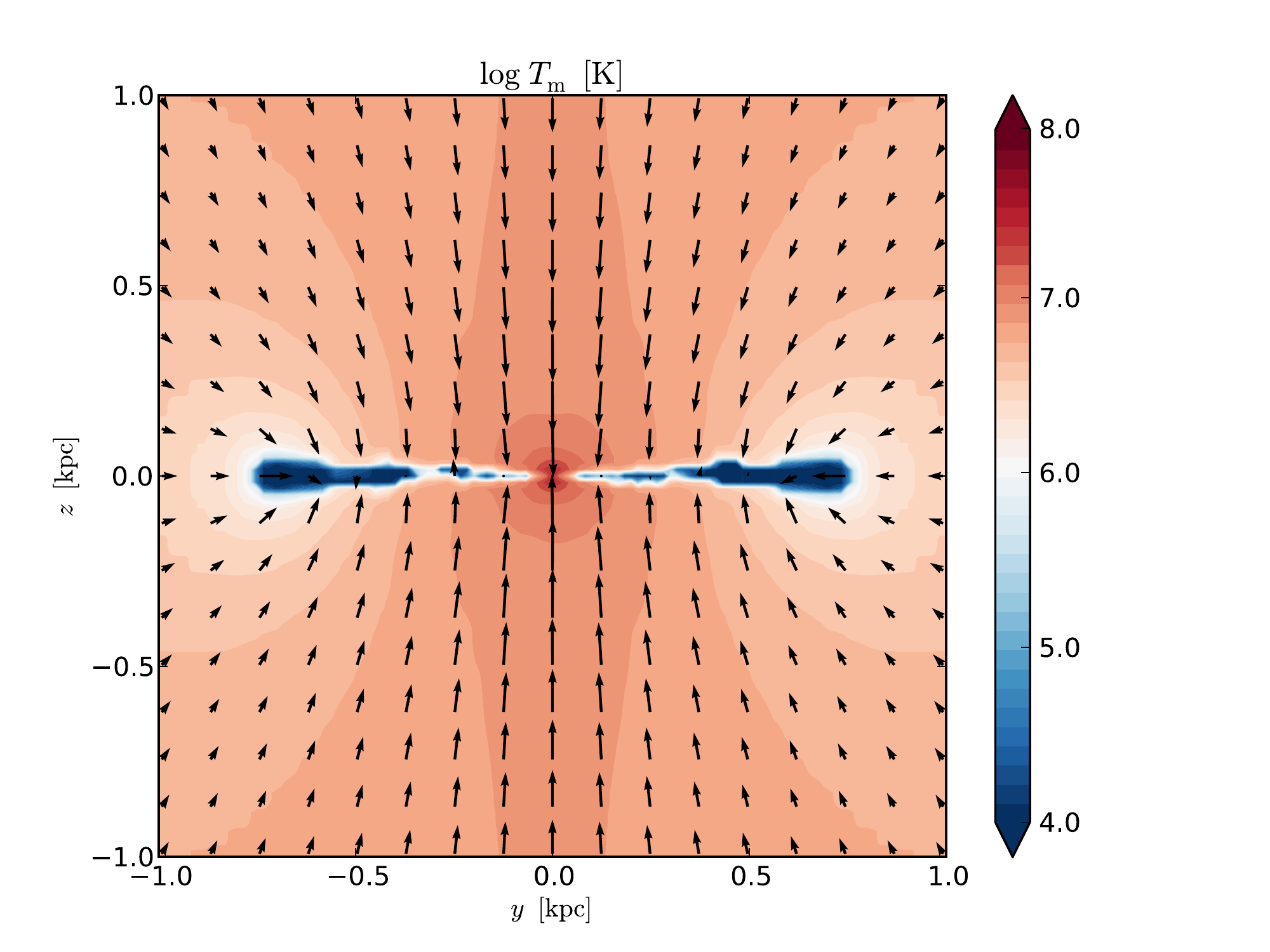}} 
     \caption{Accretion with cooling and $e_{\rm rot} = 0.3$: temperature cross-sections (2x2 kpc$^2$) through $z=0$ (top) and $x=0$ (bottom), after 2 and 4 $t_{\rm cool}$, respectively. 
     The velocity field normalization is $6000$ km s$^{-1}$ (unit arrow is 1/8 of the image width). The maps show the formation of a symmetric cold thin disk, which quickly condenses out of the hot phase and inhibits accretion.}
         \label{f:cool_T}
\end{figure}  

      
\begin{figure}
      \begin{center}
      \subfigure{\includegraphics[scale=0.28]{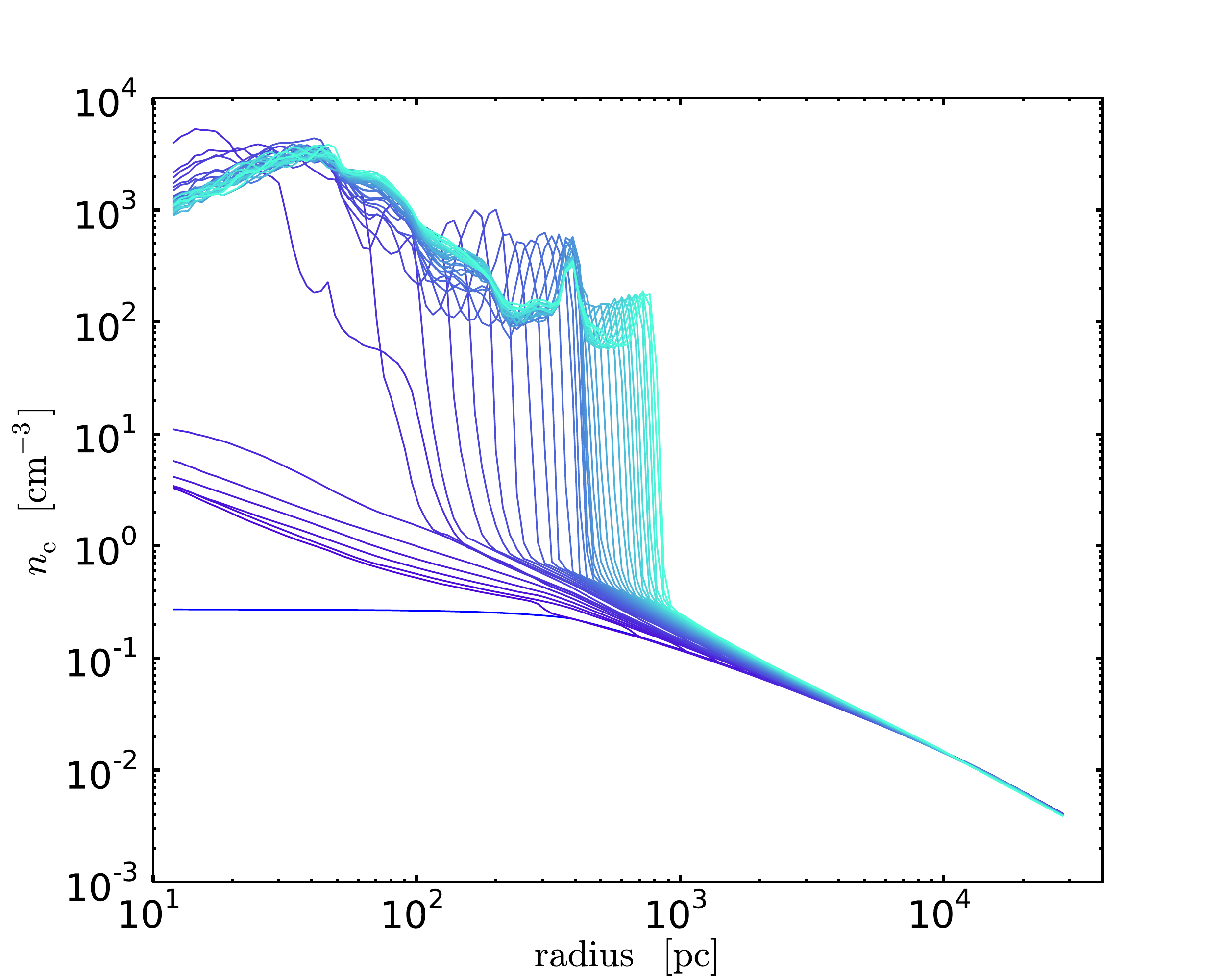}}
      \subfigure{\includegraphics[scale=0.28]{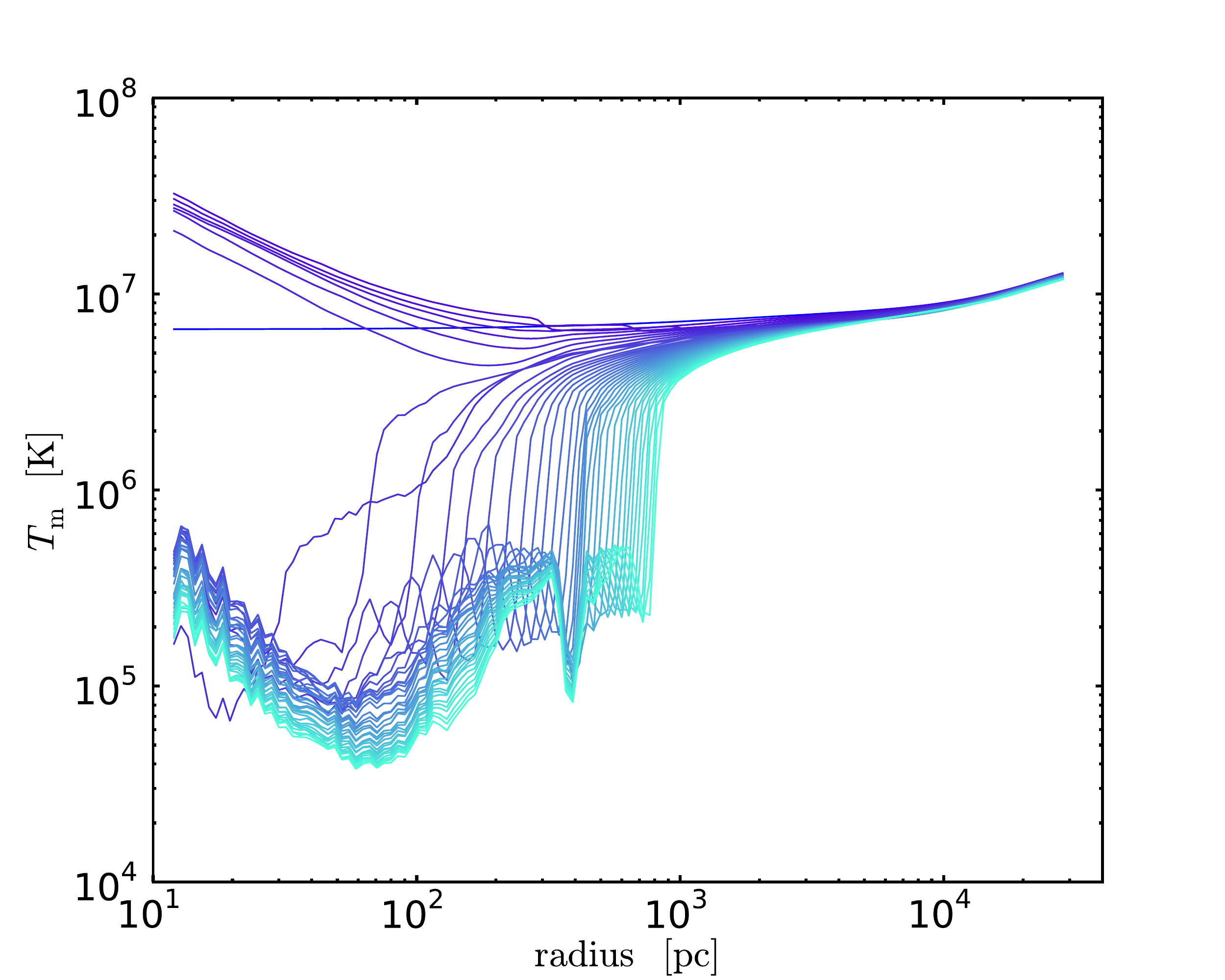}}
      \subfigure{\includegraphics[scale=0.28]{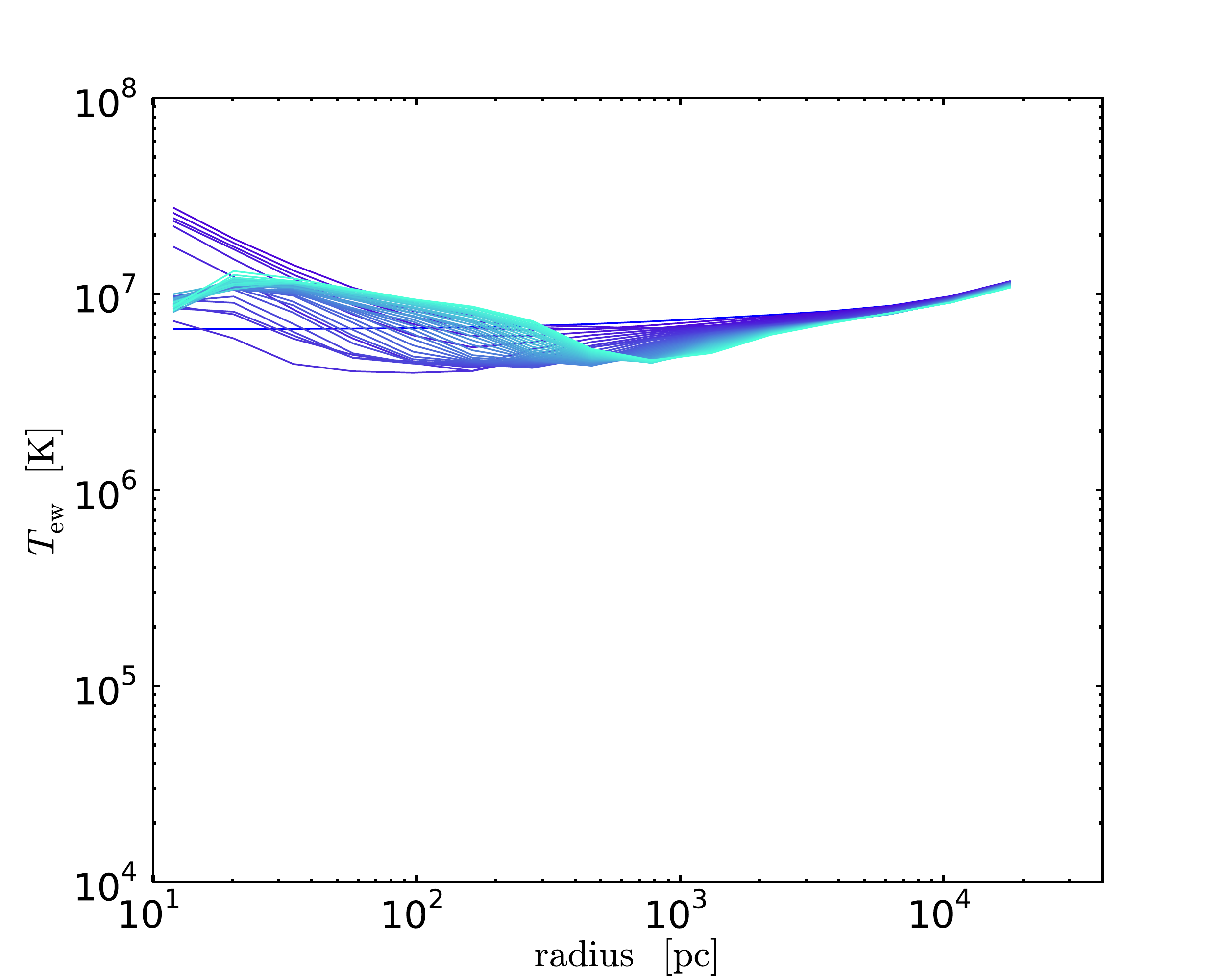}} 
     \end{center}
      \caption{Accretion with cooling and $e_{\rm rot} = 0.3$: 3D mass- and emission-weighted radial profiles of density and temperature (cf.~Fig.~\ref{f:pure_prof}). The $T_{\rm ew}$ profile has X-ray threshold of 0.3 keV and is computed in larger radial bins (emulating a {\it Chandra} observation). The cold thin disk keeps growing via condensation.
      The X-ray temperature is overall insensitive to the presence of the cold disk.
       \label{f:cool_prof}}      
\end{figure}  


\begin{figure} 
      \centering
      \subfigure{\includegraphics[scale=0.4]{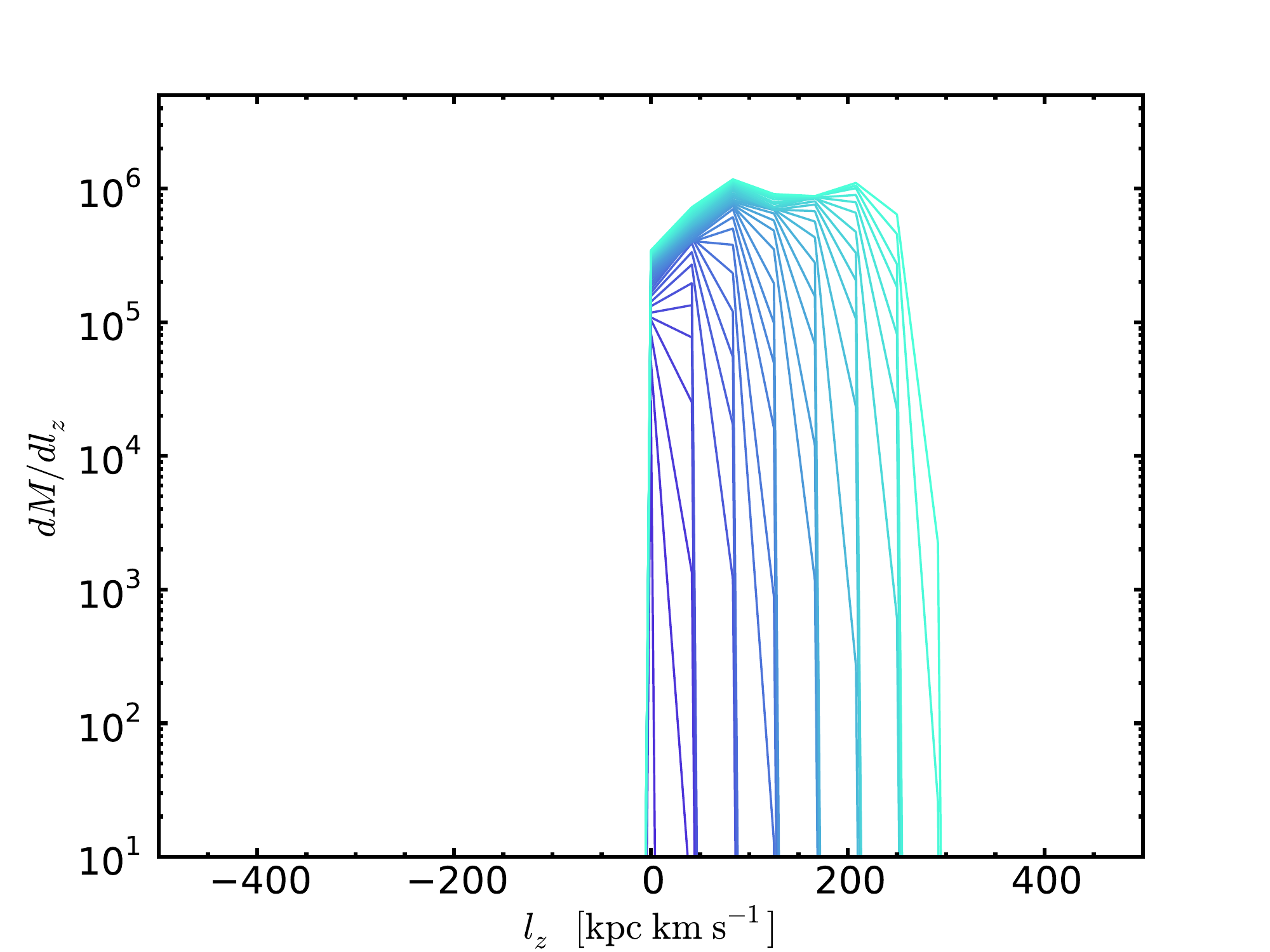}}
      \caption{Accretion with cooling and $e_{\rm rot} = 0.3$: mass PDF per bin of specific angular momentum $l_z$, for the cold phase (full evolution from blue to cyan lines). Cold gas is related to $T<10^5$\,K.
      The PDF of the hot phase (not shown) is analogous to that in Fig.~\ref{f:pure_lz}, albeit shifting by $\sim\,$15 percent to higher normalization and $l_z$. The cold phase emerges out of this distribution, adding during time higher $l_z$ gas, and thus widening the PDF.
                 }
      \label{f:cool_lz}    
\end{figure} 

\noindent      
In the next model, the reference rotating accretion flow experiences cooling due to radiation (mainly Bremsstrahlung in the X-ray band and line cooling below 1 keV) with emissivity $\mathcal{L}= n_{\rm e}n_{\rm i}\,\Lambda(T)$. 
No source of heating or turbulence affects the dynamics.
      
\subsection[]{Accretion rate} \label{s:cool_mdot}
\noindent
The central result is the decoupling between the accretion rate and the cooling rate (Fig.~\ref{f:cool_mdot}, top), no more closely tracking each other as in the spherically symmetric cooling flow (G13; Sec.~4).
The dense gas quickly loses pressure support via radiative emission (initial minimum $t_{\rm cool}$ is $\approx$\,8 Myr)
and forms a rotationally-supported thin disk.
The fast condensation affects first the inner and denser gas with lower angular momentum, 
which can be quickly accreted ($\dot M_\bullet\simeq0.5\ \msun\ {\rm yr^{-1}}$).
After 10 Myr, the cold phase arises from the hot gas with high angular momentum. Falling from $r> r_{\rm B}$ toward the center, this cooling gas rapidly increases the rotational velocity and circularizes, damping the accretion rate.
At final time, $\dot M_\bullet\simeq0.15\ \msun\ {\rm yr^{-1}}$ (blue) and the cooling rate\footnote{The cooling rate is computed as $\Delta M_{\rm cold}/\Delta t$, where $M_{\rm cold}$ is the cold gas mass in the box related to gas with $T<10^5$\, K, not yet sinked by the BH.} 
is $\dot M_{\rm cool}\simeq15\ \msun\ {\rm yr^{-1}}$ (dashed), a difference of two orders of magnitude. 
The normalized accretion rate is also low, roughly four times the Bondi rate at the kpc scale.

Compared with the adiabatic run (solid black), the accretion rate is about $6\times$ higher, since the weakened pressure support increases the effective inflow, in particular along the polar region. 
The presence of a condensing hot halo thus alters the classic thin disk picture,
where the cold gas is the only entity. 

\subsection[]{Dynamics} \label{s:cool_dyn}
\noindent
Figure \ref{f:cool_T} shows the formation of the symmetric thin disk in more detail.
The hot gas loses internal energy and thus pressure, but it is still balanced by rotation along $R$, leading to the infall toward the direction perpendicular to the $x$-$y$ plane, where the low-entropy gas finds a new rotational equilibrium. Because of the  multiphase stratification and transonic inflow, the disk experiences shocks (see, e.g., \citealt{Tejeda:2012}) 
and hydrodynamical instabilities, such as Kelvin-Helmoltz and Rayleigh-Taylor.
The overall picture is very different from a spherically symmetric cooling flow (G13), since $\dot M_\bullet \ll \dot M_{\rm cool}$. Along the polar funnel, the transonic flow is not inhibited by rotation and gas can quickly accrete. 

Massive galaxies which are dominated by cooling (due to weak or no heating)
are thus expected to show a rotationally supported cold disk, arising from the fast spin-up of the gas.
The X-ray ellipticity greater than that of the stellar component in the inner 1 kpc may be a sign of this phenomenon 
(e.g., NGC 4649), albeit moderate turbulence is still required to avoid markedly flat isophotes inconsistent with observations (\citealt{Diehl:2007}; see also \citealt{Brighenti:2000_rot}).

\subsection[]{Radial profiles and $l_z$ distribution} \label{s:cool_prof}
\noindent
The mass-weighted radial profiles in Fig.~\ref{f:cool_prof} reveal the presence of the highly dense ($10^{3}$ cm$^{-3}$) and cold thin disk, which has condensed out of the hot atmosphere. 
At variance with $T_{\rm m}$, 
the X-ray temperature ($T$ weighted by X-ray emissivity using {\it Chandra} sensitivity; bottom panel) 
only has a mild decline from large to 1 kpc radius, where it starts to stabilize around $10^7$ K.
The cold disk keeps growing through time because of the continuous condensation of higher angular momentum gas.
The rotating structure is thin, with slightly expanding dense lobes where the warm gas is still condensing
($H=c_{\rm s}/\omega$; Fig.~\ref{f:cool_T}), which can be tracked in the radial profiles.

The disk extends to $\sim\,$1 kpc after 40 Myr of evolution. 
Its growth diminishes as the cooling rate saturates
(Fig.~\ref{f:cool_mdot}). 
During periods of weak heating and turbulence, this could be a common regime, 
facilitating the development of
the rotating cold disks observed in many massive elliptical galaxies (e.g., NGC 6868, NGC 7049, and NGC 4261 -- \citealt{Werner:2014}; see also
\citealt{Mathews:2003,Young:2011,Alatalo:2013}). 
The retrieved size of the disk (e.g., via ALMA or {\it Herschel}) may point out how long the cooling-dominated phase has lasted, setting constraints on the AGN feedback duty cycle
(albeit we note that a clumpy rotational structure can be present during CCA and significant feedback; \S\ref{s:heat}).

Figure \ref{f:cool_lz} shows the $l_z$ distribution of the cold phase.
The PDF of the hot phase is analogous to that in Fig.~\ref{f:pure_lz}, shifting by $\sim\,$15 percent to higher normalization and $l_z$ due to the large-scale inflow of gas losing pressure. The cold phase emerges out of the hot gas distribution, progressively accumulating higher $l_z$ gas coming from larger radii, and thus widening the PDF while the disk grows (blue to cyan). As in the previous adiabatic run, $l_z$ is only positive, the mark of coherent rotation, albeit now in the form of a cold thin disk. Again, such a distribution of angular momentum implies substantially suppressed $\dot M_\bullet$.

\section[]{Adiabatic accretion with turbulence} \label{s:stir}


\begin{figure} 
      \begin{center}
      \subfigure{\includegraphics[scale=0.31]{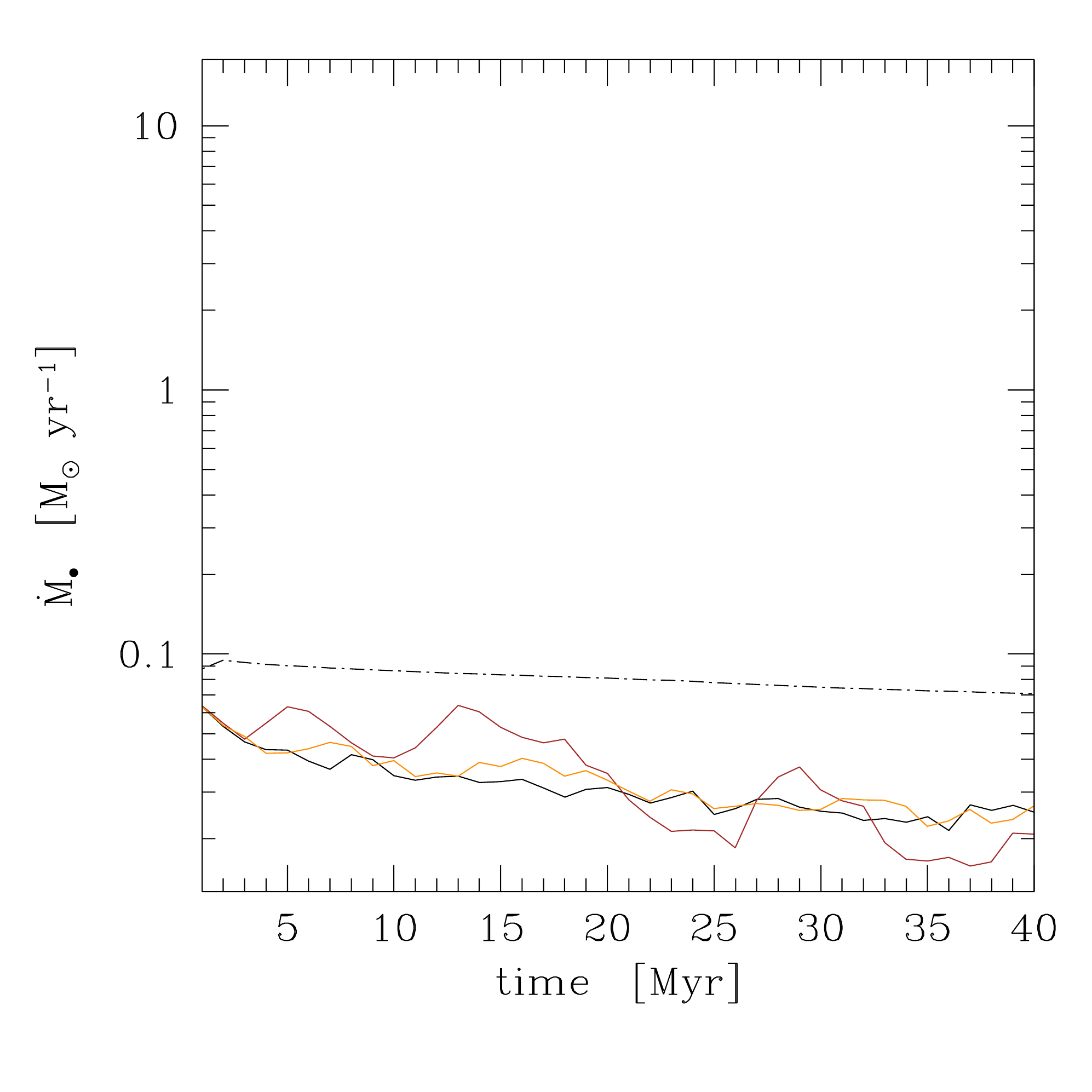}}
      \end{center}
      \caption{Adiabatic accretion with subsonic turbulence and $e_{\rm rot}=0.3$ (brown: ${\rm Ma}$\,$\sim$\,0.35, i.e., ${\rm Ta_t}$\,$\sim$\,0.75; orange: 1/4 lower Mach number, i.e., ${\rm Ta_t}$\,$\sim$\,3): evolution of the accretion rate.
      The accretion rate is suppressed by a factor $\sim\,$3 compared with the nonrotating model (dot-dashed), as in the purely rotating run (solid black; \S\ref{s:adi}), since turbulence still induces local (but not global) vorticity. The chaotic eddies generate however higher variability, as long as ${\rm Ta_t < 1}$. In the opposite regime, rotation drives the dynamics. }    
      \label{f:stir_mdot}
\end{figure}  


\begin{figure} 
      \begin{center}
       \subfigure{\includegraphics[scale=0.45]{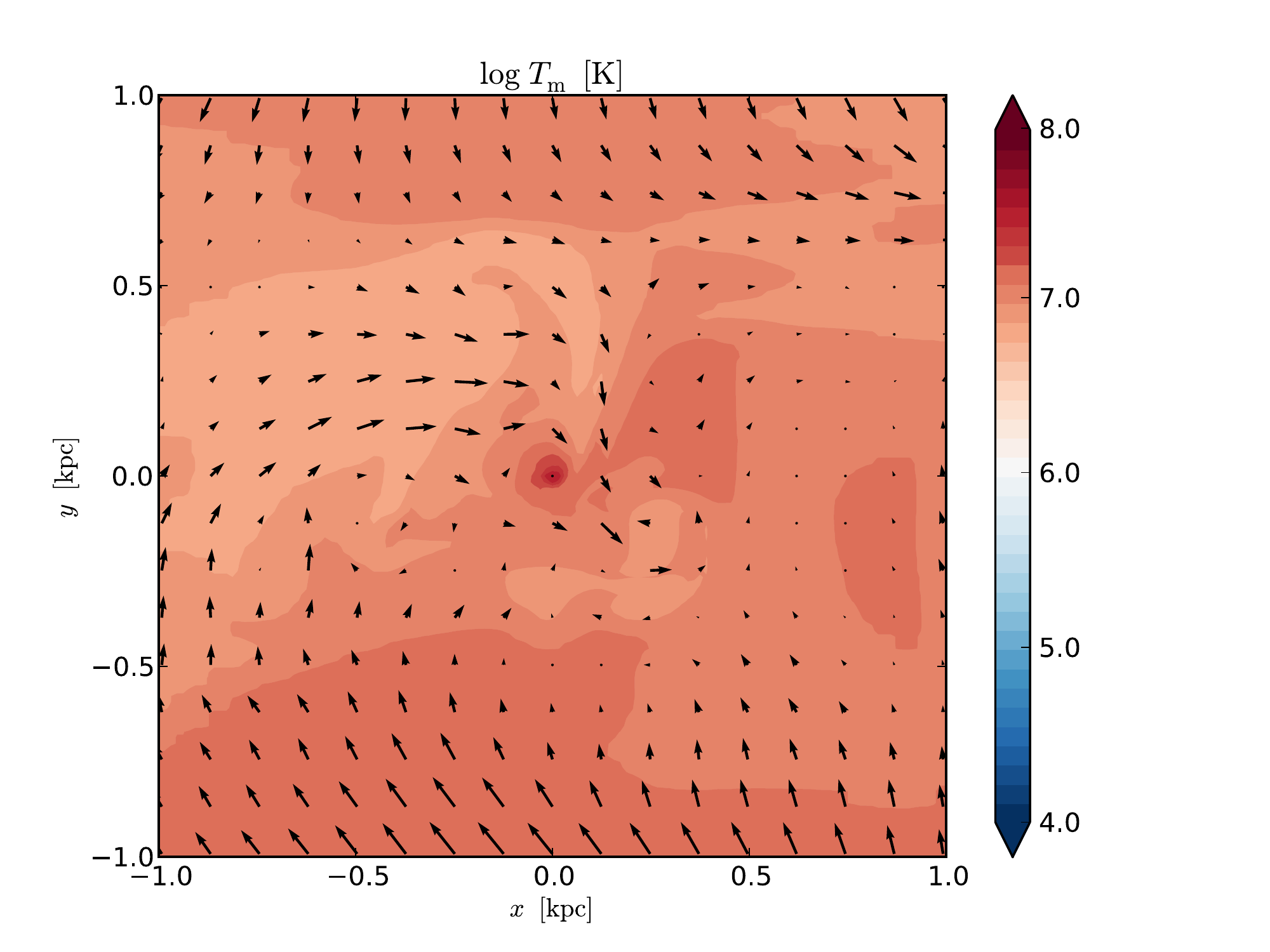}}
      \end{center}
      \caption{Adiabatic accretion with subsonic turbulence (reference ${\rm Ma}$\,$\sim$\,0.35) and $e_{\rm rot}=0.3$: mid-plane temperature cut (cf.~Fig.~\ref{f:pure_T}). The equatorial plane does not display coherent rotation, since the gas cannot properly circularize in the presence of significant turbulence ($\sigma_v > v_{\rm rot}$).
      The central vortical motions stifle again accretion.   
      \label{f:stir_T}}
\end{figure}   

\noindent
Cosmic systems are rarely (if ever) in perfect hydrostatic equilibrium. The hot gas in galaxies, groups, and clusters is continuously stirred by the action of AGN, supernovae feedback, galaxy motions, and mergers. 
We thus probe the effect of turbulence on the rotating and adiabatic flow, testing intrinsic\footnote{The rotational velocity can be cleanly removed calculating the mean in cylindrical coordinates, $v_{\rm rot} = -v_x \,y/R + v_y\, x/R$.} velocity dispersion in the range of $\sigma_v$\,$\sim$\,40\,-\,150 km s$^{-1}$.

\subsection[]{Accretion rate and dynamics} \label{s:stir_mdot}
\begin{figure} 
      \centering
      \subfigure{\includegraphics[scale=0.4]{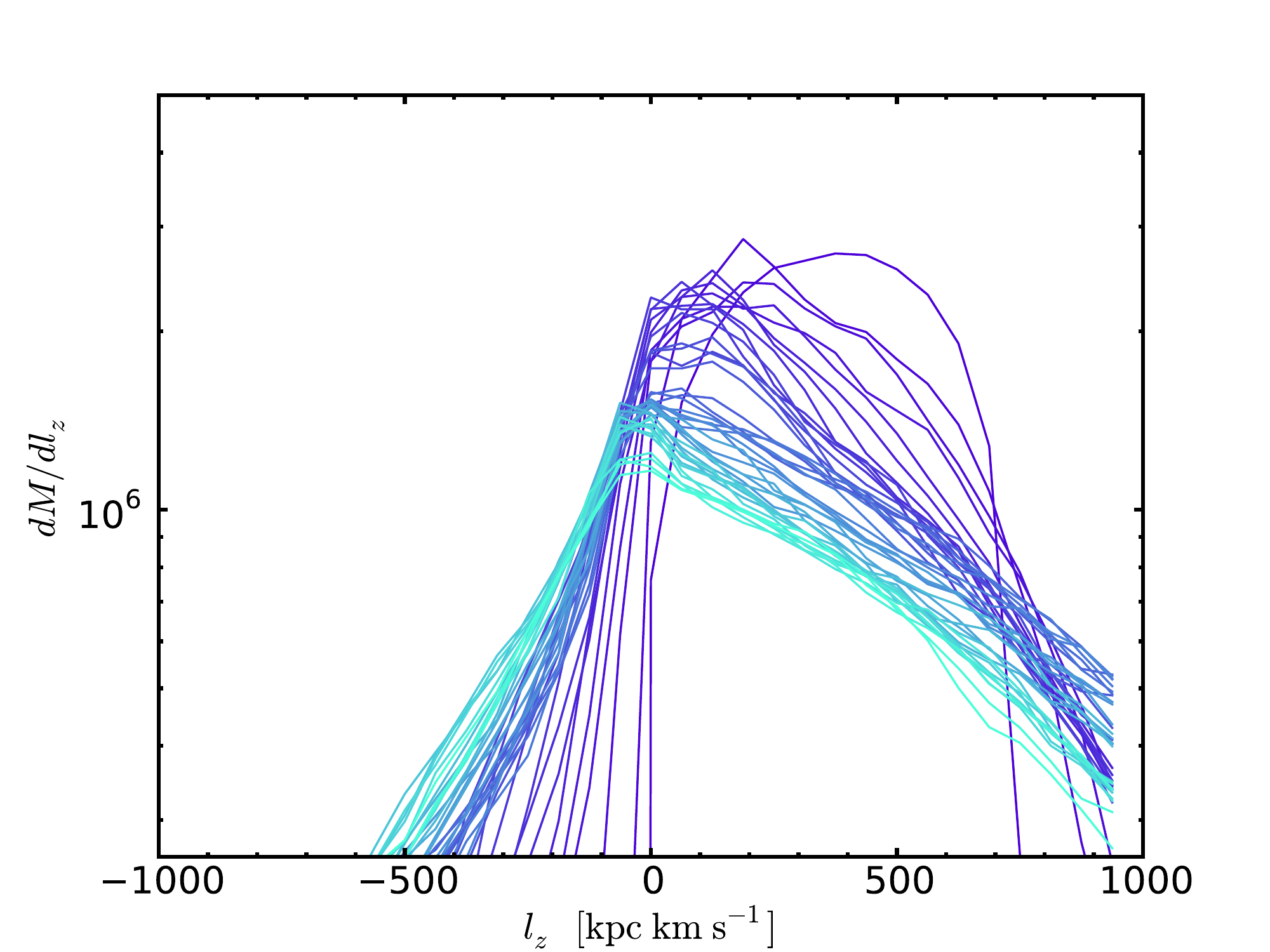}}
      \subfigure{\includegraphics[scale=0.4]{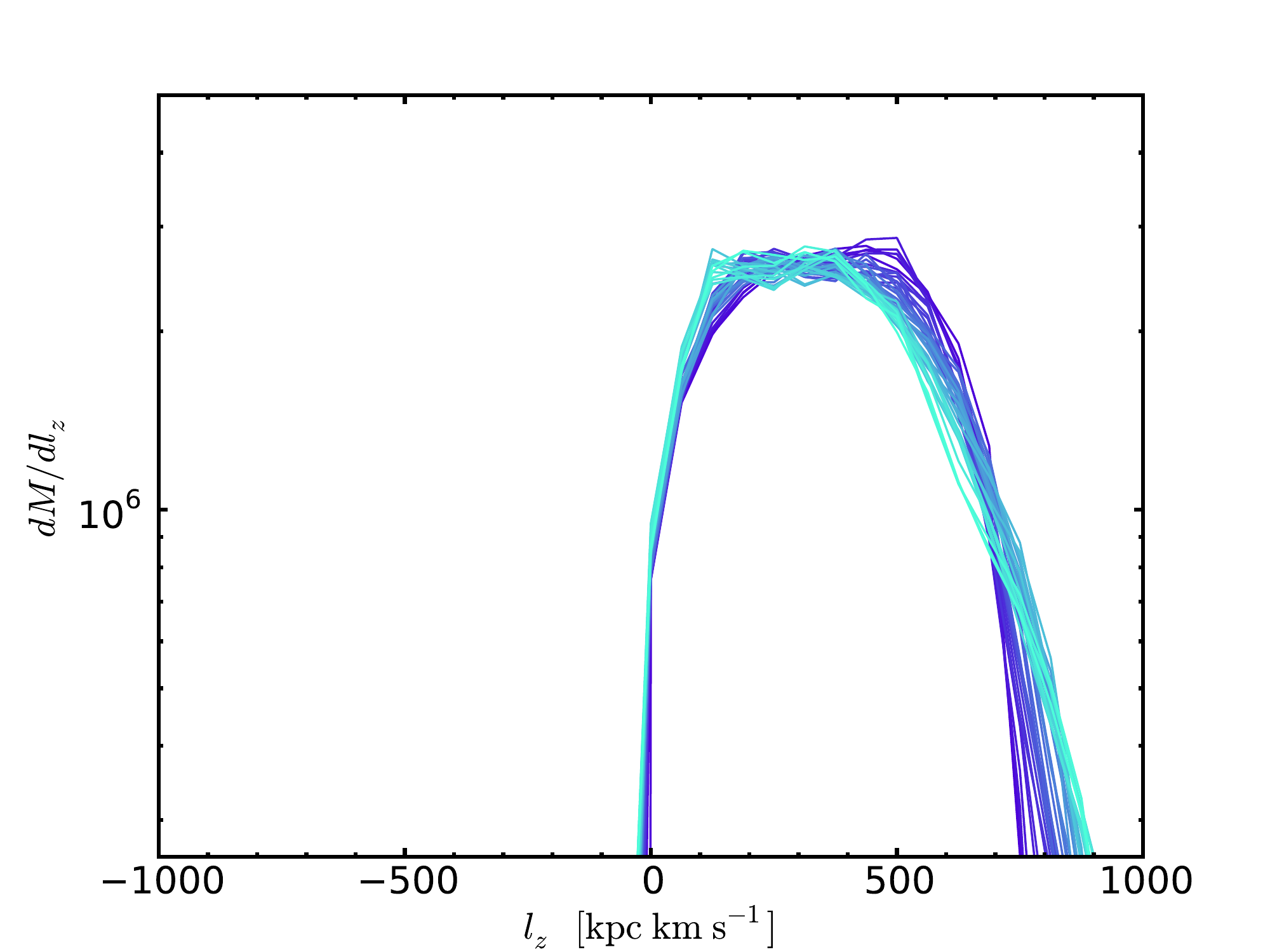}}
      \caption{Adiabatic accretion with subsonic turbulence (top: ${\rm Ma}$\,$\sim$\,0.35; bottom: 1/4 lower Mach number, i.e., ${\rm Ta_t}$\,$\sim$\,3)  and $e_{\rm rot} = 0.3$: mass PDF per bin of specific angular momentum $l_z$, for the gas within $r<8$ kpc (cf.~Fig.~\ref{f:pure_lz}). 
      If ${\rm Ta_t < 1}$, turbulence widens the PDF, inducing both retrograde and prograde motions.
      In the opposite regime, ${\rm Ta_t > 1}$, turbulence is too weak to induce retrograde motions: the underlying counter-clockwise rotation remains intact overall, with minor fluctuations superimposed.
                 }
      \label{f:stir_lz}    
\end{figure}

\noindent
Figure \ref{f:stir_mdot} shows the evolution of the accretion rate for the reference ${\rm Ma}$\,$\sim$\,0.35 (brown)
and for 1/4 lower Mach number (${\rm Ta_t}\sim3$; orange). In both runs, the suppression of $\dot M_\bullet$ is $\sim\,$1/3 compared with the nonrotating Bondi flow (dashed line; see also G13). Remarkably, the
suppression is analogous to that retrieved in the purely rotating run (black; \S\ref{s:adi}). 
While turbulence with $\sigma_v > v_{\rm rot}$ is able to disrupt the coherent rotation and to prevent the full circularization, its basic action is to transport momentum.
No total angular momentum is created, however, turbulence induces prograde or retrograde vorticity {\it locally} (Fig.~\ref{f:stir_lz}) via eddies generated during the Kolmogorov cascade. The (subsonic) turbulent accretion flow is thus analogous to a gradually varying rotating flow near the accretor (Fig.~\ref{f:stir_T}).
As discussed in $\S\ref{s:adi}$, for a pressure-supported and slowly rotating flow, the geometric funnel in which the gas can accrete is roughly invariant, linked to a suppression $\sim\,$1/3. The central spiraling motion is enhanced by the baroclinic instability because of the atmosphere stratification (\citealt{Krumholz:2006}). For subsonic turbulence, local vorticity is more relevant than the gas bulk motion relative to the accretor. 
For reference ${\rm Ma}$\,$\sim$\,0.35, the $\dot M_\bullet$ suppression related to the bulk motion (\citealt{Bondi:1944}) is just $\propto (1+M^2)^{-3/2}\simeq 0.84$. 

A difference between the turbulent run and the purely rotating, adiabatic flow is 
the increased $\dot M_\bullet$ variability.
The turbulent eddies randomly have low or high angular momentum, generating the peaks and valleys
observed in the accretion rate, which oscillate by a factor of $\sim\,$2. 
The key role of the local eddy is remarked by the turbulent run with no rotation presented in G13 (Sec.~6),
showing similarly suppressed $\dot M_\bullet$.
As ${\rm Ta_t > 1}$ (Fig.~\ref{f:stir_mdot}; orange), the flow is driven by coherent rotation, with only minor turbulent perturbations, thereby reverting to the evolution described in $\S\ref{s:adi}$.


\begin{figure}
      \begin{center}
      \subfigure{\includegraphics[scale=0.28]{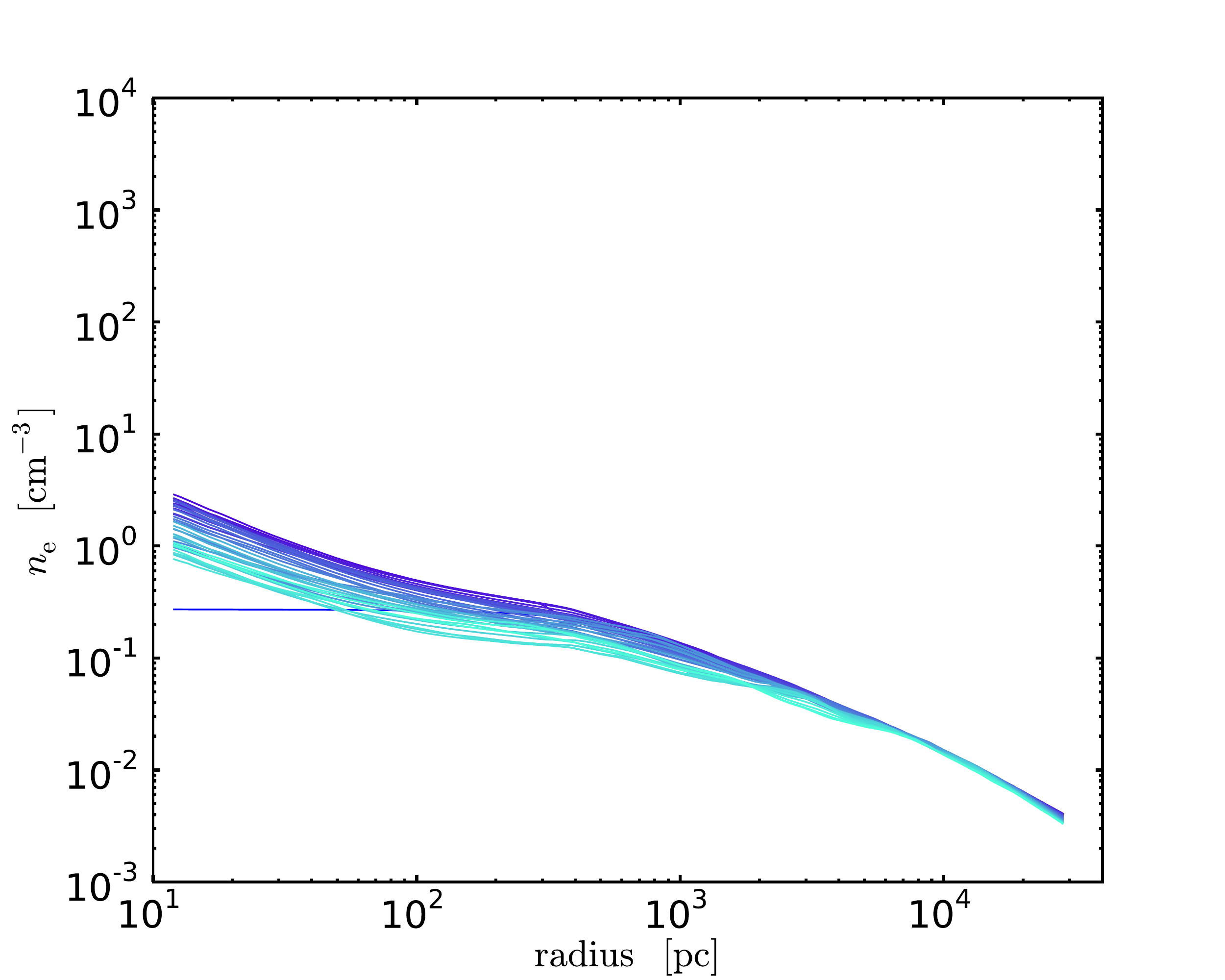}}
      \subfigure{\includegraphics[scale=0.28]{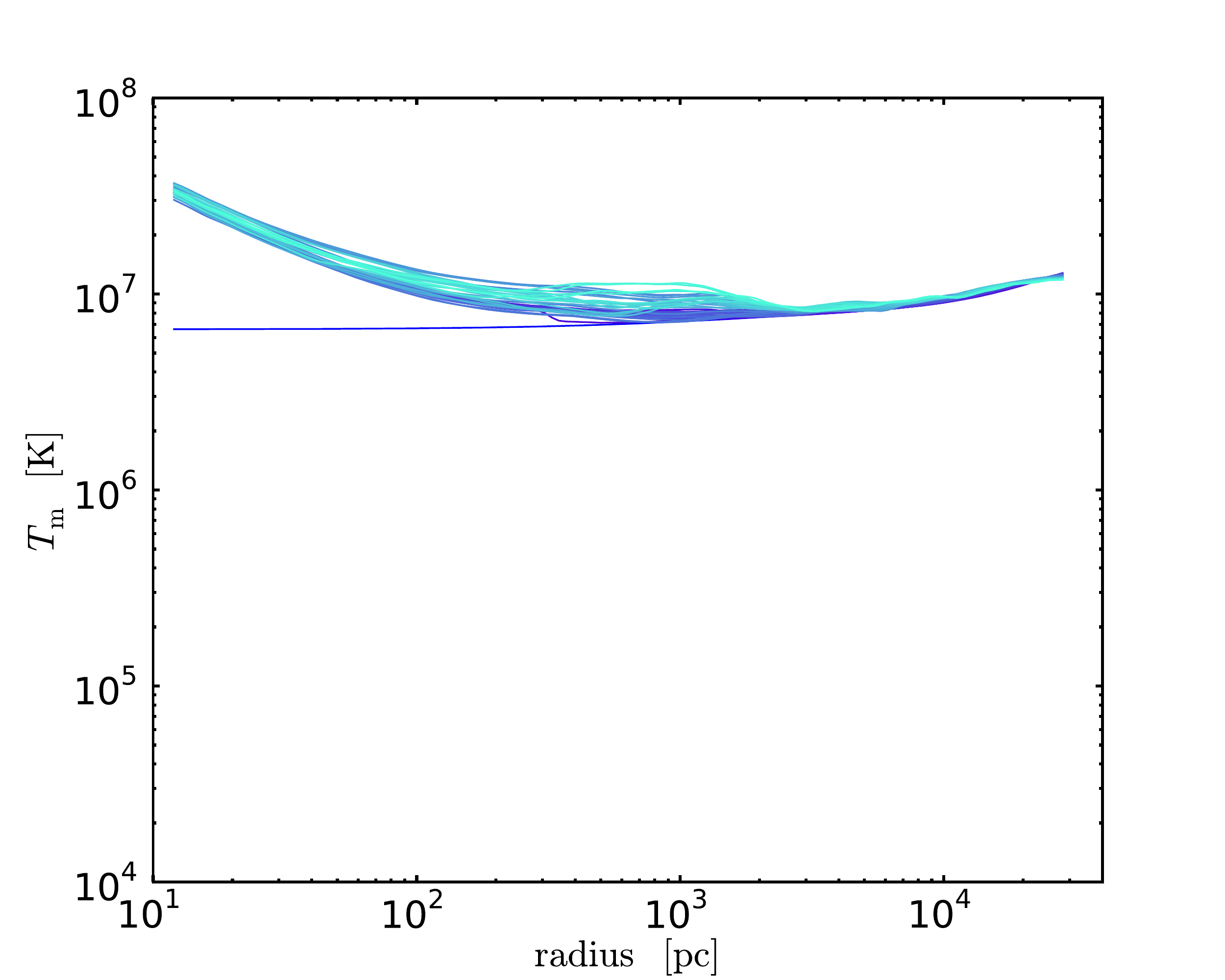}}
      \end{center}
      \caption{Adiabatic accretion with subsonic turbulence (${\rm Ma}$\,$\sim$\,0.35) and $e_{\rm rot}=0.3$: mass-weighted profiles of density and temperature (cf.~Fig.~\ref{f:pure_prof}). The profiles are cuspy, 
      as in the classic Bondi flow, albeit
      becoming progressively shallower because of turbulent diffusion (dissipational heating is negligible, as indicated by the large-scale $T$ profile).  
      \label{f:stir_prof}  }    
\end{figure}  

\subsection[]{Distribution of $l_z$ and radial profiles} \label{s:stir_prof}
\noindent
The importance of the threshold ${\rm Ta_t} \equiv v_{\rm rot}/\sigma_v \sim 1$ can be better appreciated in the distribution of angular momentum (Fig.~\ref{f:stir_lz}). If ${\rm Ta_t}  < 1$ (top), stirring
can substantially reshape PDF($l_z$).
Turbulence can be approximated as a diffusion process (with diffusivity $\sim\,$$\sigma_v L$; \citealt{Gaspari:2013_coma}), spreading linear and thus\footnote{The radial displacement has uniformly random distribution.} angular momentum (an effective viscosity). 
The initially peaked and solely positive $l_z$ distribution is progressively morphed into 
a broader PDF including negative values, with variance $\propto \sigma_v$.
The PDF is skewed toward the right tail because of the initial rotation.
In the $\sigma_v \gg v_{\rm rot}$ regime, the PDF would be symmetric around zero.

The end product of real viscosity is homogeneous velocity (Dirac delta PDF), while the steady state of turbulence is always local perturbations ($\propto\,\sigma_v$). Turbulent diffusion not only acts on momentum, but also on $K$, $\rho$, $T$, as shown by the gradually shallower radial profiles (Fig.~\ref{f:stir_prof}), and by the power spectrum analysis presented in \citet{Gaspari:2014_coma2}. The fact that turbulence mimics a transport mechanism, while
inducing significant local fluctuations,
is a key element to develop chaotic cold accretion (\S\ref{s:heat}).
The broadening of the angular momentum distribution alone does not stimulate boosted accretion. 
In contrast to the cold clouds experiencing major inelastic collisions, the hot diffuse halo is in global hydrostatic equilibrium
because of the pressure support, thereby the subsonic eddies do not mostly cancel angular momentum.

In the opposite regime ${\rm Ta_t} > 1$ (Fig.~\ref{f:stir_lz}, bottom), turbulence is too weak to induce retrograde motions: the underlying counter-clockwise $l_z$ distribution overall remains intact and the coherent rotation drives the dynamics. 
The reduced effective diffusivity is also evident in the radial profiles (not shown), where the central density decreases only to $n_{\rm e}\simeq1.5$ cm$^{-3}$.

\section[]{Accretion with cooling and turbulence}  \label{s:stir_cool}


\begin{figure} 
     \centering
     \subfigure{\includegraphics[scale=0.45]{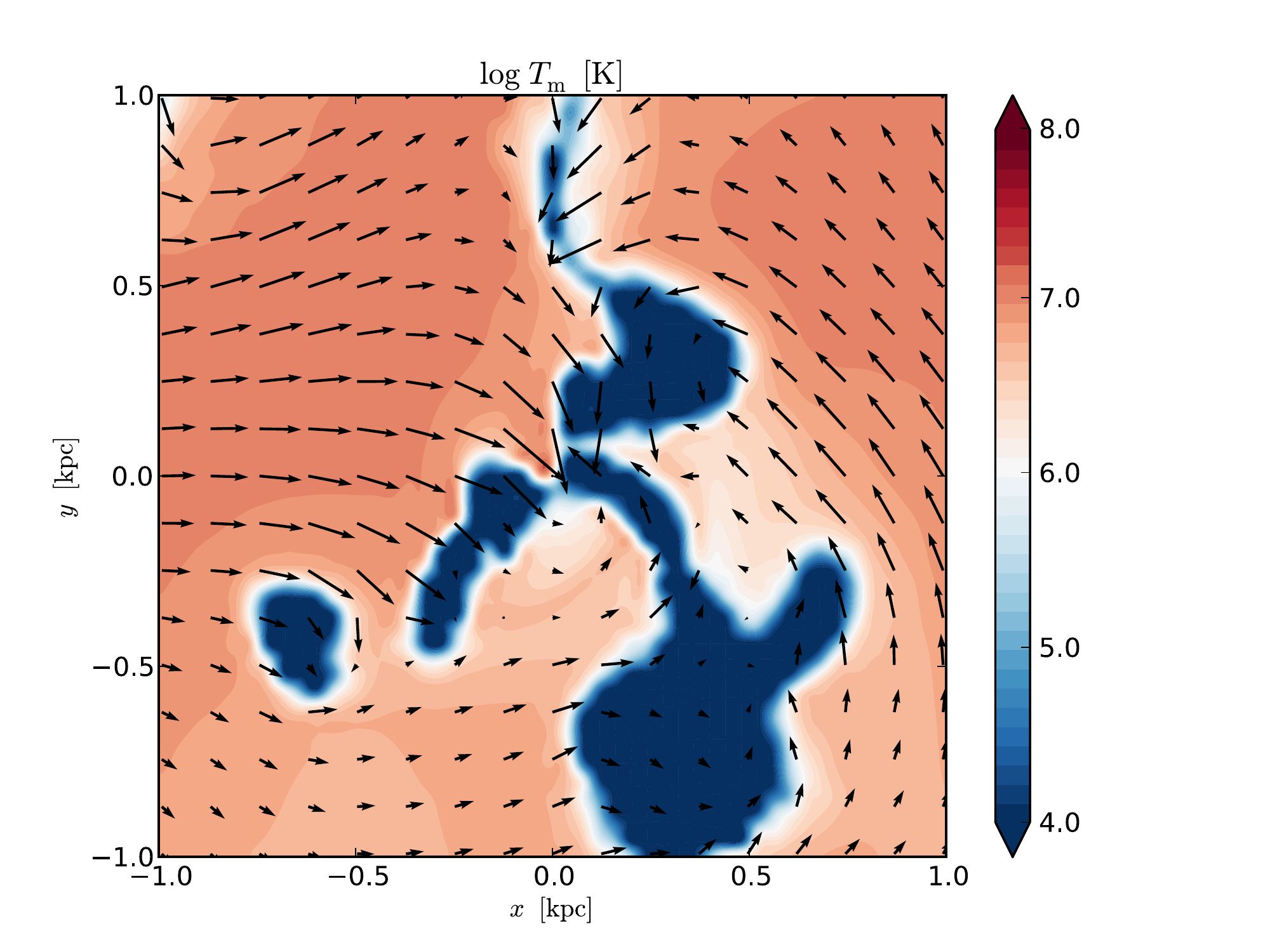}}
     \caption{Accretion with cooling, turbulence (${\rm Ma}$\,$\sim$\,0.35), and $e_{\rm rot} = 0.3$: mid-plane $T_{\rm m}$ cross-section through $z=0$ (final time). Without heating, the system experiences a massive CCA flow. Extended cold filaments and clouds condense out of the hot phase via TI, and through recurrent chaotic collisions are quickly accreted by the BH.
     No steady thin disk can be formed. }
      \label{f:stir_cool_T}  
\end{figure} 


\begin{figure}
      \center
      \subfigure{\includegraphics[scale=0.31]{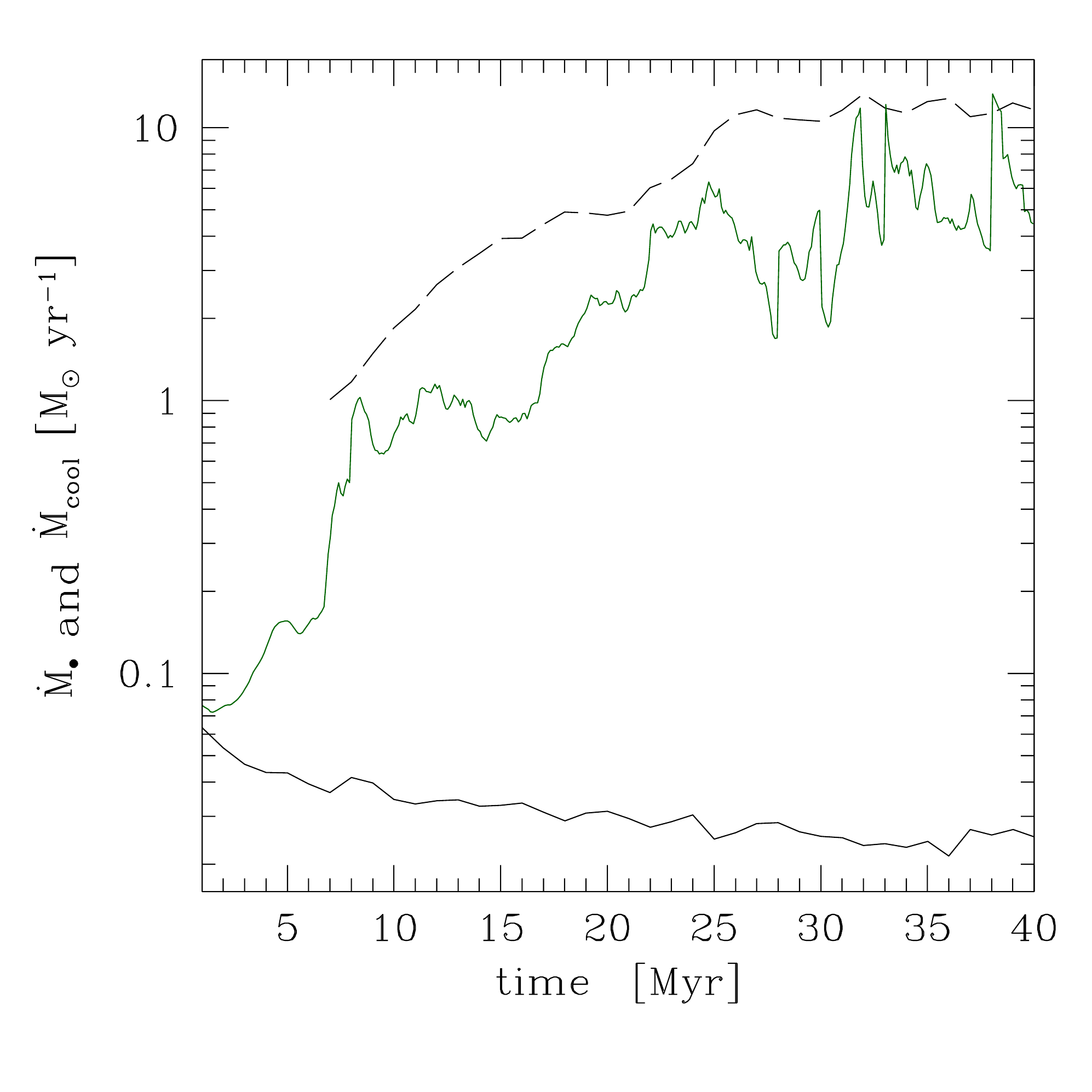}}
      \subfigure{\includegraphics[scale=0.31]{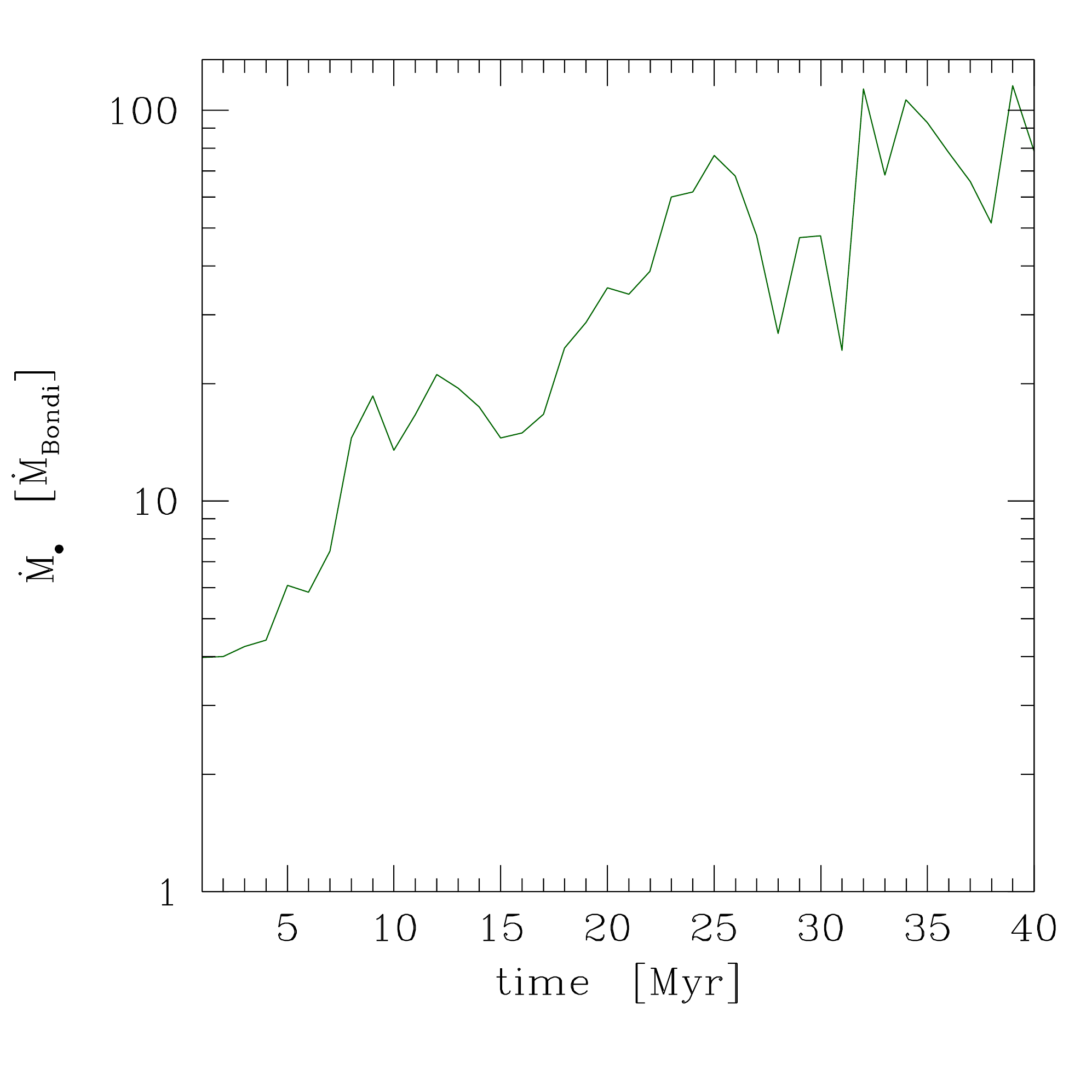}}
      \caption{Accretion with cooling, turbulence (${\rm Ma}$\,$\sim$\,0.35), and $e_{\rm rot} = 0.3$ (green): evolution of the physical and normalized accretion rate (cf.~Fig.~\ref{f:cool_mdot}; the average of $\dot{M}_\bullet$ in the top panel has 0.1 Myr step). 
      The dashed line is the average cooling rate (1 Myr step). 
      The solid black line is the adiabatic rotating model (\S\ref{s:adi}).
       Chaotic cold accretion drives the dynamics as long as ${\rm Ta < 1}$. 
       Recurrent collisions in the cold phase cancel angular momentum and boost the accretion rate up to two orders of magnitude with respect to the Bondi rate. The accretion rate is linearly tied to $\dot M_{\rm cool}$ again. }
      \label{f:stir_cool_mdot}  
\end{figure}    


\begin{figure} 
      \centering
      \subfigure{\includegraphics[scale=0.4]{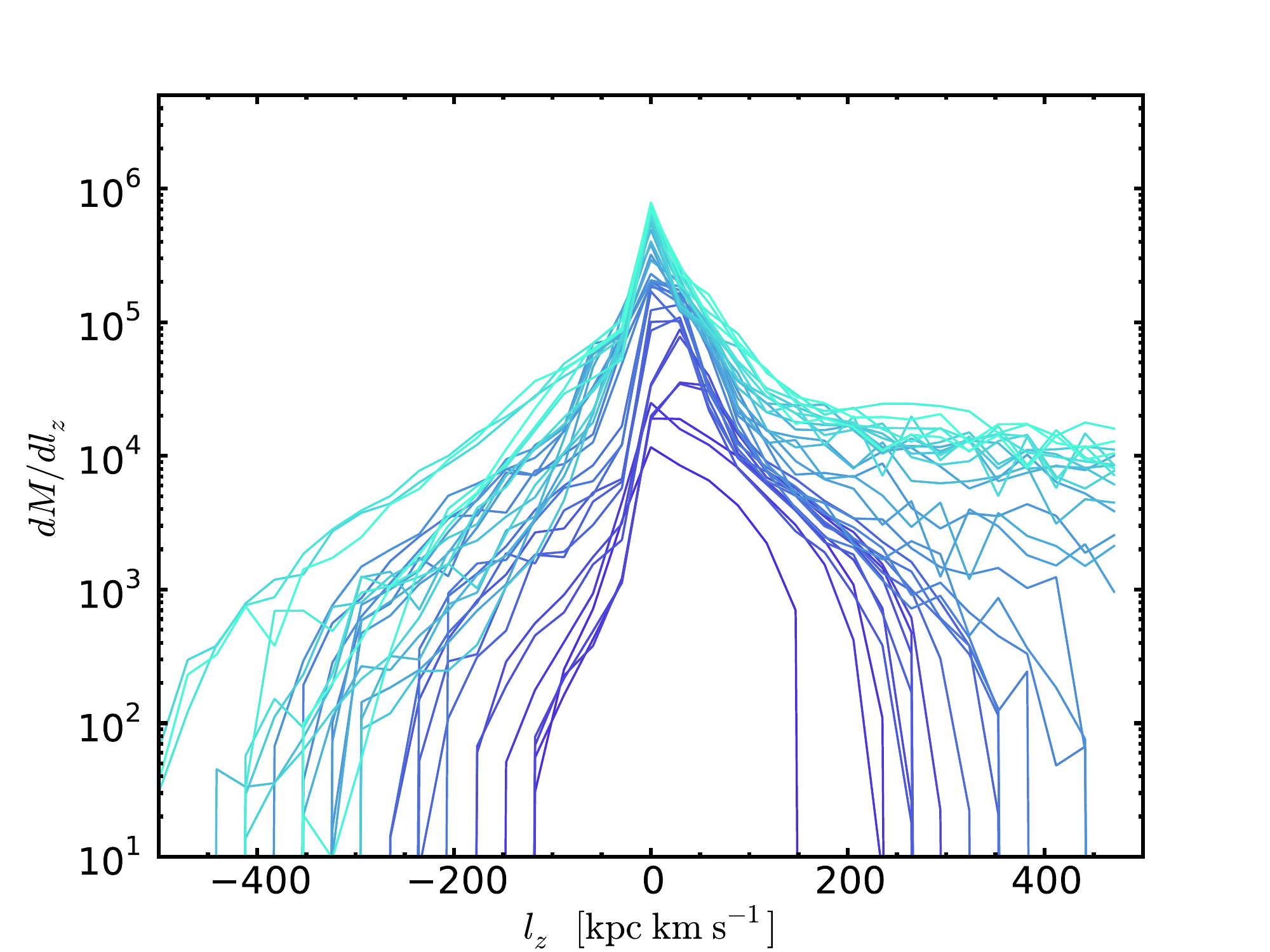}}
      \caption{Accretion with cooling, turbulence (${\rm Ma}$\,$\sim$\,0.35), and $e_{\rm rot} = 0.3$: mass PDF per bin of specific angular momentum $l_z$, for the cold phase (the hot phase has PDF analogous to Fig.~\ref{f:stir_lz}, top). At variance with the purely radiative run, the cold phase randomly condenses out of a broad $l_z$ distribution, including both prograde and retrograde motions. This permits frequent major collisions, canceling angular momentum. The mass of unaccreted cold gas tends to significantly rise with time due to the unheated cooling flow. }
      \label{f:stir_cool_lz}      
\end{figure}


\begin{figure}
      \begin{center}
      \subfigure{\includegraphics[scale=0.28]{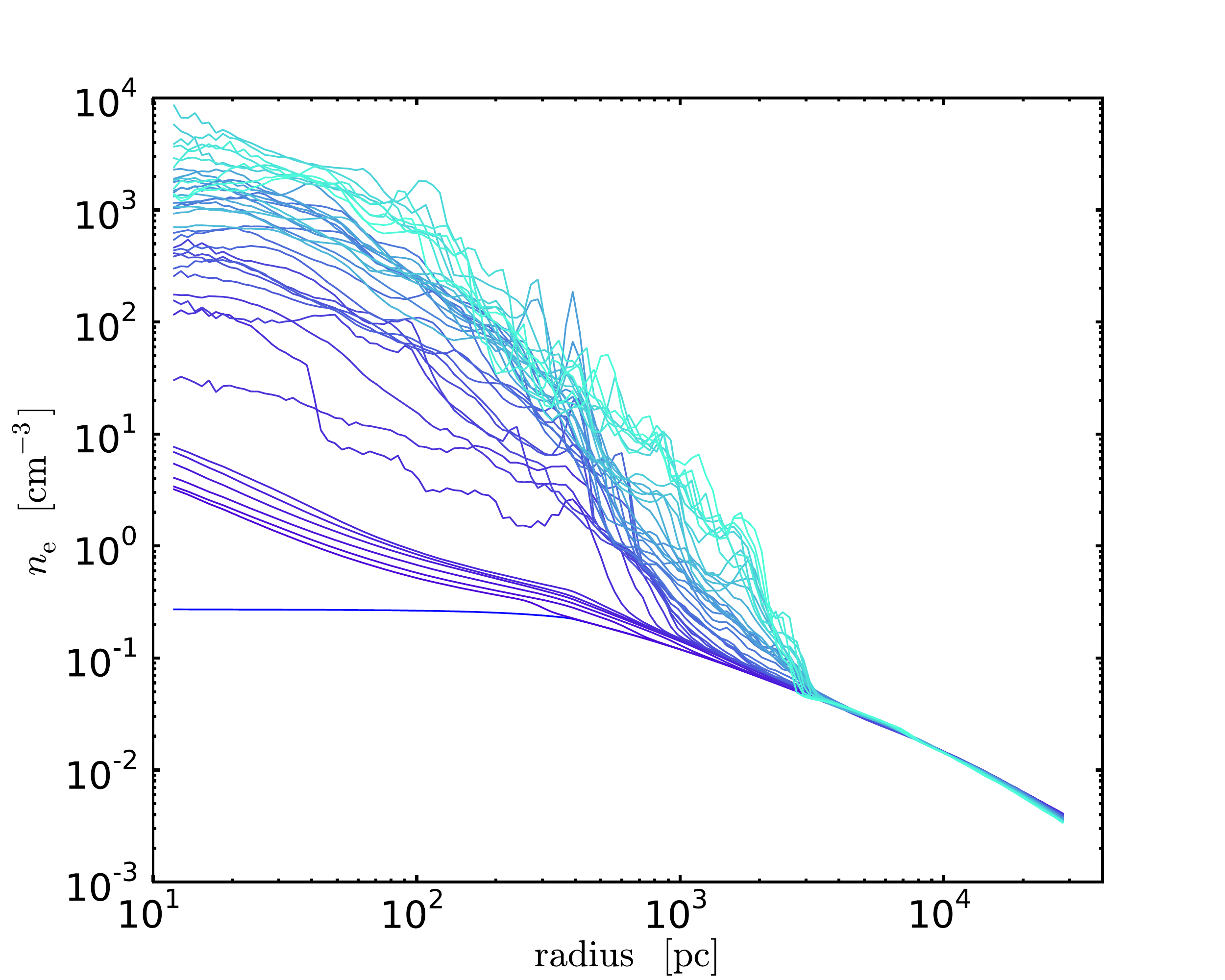}}
      \subfigure{\includegraphics[scale=0.28]{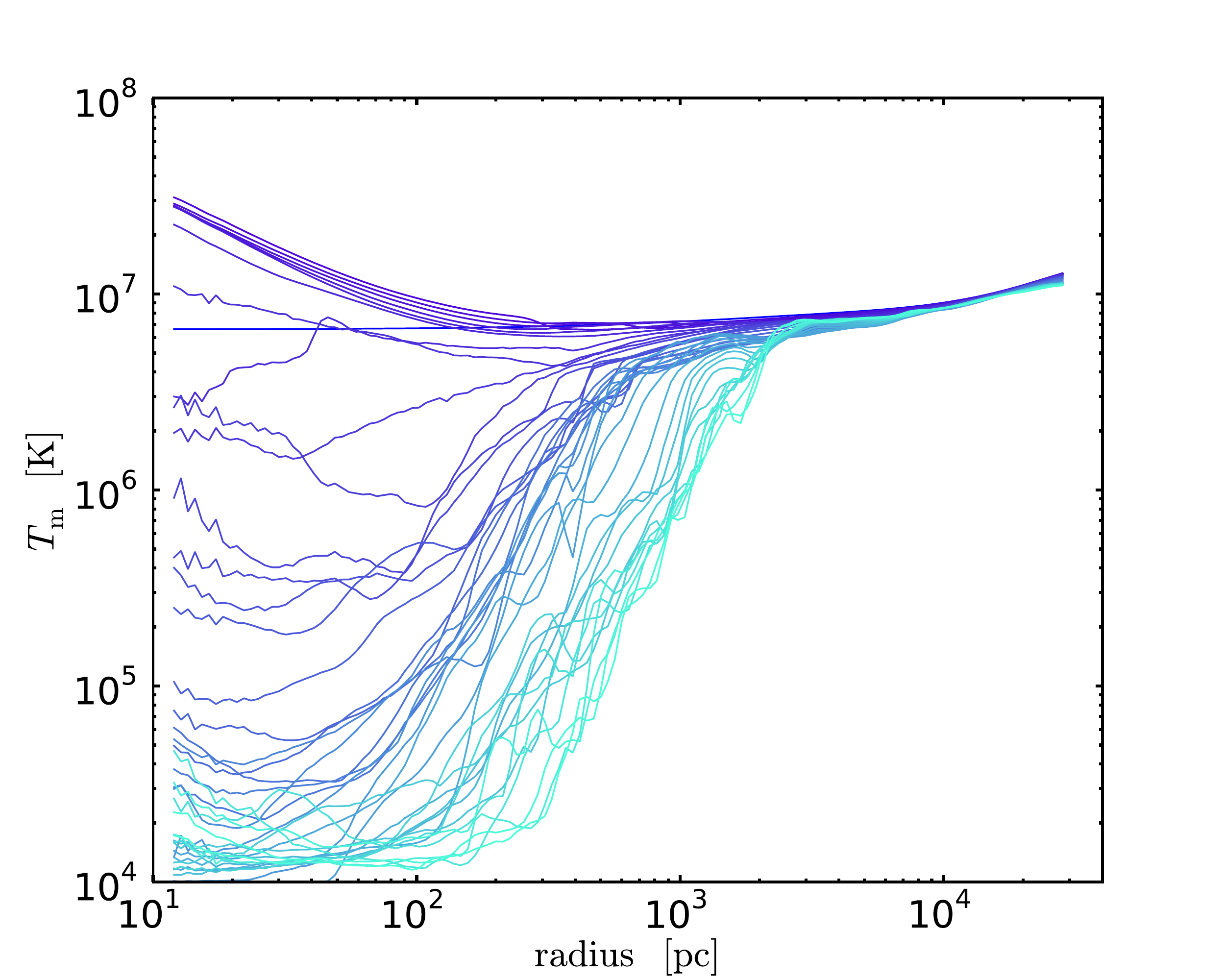}}
      \subfigure{\includegraphics[scale=0.28]{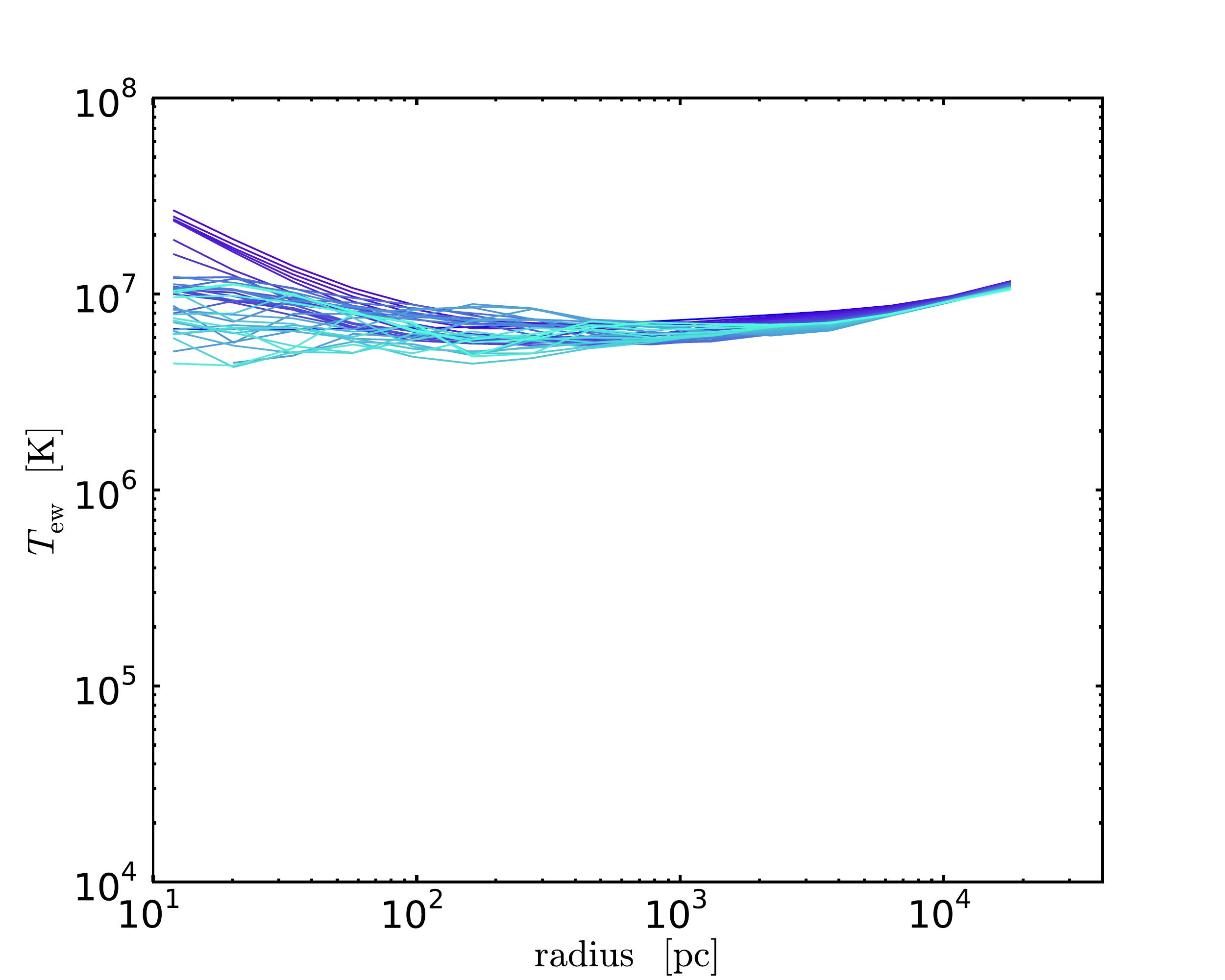}} 
      \end{center}
      \caption{Accretion with cooling, turbulence (${\rm Ma}$\,$\sim$\,0.35), and $e_{\rm rot} = 0.3$: mass-- and emission--weighted radial profiles of density and temperature (cf.~Fig.~\ref{f:cool_prof}). The profiles show the condensation of warm gas out of the hot phase via TI up to several kpc. The X-ray temperature is flat, with no cuspy core, which is a characteristic mark of cold accretion. } 
      \label{f:stir_cool_prof}  
\end{figure}  

\noindent
In the next model, we turn on radiative cooling, while the reference subsonic turbulence (${\rm Ma}$\,$\sim$\,0.35) stirs the hot atmosphere.
The development of a clumpy multiphase medium completely changes the behavior of accretion.
This simulation also warns how mild chaotic motions can profoundly modify the picture presented by analytic models (e.g., the classic thin disk).

\subsection[]{Dynamics}  \label{s:stir_cool_dyn}
\noindent
In Fig.~\ref{f:stir_cool_T}, the temperature map reveals the absence of major coherent motions. 
No source of heating is present, nevertheless the perturbations seeded by subsonic turbulence grow nonlinear via thermal instability (TI) and produce extended multiphase structures (see G13 and \citealt{McCourt:2012} for further discussions on the formed TI).     
The dynamics of the gas is chaotic
because of high sensitivity of the long-term dynamics on local TI.
The recurrent chaotic collisions 
between the condensed clouds, filaments, and a clumpy torus in the inner region
promote the cancellation of angular momentum (see also \citealt{Nayakshin:2007,Pizzolato:2010})
boosting the accretion rate.
On top of turbulent diffusion, 
collisions promote further disruption of coherent motions and rotating structures.
No steady thin disk can be formed (compare the evolution with that in \S\ref{s:cool}), although at later times a volatile and clumpy cold torus emerges out of the residual gas with large $l_z$ due to incomplete cancellation (\S\ref{s:stir_cool_prof}).
The multiphase gas halo in galaxies should be thus treated as 
a collisional, hydrodynamical system. The cold phase is not well described by ballistic orbits. 
The collisions are frequent in the inner 1 kpc core (with Myr variability), as major filaments have lengths 
comparable to this radius, initiating further interactions between cold elements.

\subsection[]{Accretion rate}\label{s:stir_cool_mdot}
\noindent
Chaotic cold accretion drives the dynamics, in a similar manner as found in G13 (as long as 
${\rm Ta_t} < 1$; see next \S\ref{s:heat}).
The cold clouds condense out of the stirred hot phase, 
thereby experiencing chaotic streamlines.
The rapid prograde versus retrograde collisions cancel angular momentum, and consequently boost the accretion rate up to two orders of magnitude with respect to the kpc-scale Bondi rate (Fig.~\ref{f:stir_cool_mdot}, bottom). Compared with the adiabatic run (either turbulent or rotating; solid black, top), the increase is almost a factor $10^{3}$.
At variance with the thin disk evolution (\S\ref{s:cool}), the accretion rate is linearly tied to $\dot M_{\rm cool}$ (dashed), albeit experiencing substantial chaotic variability up to $\sim\,$1 dex. 
The cooling rate saturates slightly faster around $15\ \msun$ yr$^{-1}$ because of the turbulent mixing of entropy.
At $t>25$ Myr (top), the residual cold gas with high $l_z>0$ induces deeper $\dot M_\bullet$ valleys (via a clumpy torus).

This regime with no heating produces unrealistically high cooling rates and condensation. 
The X-ray spectra indicate $\dot M_{\rm cool}$ lower by at least an order of magnitude (e.g., \citealt{Tamura:2003, Peterson:2006} for a review). Nevertheless, it is instructive to understand chaotic cold accretion embedded in a pure cooling flow. 
Some systems may experience a delayed AGN feedback (in particular at high redshift), enabling a massive CCA for a transient time (e.g., Phoenix cluster; \citealt{McDonald:2012_Phoenix_HST}).

\subsection[]{Distribution of $l_z$ and radial profiles}  \label{s:stir_cool_prof}
\noindent
At variance with the purely radiative run, the cold phase randomly condenses out of the $l_z$ distribution shown in Fig.~\ref{f:stir_lz} (top), i.e., the cold clouds and filaments can be generated with both prograde and retrograde motions. 
Initially, the inner gas with lower $l_z$ cools faster, growing a modest PDF (dark blue line).
After 10 Myr, the wings of the broader PDF start to experience significant recurrent narrowing, as a result of violent collisions canceling angular momentum (occurring in a period of a few Myr). 
However, the mass of unaccreted cold gas tends to substantially rise with time (leading to a PDF with larger width and normalization), progressively obfuscating the previous effect.      
In the subsequent heated run (Fig.~\ref{f:heat_lz}), it will be easier to isolate the action of collisions, lacking the formation of a massive cooling flow. 

An unbalance toward the right wing ($l_z >200$ kpc km s$^{-1}$) persists because of the initial halo rotation, implying that a prograde (though clumpy) torus-like structure is a recurrent phenomenon. As ${\rm Ta_t} \ll 1$, the role of rotation becomes negligible and the prograde bias disappears (G13). Conversely, as ${\rm Ta_t} > 3$, the cold phase can only be generated with positive $l_z$ (Fig.~\ref{f:stir_lz}, bottom) and the accretion follows that of the coherent thin disk (\S\ref{s:cool}).

In Fig.~\ref{f:stir_cool_prof}, the radial profiles are analogous to that found in G13, showing the massive condensation of warm or cold gas out of the hot phase via thermal instability, and the fluctuations imparted by turbulence. 
After full condensation, the cold gas typically populates the region within 3 kpc.
Maximum density is slightly higher compared with G13 ($n_{\rm e}\sim10^4$ cm$^{-3}$), since the initial hydrostatic atmosphere has slightly shallower density gradient.
The core X-ray temperature is flat, a characteristic mark of cold accretion dominating over the hot mode.

\section[]{Accretion with heating, cooling, and turbulence: chaotic cold accretion (CCA)}  \label{s:heat}
\noindent
In the last set of models, we focus on the typical state of the hot plasma in a massive galaxy, group, or cluster.
The cooling flow is now quenched via heating (\S\ref{s:init2}) 
by 10\,-\,20 fold, in agreement with {\it XMM-Newton} spectral data (\citealt{Tamura:2003, Peterson:2006}).
The source of heating is mainly attributed to AGN feedback, albeit supernovae, thermal conduction, and mergers
can possibly contribute.

\subsection[]{Accretion rate} \label{s:heat_acc}
\begin{figure} 
      \begin{center}
      \subfigure{\includegraphics[scale=0.31]{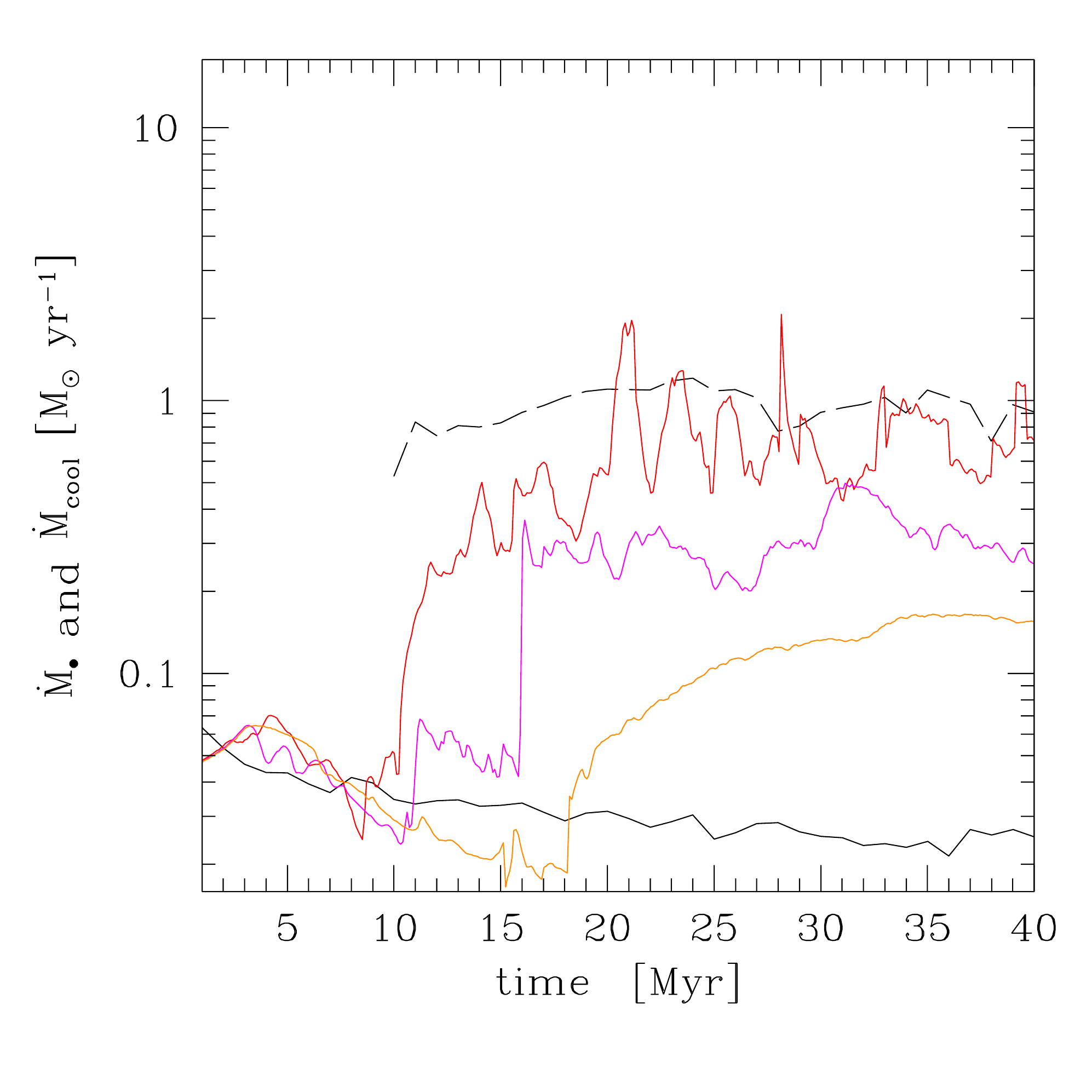}}
      \subfigure{\includegraphics[scale=0.31]{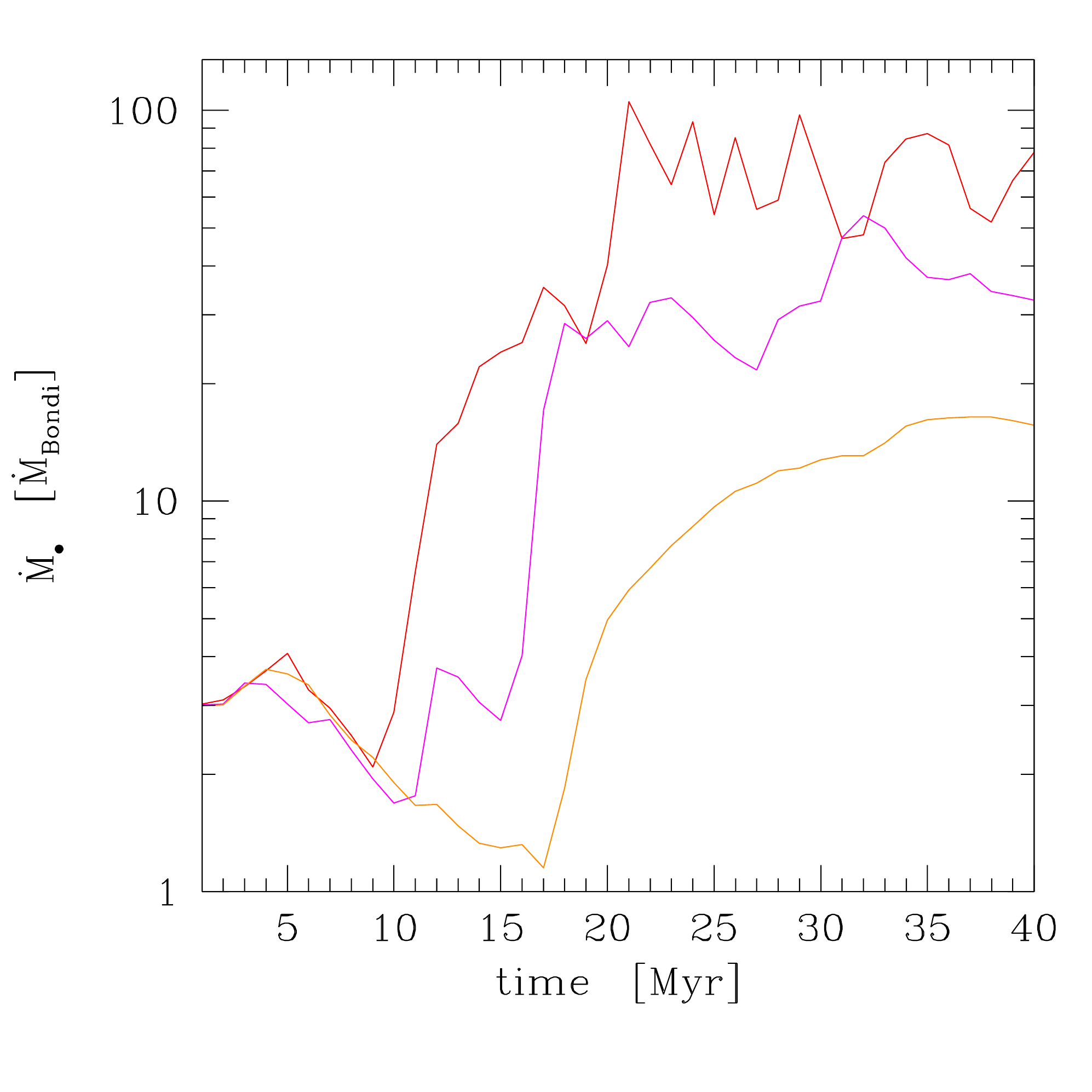}}
      \end{center}
      \caption{Accretion with heating, cooling, $e_{\rm rot}=0.3$, and varying levels of turbulence: evolution of the accretion rate (the $\dot{M}_\bullet$ average in the top panel has 0.1 Myr step).  
      The Mach number varies from the reference ${\rm Ma}$\,$\sim$\,0.35 (red) to 1/2 (magenta) and 1/4 (orange) of this value, 
      i.e., ${\rm Ta_t}\simeq 0.7, 1.5, 3$, respectively. The dashed line is the average net cooling rate (1 Myr step).
      As before, the solid black line is the adiabatic rotating model, and the runtime Bondi rate for the normalized plot is computed at $r\approx\,$1-2 kpc (\S\ref{s:adi}).
      In the atmosphere with ${\rm Ta_t}<1$, chaotic cold accretion drives the dynamics, boosting the accretion rate up to $100\times$ the Bondi rate,      
      which is consistent with the nonrotating CCA evolution shown in G13.
      As ${\rm Ta_t}>1$, the accretion flow shifts from turbulence-driven (linked to extended filaments and boosted accretion) to rotationally-driven (tied to a coherent disk and suppressed accretion). }
       \label{f:heat_mdot}  
\end{figure}

\noindent
The major result is that, even in the presence of significant rotation ($v_{\rm rot}\approx100$ km s$^{-1}$), the accretion
rate is boosted up to $100\times$ the Bondi rate (Fig.~\ref{f:heat_mdot}, red), which is consistent with the $e_{\rm rot}=0$ model presented in G13 (Sec.~7).
The peaks in the accretion rate are comparable to the quenched cooling rate, $\dot M_{\rm cool}\sim1\ \msun\,{\rm yr}^{-1}$ (dashed).\footnote{Star formation is observed to be inefficient in massive elliptical galaxies, $\lta\,$1 percent of the pure cooling rate (\citealt{McDonald:2014}).}
Adopting $\dot M_\bullet\sim\dot M_{\rm cool}$ is an effective subgrid accretion model for large-scale simulations and analytic calculations.
Initially, accretion is driven by the rotating hot flow. 
After 10 Myr,
chaotic cold accretion drives the dynamics.
At variance with the runs in \S\ref{s:stir_cool}, the presence of heating prevents the formation of a catastrophic cooling flow: the average $\dot M_\bullet$ and $\dot M_{\rm cool}$ do not increase with time.
In the regions where $t_{\rm cool}/t_{\rm ff}<10$ (between $r\sim$\,100 pc and several kpc),
thermal instability grows nonlinearly (see also \citealt{Gaspari:2012a}), the cold gas condenses out of the hot phase, and the chaotic collisions promote cancellation of angular momentum (\S\ref{s:heat_dyn}). 
Multiwavelength observations support the TI and CCA scenario,
detecting extended multiphase gas in the core of many massive galaxies which is cospatial
in X-ray, FUV, H$\alpha$, and molecular band (\citealt{McDonald:2010,McDonald:2011a,Werner:2013,Werner:2014}; \S\ref{s:comp}). 

CCA dominates as long as the following criterium is met:
\begin{equation} 
{\rm Ta_t} < 1.
\end{equation} 
In Fig.~\ref{f:heat_mdot}, we show the models with 1/2 and 1/4 lower turbulence with respect to the reference ${\rm Ma}$\,$\sim$\,0.35, which correspond to ${\rm Ta_t}\sim1.5$ (magenta) and 3 (orange), respectively. As the rotational velocity exceeds the turbulent velocity dispersion, the accretion rate is progressively suppressed by a factor $\propto {\rm Ta_t}$.
The accretion flow shifts from turbulence-driven, with chaotic filaments and boosted $\dot M_\bullet$, to rotationally-driven,
displaying a coherent disk and reduced $\dot M_\bullet$. In the regime ${\rm Ta_t}\gg1$, $\dot M_\bullet$ saturates around $\sim\,$$0.1\ \msun$ yr$^{-1}$, as perturbations induced by turbulence are not influencing the evolution of the thin disk. 

In a complementary run (not shown), we doubled
the rotational velocity ($\approx\,$200 km s$^{-1}$) while fixing the reference ${\rm Ma}$\,$\sim$\,0.35 (again ${\rm Ta_t}$$\,\sim\,$1.5).
The results are analogous to the previous models after comparing identical cooling rate
(a stronger flattening implies higher $\rho$ at large $r$, thus larger cooling rates).

Comparing to observations, accretion rates can be estimated from the jet or cavity power, which for common massive ellipticals spans $P_{\rm cav}\approx10^{42}$\,-\,$10^{44}$ erg\,s$^{-1}$ (\citealt{Allen:2006}). 
From the quiescent to strong feedback state $\dot M_\bullet = P_{\rm cav}/(\varepsilon\,c^2)\approx 2\times10^{-2}$\,-\,$2\ \msun\,{\rm yr^{-1}}$, where $\varepsilon\sim10^{-3}$ is the typical mechanical efficiency for massive ellipticals (\citealt{Gaspari:2012b}).
The simulated models cover this range (Fig.~\ref{f:heat_mdot}), with CCA representing the strong impulsive feedback stage, while the disk-dominated phase is associated with the more quiescent galaxies. We notice that the accretion rates are sub-Eddington, $\dot M_\bullet/\dot M_{\rm Edd}\approx 3\times10^{-4}$\,-\,$3\times10^{-2}$ (where $\dot M_{\rm Edd} = 66\ \msun\,{\rm yr^{-1}}$), i.e., the common regime where AGN are observed to be radiatively inefficient and dominated by mechanical energy input (\citealt{Russell:2013}).


\begin{figure} 
     \subfigure{\includegraphics[scale=0.46]{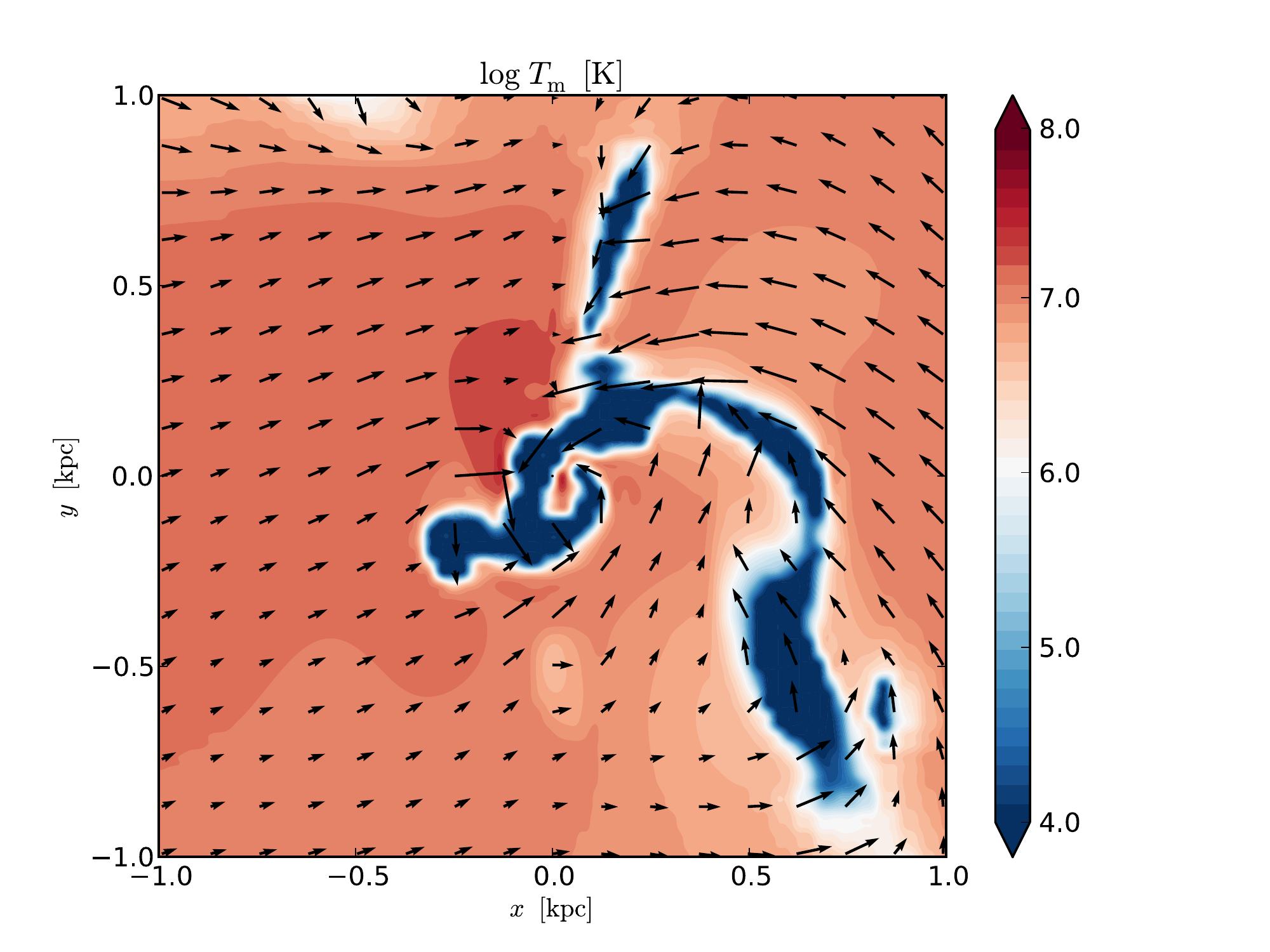}}
    \caption{Accretion with heating, cooling, turbulence (${\rm Ma}$\,$\sim$\,0.35), and $e_{\rm rot}=0.3$: temperature cross-section, with the velocity field overlaid. 
    Full circularization is not possible in a turbulent environment characterized by ${\rm Ta_t <1}$.
    The recurrent chaotic collisions between the cold filaments, clouds, and the clumpy torus,
    cancel the angular momentum of the cold gas, leading to the rapid peaks in the accretion rate, $\dot M_\bullet$\,$\sim$\,$\dot M_{\rm cool}$.}
      \label{f:heat_T}  
\end{figure}

\subsection[]{CCA dynamics and $l_z$ distribution}  \label{s:heat_dyn}
\begin{figure} 
      \centering
      \subfigure{\includegraphics[scale=0.4]{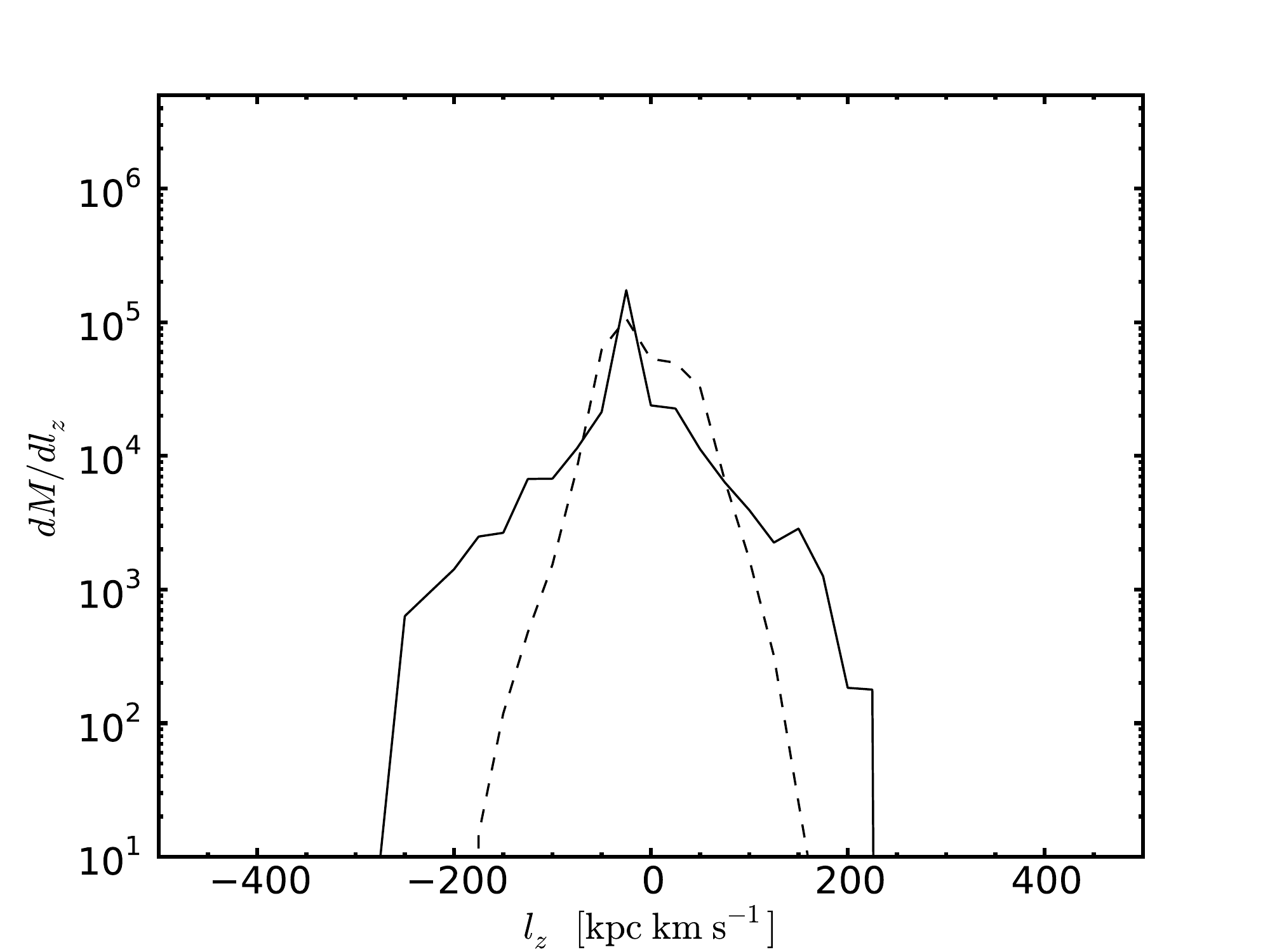}}
      \caption{Accretion with heating, cooling, turbulence (${\rm Ma}$\,$\sim$\,0.35), and $e_{\rm rot}=0.3$: mass PDF per bin of specific angular momentum $l_z$, for the cold phase (the hot phase has PDF analogous to Fig.~\ref{f:stir_lz}, top). After three rapid $\dot M_\bullet$ bursts, the PDF narrows on average from 20 Myr (solid) to 27 Myr (dashed) due to major collisions canceling positive and negative $l_z$.      
      The action of collisions is periodically contrasted by fresh condensation, favoring a broader distribution. } 
      \label{f:heat_lz}    
\end{figure}

\noindent
Chaotic cold accretion is driven by the following processes.
First, turbulence broadens the distribution of angular momentum related to the hot atmosphere, as
shown in Fig.~\ref{f:stir_lz} (top; see discussion in \S\ref{s:stir_prof}). The broadening is $\propto \sigma_v$ and induces both prograde and retrograde chaotic motions.
The broadening of the PDF($l_z$) alone does not imply boosted accretion, since the hot gas is supported by pressure.
The second step is the condensation of cold gas. The cooling gas retains the imprint of the hot phase, emerging 
out of the broad $l_z$ distribution with both positive and negative values. 
During infall\footnote{Nonlinear condensation is too fast and clouds are too massive to be directly affected by the driving, which is slowly injected at large scale.}, the clouds start to significantly collide, mainly in the inner 1 kpc (Fig.~\ref{f:heat_T}), inducing major angular momentum cancellation (see also \citealt{Pizzolato:2010}). 
In Fig.~\ref{f:heat_lz}, we focus on the PDF($l_z$) evolution between 20 and 27 Myr, as three major $\dot M_\bullet$ bursts and collisions occur in rapid succession. Despite the continuous cold gas condensation,
a rapid narrowing of the $l_z$ distribution occurs, leading to the $\dot M_\bullet$ bursts seen in Fig.~\ref{f:heat_mdot}.
At other times (not shown), gas condensation can regenerate one or both tails. 
Since the cooling rate and cold mass is quenched by more than an order of magnitude compared with the pure cooling flow (consistently with observations; \citealt{Tamura:2003}), collisional effects are not as overwhelmed by condensation through time as in the nonheated run (\S\ref{s:stir_cool}). The residual cold phase achieves statistical steady state after $\sim\,$$2 \,t_{\rm cool}$, while being shaped by the recurrent broadening and narrowing of the angular momentum distribution with Myr variability (a prograde bias is still present after 40 Myr).

Both turbulence and collisions characterize CCA. Without turbulence the angular momentum of the newborn cold phase would only have positive $l_z$ (cf.~Fig.~\ref{f:stir_lz}, bottom), leading to a coherent disk. Without collisions, the cold phase might not strongly boost accretion (no rapid $l_z$ cancellation). Both processes are tightly related to $\sigma_v$.
If ${\rm Ta_t}<1$, as in the reference run, turbulent diffusion
dominates over the advection because of coherent rotation, leading to substantial PDF broadening and head-on cloud collisions.

However, if ${\rm Ta_t}>1$, the broadening is too weak and collisions cannot cancel angular momentum. 
The suppression in the accretion rate is relatively smooth, $\dot M_\bullet \propto {\rm Ta_t}^{-1}$ (Fig.~\ref{f:heat_mdot}), until the thin disk evolution dominates for ${\rm Ta_t} > 3$ (i.e., at least an order of magnitude in classic Taylor number; \S\ref{s:intro}).


The linear transition can be crudely understood in terms of mixing length approximation.
Accretion in a rotating atmosphere is ultimately limited by the diffusion time related to collisions (the PDF broadening, albeit directly tied to ${\rm Ta_t}$, is a necessary, but not sufficient, condition).
Using as
effective collisional viscosity\footnote{Notice that CCA diffusivity, based on collisions and $l$ cancellation, is different from the shear viscosity and associated steady inward or outward transport of $l_z$ postulated in Keplerian thin disks (\citealt{Shakura:1973}). In the latter model, the velocity is either prograde or retrograde only, and the cold rings experience friction due to internal micro-turbulence (e.g., driven by the magnetorotational instability; \citealt{Balbus:1998}).} 
$\nu\sim \sigma_v\, \lambda$, with $\lambda$ the collisional mean free path, we can write the following scaling:
\begin{equation}\label{e:diff}
t_{\rm acc}\approx t_{\rm diff} \equiv\frac{r^2}{\nu} \sim \frac{v_{\rm rot}}{\sigma_ v}\; \frac{r}{\lambda}\; \,t_{\rm dyn}
\end{equation}
where we assumed the gas dynamical timescale is mainly associated with rotation. 
The smaller the velocity dispersion, the longer the diffusion and hence accretion timescale. 
The clouds and filaments must travel into the inner 1 kpc region to experience major collisions. Here, 
the mean free path is typically comparable to the radius (yet smaller than the global system),
yielding the scaling $t_{\rm acc}\sim {\rm Ta_t}\: t_{\rm dyn}$, in agreement with our findings (Fig.~\ref{f:heat_mdot}).
In general, $\lambda$ varies depending on the size of the condensing clouds, which can be smaller than $r$. This slows the accretion time, creating the valleys in $\dot M_\bullet$. 
While the accretion peaks are provided by major collisions, the average accretion rate is $\propto {\rm Ta_t}^{-1}$. 
As ${\rm Ta_t} < 1$, turbulent diffusion overcomes rotation (Eq.~\ref{e:diff} is no longer accurate) 
and the sinking of cold clouds, after condensation in a cooling time, occurs in a radial free-fall timescale, returning to the pure CCA evolution shown in G13.

We note that the ram pressure drag is here subdominant. The density contrast between the cold and hot phase is $\rho_{\rm c}/\rho_{\rm h}\sim10^3$. The cloud halting distance to lose all the kinetic energy can be estimated as
$d_{\rm halt}\sim(e_{\rm kin, c}/\dot{e}_{\rm kin, c})\,v_{\rm c}\sim(\rho_{\rm c}/\rho_{\rm h})\,r_{\rm c}$, using $e_{\rm kin,c}\sim \rho_{\rm c}\,v_{\rm c}^2\,r_{\rm c}^3$ and $\dot{e}_{\rm kin,c}\sim F_{\rm drag}\,v_{\rm c} \sim (\rho_{\rm h}\,v_{\rm c}^2\,r_{\rm c}^2)\,v_{\rm c}$. Therefore, even small clouds of 50 pc size would require distances larger than the box to lose most of the energy due to drag, while collisions occur in just a small fraction of the system size, mostly within $r<$\,1 kpc.


\begin{figure*} 
      \begin{center}
      \subfigure{\includegraphics[scale=0.28]{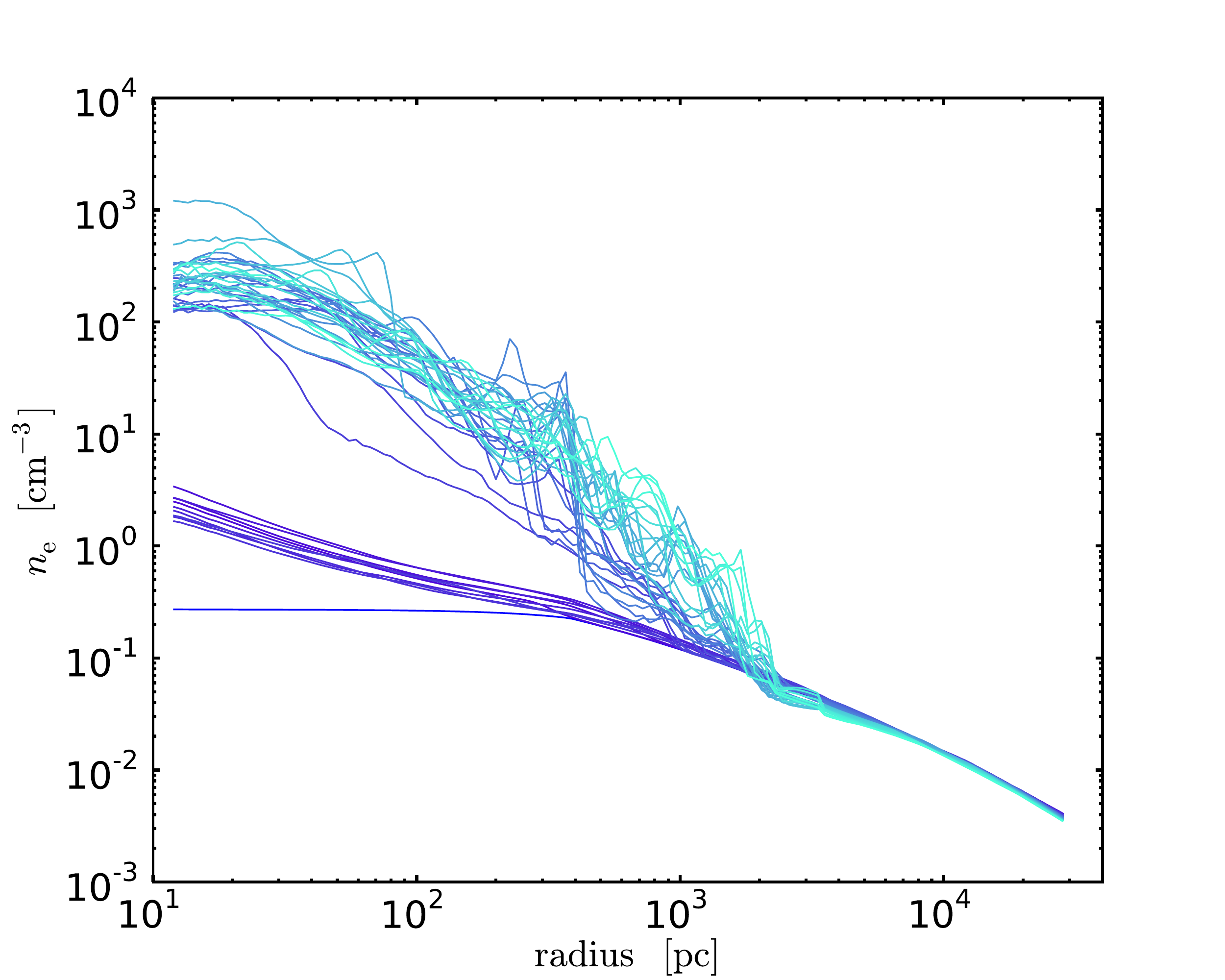}}
      \subfigure{\includegraphics[scale=0.28]{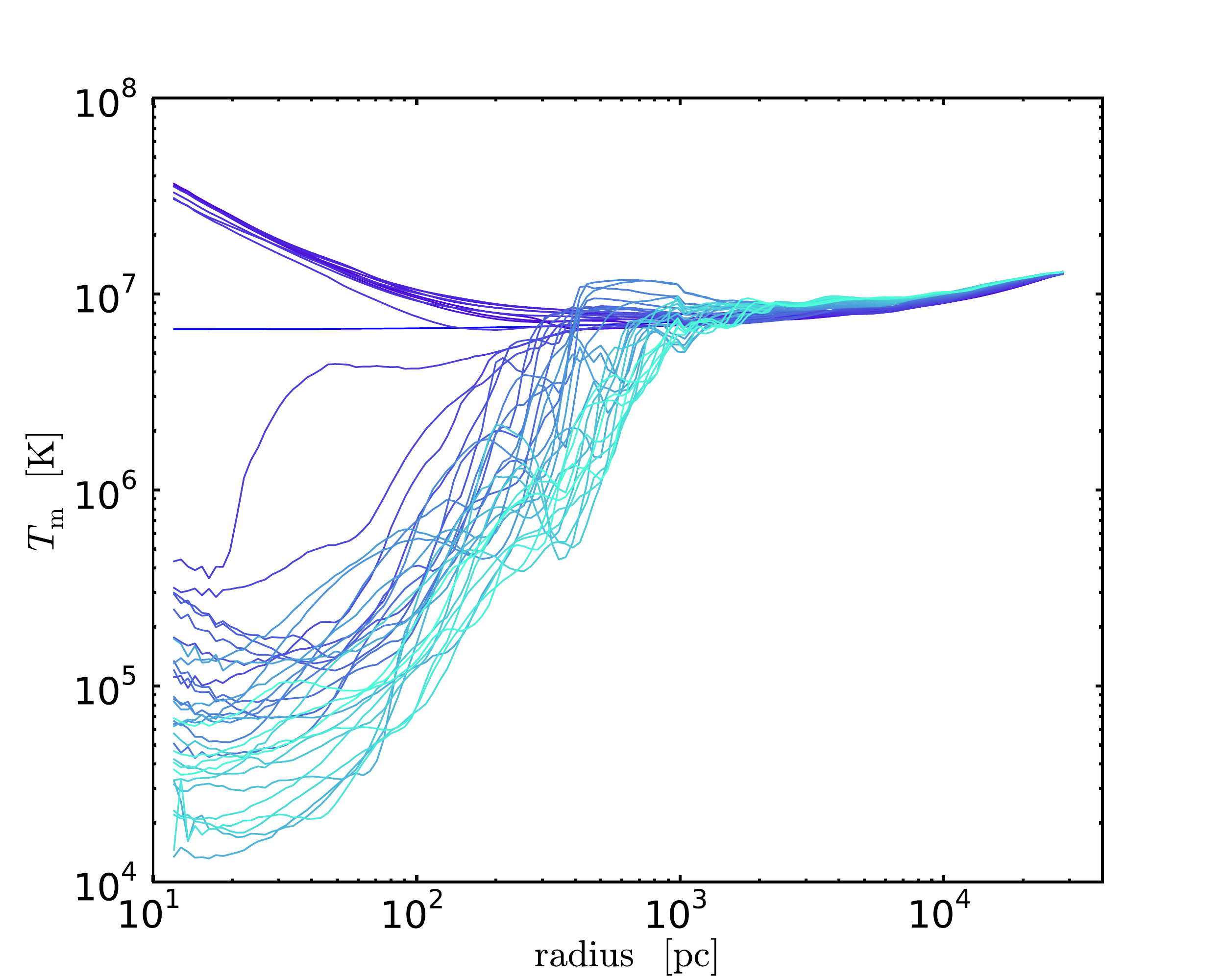}}
      \subfigure{\includegraphics[scale=0.28]{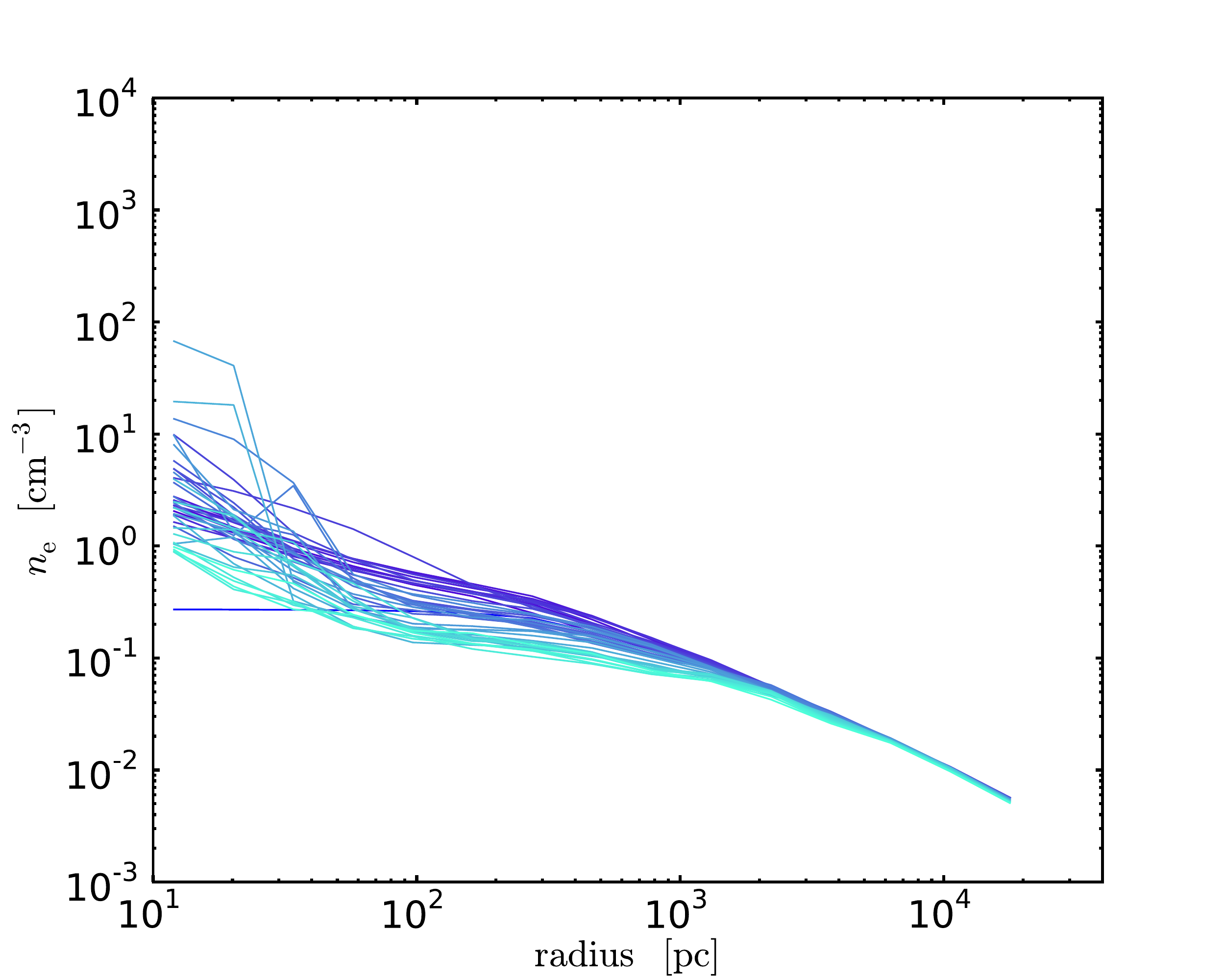}}
      \subfigure{\includegraphics[scale=0.28]{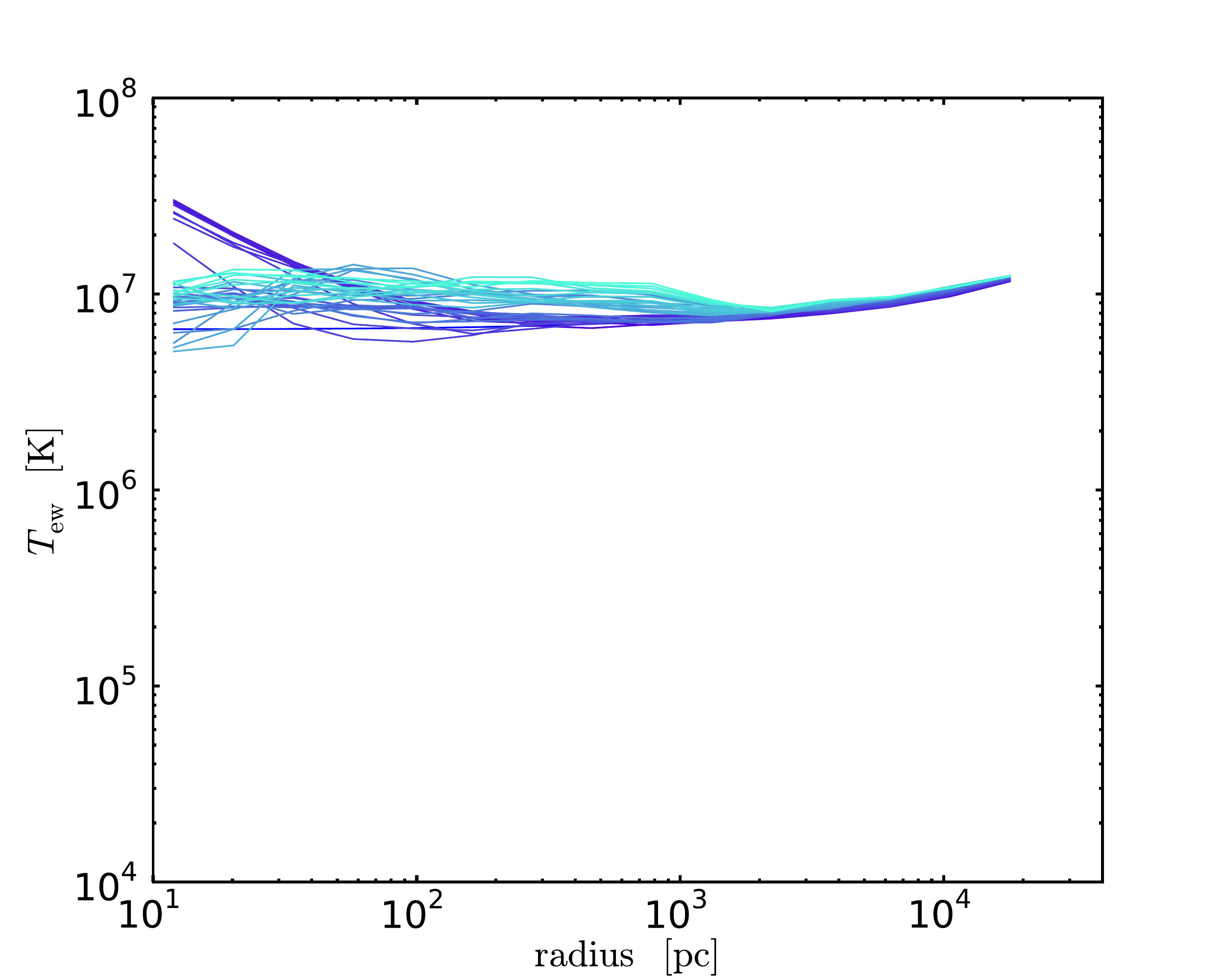}} 
      \end{center}
      \caption{Accretion with heating, cooling, turbulence (${\rm Ma}$\,$\sim$\,0.35), and $e_{\rm rot}=0.3$: 3D mass- (top) and X-ray emission-weighted ($\propto$\,$\mathcal{L}$; bottom) radial profiles of density and temperature (cf.~Fig.~\ref{f:cool_prof}). 
      The profiles show the extended multiphase structure, which is largely concealed in the X-ray band.
      The X-ray temperature profile is remarkably flat, in contrast to the hot-mode accretion, 
      which has a peaked temperature profile. This is a key observable, which can be thoroughly tested (\citealt{Wong:2014,Russell:2015}), to unveil the evolutionary stage in which the galaxy is residing.} 
      \label{f:heat_prof}
\end{figure*} 

The reference $v_{\rm rot}\approx100$ km s$^{-1}$ is near the high end of the realistic range (as we want to clearly understand the impact of rotation). 
For $e_{\rm rot}> 0.3$, the galaxy is substantially flattened and the gas $v_{\rm rot}$ would inconsistently exceed the stellar rotational velocity, which is commonly $\sim\,$100 km s$^{-1}$ (\citealt{Davies:1983}). Therefore, many massive elliptical galaxies should reside in the pure CCA regime (${\rm Ta_t}\ll1$), in particular after strong injection of turbulence, e.g., via AGN outbursts and merger events. On the other hand, a fraction of elliptical galaxies are observed to host central cold disks (\citealt{Young:2011,Alatalo:2013}), which is a by-product of incomplete angular momentum cancellation. This is associated with ${\rm Ta_t}\gta3$ as turbulence and heating start to decrease, in agreement with the observational findings presented by \citeauthor{Werner:2014} (2014; Table 1).
In the disk stage, AGN feedback can still be active (\citealt{Hamer:2014,McNamara:2014}), albeit being less episodic;
the accretion rates and feedback are dramatically suppressed only in the hot mode (\S\ref{s:stir}).

\subsection[]{Radial profiles}  \label{s:heat_prof}
\noindent
The evolution of the $t_{\rm cool}/t_{\rm ff}$ (TI-ratio) profiles is analogous to those presented in G13 (their Fig.~15).
The threshold $t_{\rm cool}/t_{\rm ff}\lta10$ indicates where the nonlinear condensation of extended multiphase gas develops (\citealt{Gaspari:2012a, McCourt:2012, Sharma:2012}). 
Within the Bondi radius and above $\sim$\,7 kpc thermal instability cannot grow nonlinearly, as buoyancy
dominates.
Most cold clouds and filaments fill the region 200 pc\,-\,1 kpc, where the TI-ratio has a minimum around 4 and the initial gas density is above 0.1 cm$^{-3}$, leading to more rapid condensation. The cloud funneling toward the BH allows more frequent interactions within such region. The dropout of cold gas results in a gradual increase of the average entropy and thus TI-ratio, as the plasma becomes more tenuous (in a later stage, feedback self-regulation prevents overheating; \citealt{Gaspari:2012a}).


\begin{figure*} 
      \begin{center}
      \subfigure{\includegraphics[scale=0.43]{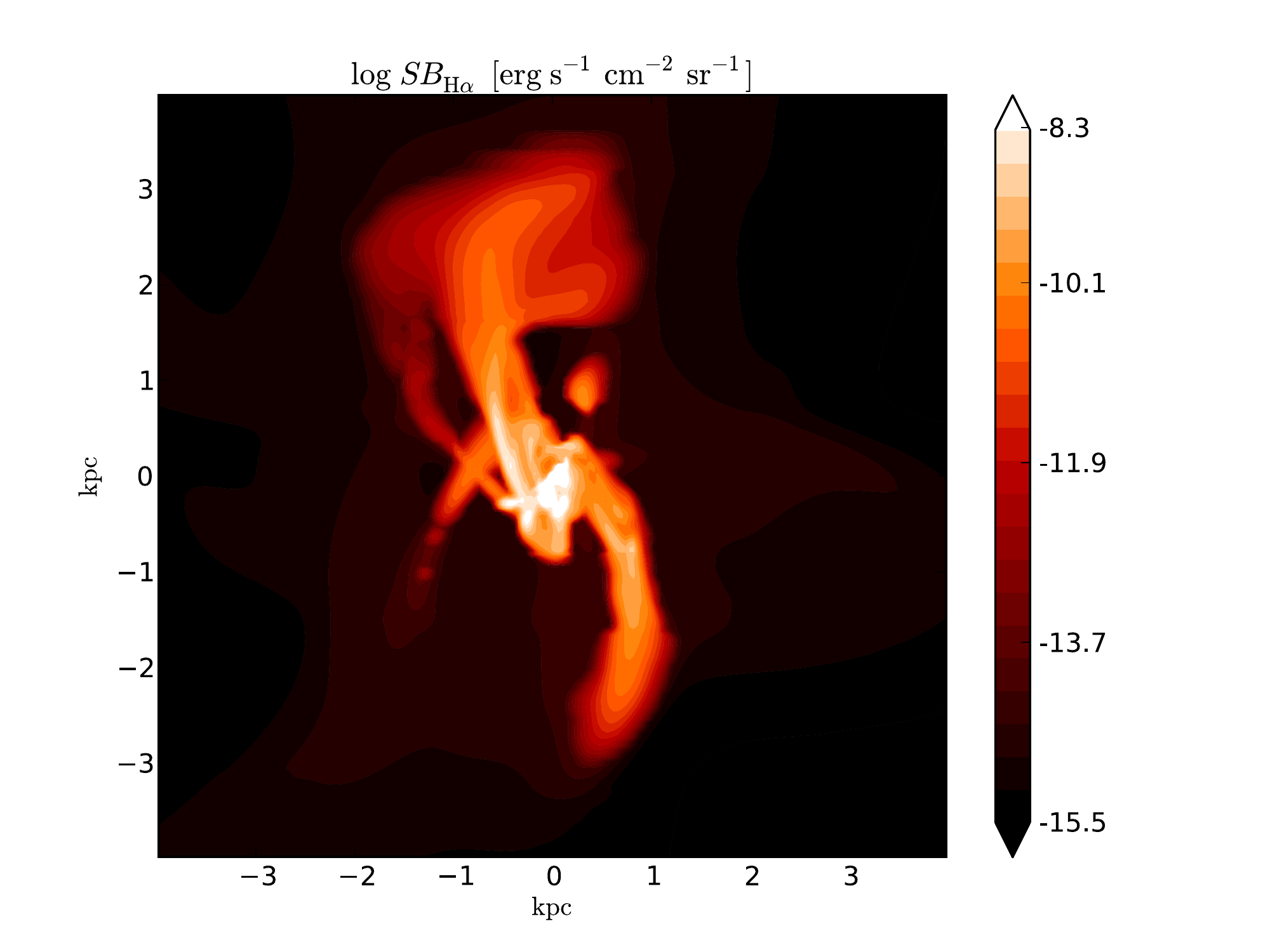}}
      \subfigure{\includegraphics[scale=0.43]{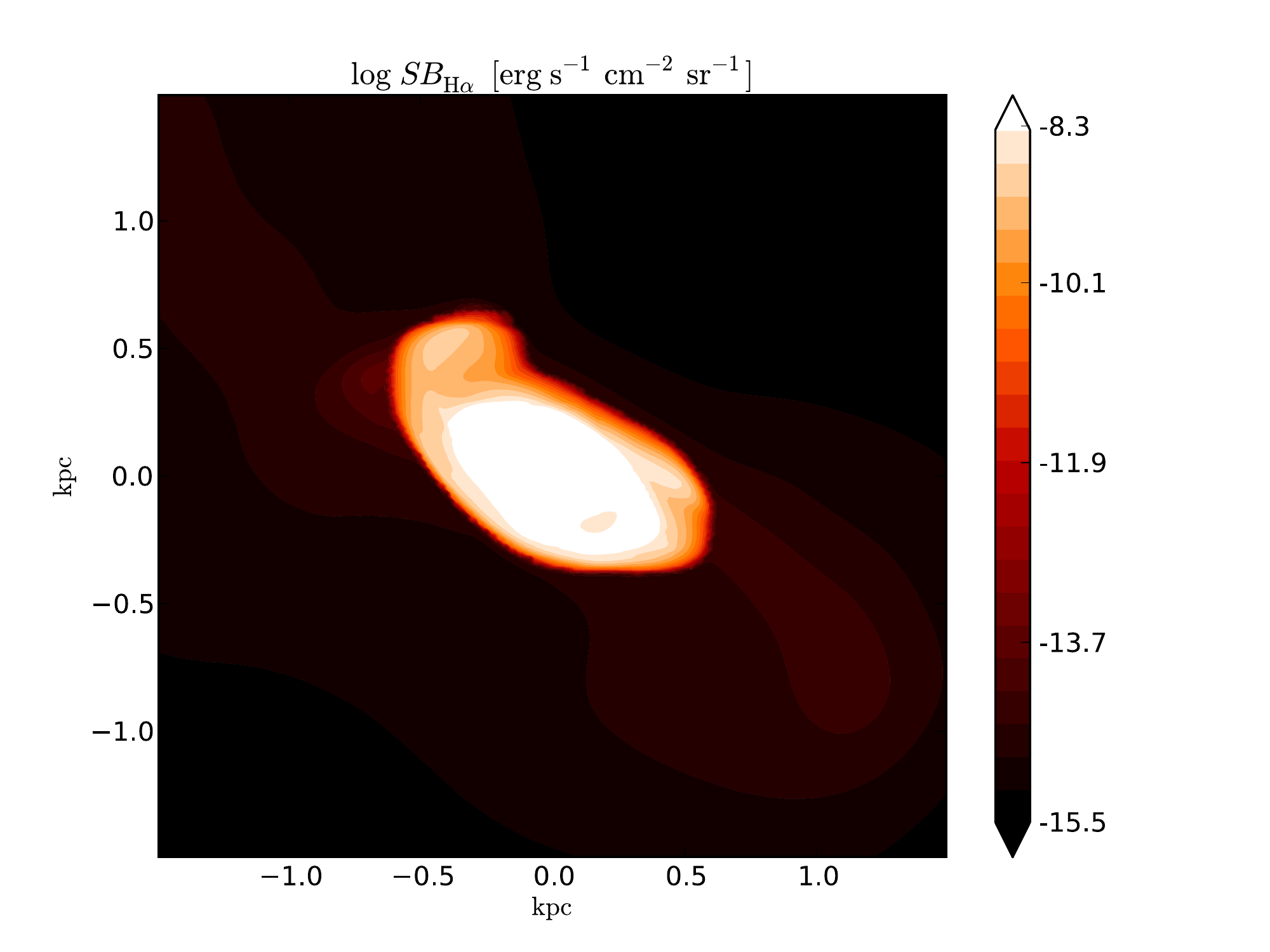}}       
      \subfigure{\includegraphics[scale=0.21]{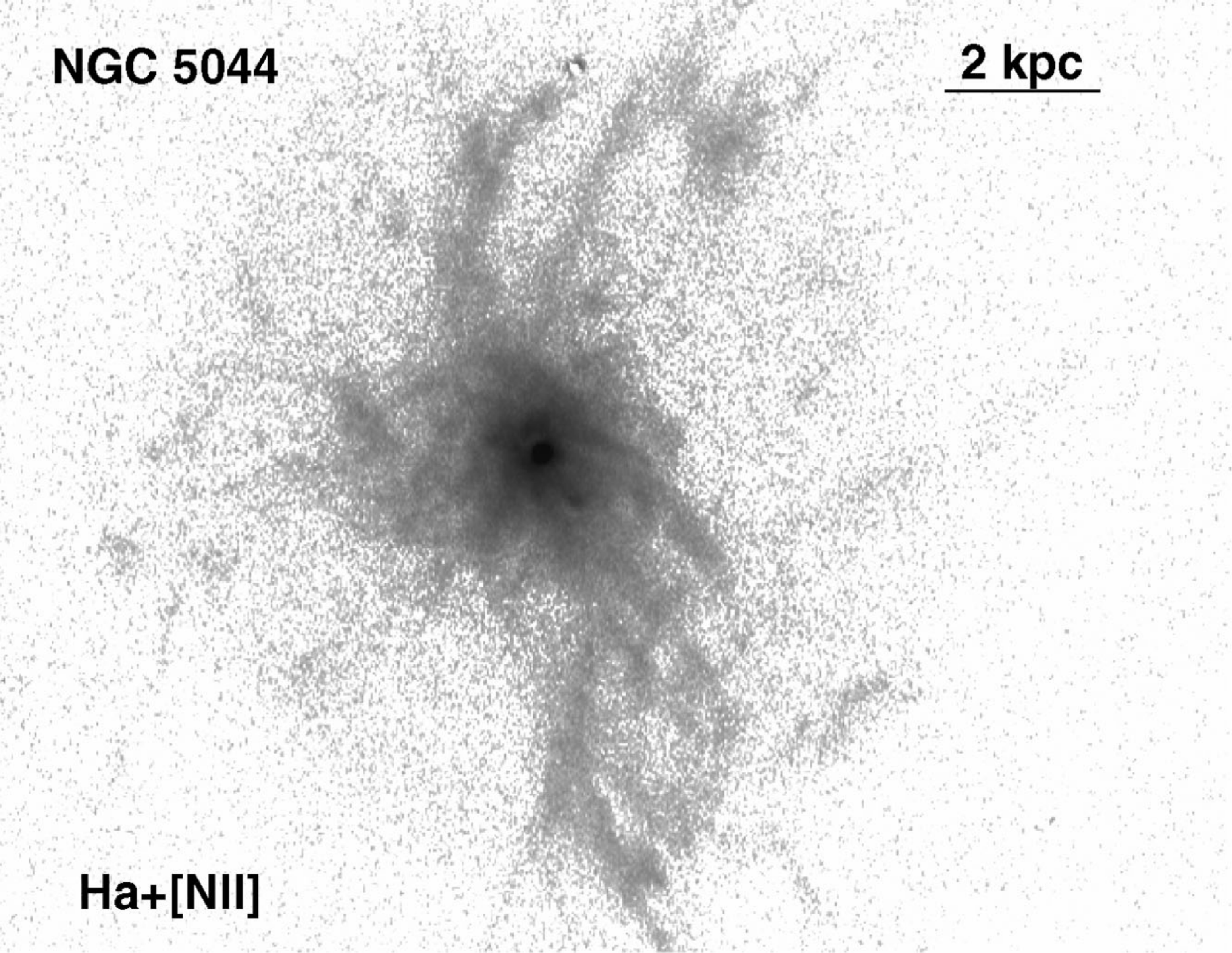}}
      \hspace{+1.7cm}
      \subfigure{\includegraphics[scale=0.21]{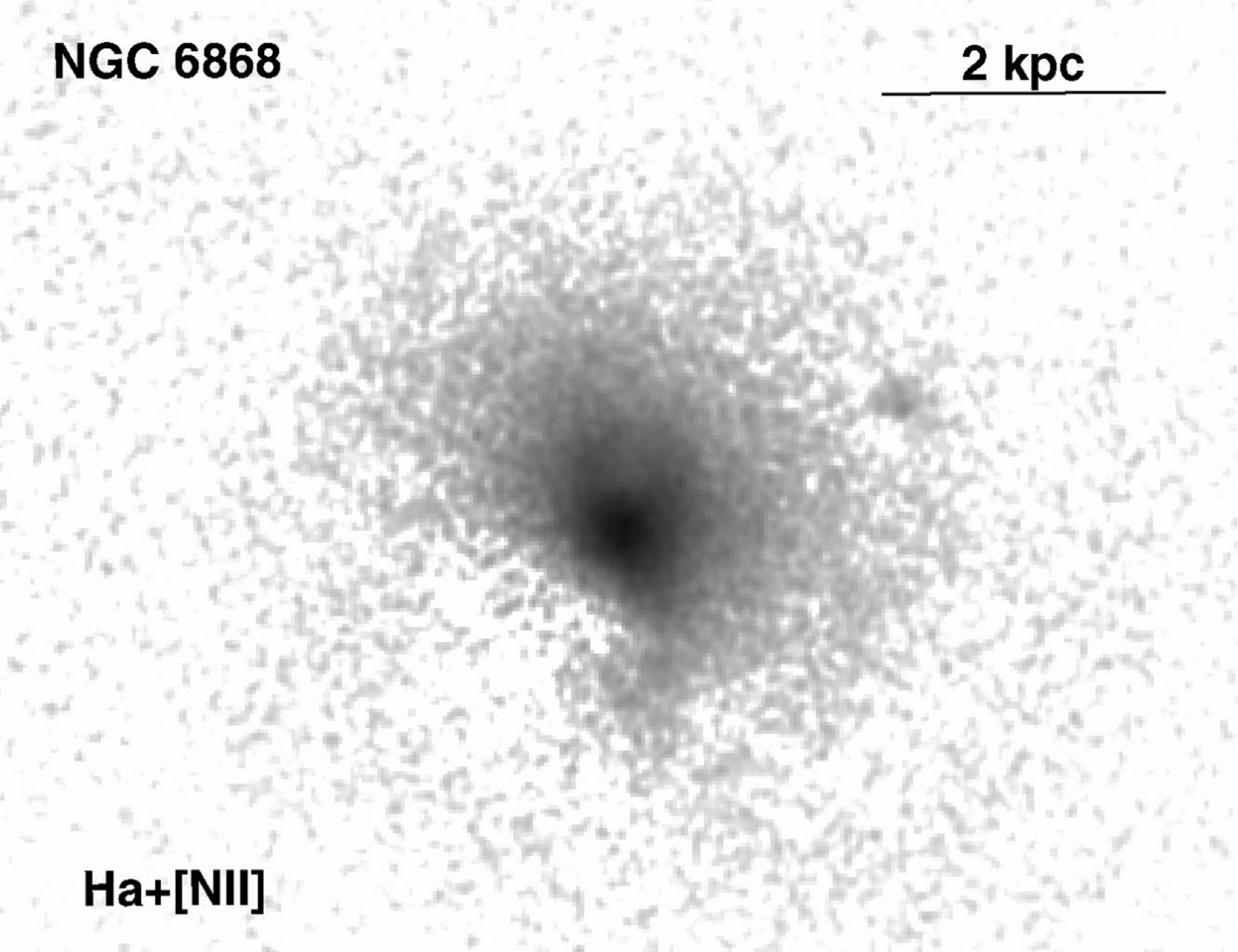}      \hspace{0.7cm}}
      \end{center}
      \caption{Top: Synthetic H$\alpha$ surface brightness image along an arbitrary line of sight during the two characteristic late stages: the filamentary phase (left; run with ${\rm Ta_t}\simeq0.7$) and the disk-dominated phase (right; ${\rm Ta_t}\simeq3$). The resolution mimics that of SOAR, $\sim\,$0.2 arcsec.
                   Bottom: Observed SOAR H$\alpha$+[NII] images from \citet{Werner:2014} showing the two analogous stages in two real massive elliptical galaxies: chaotic cold accretion (left; NGC 5044) and the rotating disk (right; NGC 6868).} 
      \label{f:Halpha}
\end{figure*} 

For systems with ${\rm Ta_t}>1$, the gas is mainly supported by coherent rotation during the condensation process, implying that radial compression is less relevant for TI growth.
With no radially compressive term, it has been shown that the TI-ratio threshold is lower by an order of magnitude, $t_{\rm cool}/t_{\rm ff} \lta 1$ (\citealt{McCourt:2012}). The free-fall time does not increase: the effective gravity is lower along $R$, but the cold gas tends to fall along the $z$ direction and settle on the equatorial plane; hence $t_{\rm ff}(z)=(2 z/g(z))^{1/2}$, which is identical to $t_{\rm ff}(r)$.
In other words, nonlinear TI and fragmented clouds are more difficult to form in a cooling gas shaped by the centrifugal force. This is confirmed by the ${\rm Ta_t}>1$ runs, having a rather stable $\dot M_\bullet$ evolution. Notice that $t_{\rm cool}/t_{\rm ff} > 1$ does not imply the cold phase will not form, but that the condensation is monolithic, leading to a disk structure instead of filaments (e.g., NGC 6868 and NGC 7049; \citealt{Werner:2014}).
The reduced fragmentation
also affects the ${\rm Ta_t}= 0.7$ run, favoring filaments instead of spherical clouds.
The major cold phase interactions can be thus more prolonged inducing broader $\dot M_\bullet$ peaks compared with G13 model.

The mass-weighted radial profiles (Fig.~\ref{f:heat_prof}, top panels) highlight 
the extended multiphase structure and Myr variability of the accretion flow. 
The extension of multiphase filaments up to a few kpc is consistent with ${\rm H}\alpha$ observations (Fig.~1 in \citealt{Werner:2014}; \S\ref{s:comp}).
Performing an X-ray observation (bottom) instead conceals them, allowing the hot component to emerge ($T>0.3$ keV).
As in G13, the X-ray emission-weighted temperature profile is flat, in contrast to the peaked profile
of the adiabatic flow (Fig.~\ref{f:stir_prof}).
This is a key observable and prediction, which can be tested, in particular, with future X-ray missions (e.g., {\it Athena}).
Recently, the Megasecond {\it Chandra} observation of NGC 3115 (\citealt{Wong:2014}) discovered a flat temperature core with multiphase structure within the Bondi radius. Their retrieved density profile within 200 pc has moderate slope $\propto r^{-1}$, which is consistent with our findings (Fig.~\ref{f:heat_prof}, bottom left).
\citet{Russell:2015} also discovered a flat X-ray temperature and $r^{-1}$ density profile
within the Bondi radius of M 87 with deep {\it Chandra} data.
Both observations remarkably corroborate the CCA predictions. 

The mass-weighted density profiles show that an inner ($\sim\,$100 pc) clumpy torus is relatively common, though continuously formed and dismantled by the chaotic dynamics. 
In fact, even in CCA a prograde bias is still present in the long term (\S\ref{s:stir_cool_prof}),
as turbulence generates vorticity locally but not globally.
The presence of an inner obscuring torus is supported by extensive AGN observations literature (\citealt{Bianchi:2012} for a review).
The central volatile structure derives from the accumulation of multiple filaments, which have not yet completely canceled angular momentum. On the other hand, its rising cross-section increases the collisional rate with the incoming clouds, promoting efficient angular momentum cancellation. The PDF($l_z$) of the cold gas within 100 pc is self-similar to Fig.~\ref{f:heat_lz}, continuously broadening and narrowing through time, albeit in a smaller range, $|l_z| < 40$ kpc km s$^{-1}$ (i.e., below circular angular momentum at 100 pc).
As ${\rm Ta_t} > 1$, collisions become less efficient (Eq.~\ref{e:diff}), and
the torus can stabilize in a more coherent and extended disk. 

The cold phase temperature can fluctuate between $10^4$\,-\,$10^5\ {\rm K}$, implying that, in a heated and turbulent environment, the cooling gas does not have to necessarily collapse to the floor temperature, but it can remain relatively warm. Multiwavelength data similarly show cold filaments with a complex multitemperature transition layer (\citealt{McDonald:2009,McDonald:2010,McDonald:2011a}). 
Subsonic turbulence in the hot phase becomes supersonic through the cold medium, thus increasing the efficiency of both turbulent mixing and dissipation, consequently reheating the filaments. 
The observed filamentary warm gas frequently shows significant velocity dispersions ($\sim\,$100 km s$^{-1}$; e.g., \citealt{Canning:2014,Werner:2014}) corroborating the importance of turbulent motions.

\section{Comparison with H$\alpha$ observations}  \label{s:comp}

As reviewed in \S\ref{s:intro}, massive elliptical galaxies have been observed to host a significant reservoir of cold gas
within the central few kpc. In the CCA mechanism, the extended multiphase gas is a natural outcome of the TI condensation.
A key observable of the condensed medium is H$\alpha$ emission, which is mostly associated with 
gas at $10^4$ K.
We thus compute synthetic surface brightness maps of H$\alpha$ line from the last runs (\S\ref{s:heat})
and compare them with the latest observations (Fig.~\ref{f:Halpha}).

The H$\alpha$ emission (6564.6\,\AA; $n=3\rightarrow2$ hydrogen transition) is dominated by the recombination of ionized hydrogen H$^+$ (\citealt{Dong:2011,Draine:2011}). The recombination luminosity is
\begin{align}\label{e:Halpha}
dL_{\rm H\alpha} &= 4\pi\,j_{\rm H\alpha}\,dV \notag\\
&=4\pi\times 2.82 \times 10^{-26}\,T_4^{-0.942-0.031\,\ln T_4} \,n_{\rm e} n_{\rm H^+}\,dV,
\end{align} 
where $T_4\equiv T/10^4\ K$. 
The $10^4$ K gas is optically thick to radiation above 13.6\,eV, i.e., Lyman photons are quickly re-absorbed.
The recombination rate  $j_{\rm H\alpha}$
includes such absorption (the so-called `case B' recombination; see \citealt{Draine:2011}). The ionization fraction $f\equiv n_{\rm H^+}/n_{\rm H}$ is typically low at $10^4$ K. In abundance equilibrium, the recombination and collisional ionization rates (see \citealt{Katz:1996}) 
determine $f$, such that $n_{\rm H^+}/n_{\rm H}=1+\alpha_{\rm R,\,H^+}/(\alpha_{\rm R,H^+}+\Gamma_{\rm C,H^0}$). At $T=10^4$ K the ionization fraction is $f\simeq10^{-3}$. 
Photoionization driven by 
OB and pAGB stars can increase this level.  As adopted in reference literature studies (e.g., \citealt{Joung:2006}), 
we use $f\sim10^{-2}$ (see also \citealt{Dale:2015}).
In a subsequent series of papers,
we will retrieve and study the ionization fraction in a more realistic way with
the inclusion of chemical reaction networks and stellar heating.

In Fig.~\ref{f:Halpha} (top), we present the synthetic H$\alpha$ maps at two characteristic moments, the filamentary and disk-dominated phase. The synthetic maps are created by summing the contribution from gas in each cell using Eq.~\ref{e:Halpha}; the latter equation naturally selects gas with $T\lta10^4$ K. 
The resolution mimics that of SOAR, $\sim\,$0.2 arcsec. 
In the bottom panels, we show for qualitative comparison the observed maps retrieved with the 4.1 m SOAR telescope (from \citealt{Werner:2014}).
In the top left image (${\rm Ta_t}\simeq0.7$ run), a complex network of cold filaments and clouds has condensed out of the hot halo, reaching $r\simeq3.3$ kpc. The system is in full chaotic cold accretion mode. Within 1 kpc, the frequent interactions between the filaments and clouds make the H$\alpha$ distribution approach spherical symmetry.
The observed H$\alpha$ emission (bottom left) shows similar filamentary and core morphology. The real emission appears more extended, albeit the synthetic maps are not contaminated by the stellar and [NII] contribution. At a distance $d_{\rm L}\simeq40.3$ Mpc, the total synthetic H$\alpha$ flux is $F_{\rm H\alpha}\simeq4.0\times10^{-13}\ {\rm erg\,s^{-1}\,cm^{-2}}$ ($L_{\rm H\alpha}\simeq7.8\times10^{40}$\,erg\,s$^{-1}$), which is comparable to the real NGC 5044 flux, $F_{\rm H\alpha+[NII]}\simeq7.6\times10^{-13}\ {\rm erg\,s^{-1}\,cm^{-2}}$ (\citealt{Werner:2014}).
It is worth noting that NGC 5044 H$\alpha$ luminosity marks the upper envelope of observed values; several elliptical galaxies have $\sim\,$1 dex lower luminosity.\footnote{Dust, which we do not model, can in part absorb H$\alpha$ emission.}

In the top right panel, the $\rm {Ta_t}\simeq3$ run displays a different behavior. The system is no longer in filamentary CCA mode; the rotating disk drives the dynamics. The disk is contained within 500 pc and is slightly perturbed by the relatively weak turbulence. The observed H$\alpha$ image of the massive elliptical NGC 6868 (bottom right) shows a very similar smooth pattern associated with a rotating structure (which has been confirmed through the [CII] velocity distribution; \citealt{Werner:2014}). 
Remarkably, NGC 6868 does not show major signs of AGN feedback, as radio jets or outflows,
while NGC 5044 which resides in filamentary CCA mode is strongly perturbed by AGN activity (\citealt{Gastaldello:2009}).
Considering the disk size, this quiescent phase could last since $\gta\,$50 Myr. The NGC 6868 cooling time is indeed relatively long, $t_{\rm cool}\approx78$\,Myr.
At a distance of 41.2 Mpc, the synthetic flux is $F_{\rm H\alpha}\simeq1.1\times10^{-13}\ {\rm erg\,s^{-1}\,cm^{-2}}$, which is consistent with the observed flux $F_{\rm H\alpha+[NII]}\simeq2.7\times10^{-13}\ {\rm erg\,s^{-1}\,cm^{-2}}$ (\citealt{Macchetto:1996}). 

To conclude, despite the fact that it is not
our goal to match one-to-one the observed image of a single galaxy,
as this would require artificial fine-turning of run parameters and initial conditions, 
the simulated models capture the essential features of real elliptical galaxies showing a dual morphology of filamentary and rotating multiphase nebulae within the core.

\section{Summary and conclusions}  \label{s:disk}
\noindent
We carried out 3D high-resolution simulations to perform controlled astrophysical experiments of the accretion flow
on to a supermassive black hole, resolving the galactic 50 kpc scale down to the inner sub-pc scale.
We gradually increased the realistic complexity of the hydrodynamic flow, including 
cooling, turbulence, and heating. 
We focused on the role of rotation (reference $e_{\rm rot}=0.3$, i.e., $v_{\rm rot}\approx100$ km s$^{-1}$) in altering accretion rates, in comparison with the nonrotating models presented in \citealt{Gaspari:2013_cca} (G13). The main features of each accretion flow can be summarized as follows.\\

\renewcommand{\labelitemi}{$\bullet$}
\begin{itemize}
\item {\it Adiabatic rotating flow.}\\
The hot pressure-dominated flow forms a central rotational barrier ($\lta\,$$r_{\rm B}$), with a thick toroidal geometry and mild variability. The gas can only accrete along a polar funnel with half-opening angle $\sim\,$$\pi/4$. Compared with the spherically symmetric model, the accretion rate is suppressed by a characteristic factor of $\sim\,$3.
The accretion rate is comparable to the reference $\dot M_{\rm Bondi}$ computed at $r$\,$\sim$\,1-\,2 kpc. 
The stratification of the atmosphere slightly decreases the central density and accretion through time.
The characteristic mark of the hot flow is the cuspy (X-ray or mass-weighted) temperature profile.
The coherent prograde rotation preserves the initial positive angular momentum distribution.\\ 

\item {\it Radiative rotating flow.}\\
The radiatively cooling gas loses pressure support in a cooling time and circularizes on the equatorial plane, forming a cold thin disk. At variance with the classic cooling flow, the accretion rate is suppressed and decoupled from the cooling rate, $\dot M_\bullet \lta 10^{-2}\, \dot M_{\rm cool}$. However, $\dot M_\bullet$ is still a decade higher compared with the adiabatic flow due to the halo condensation. 
The cold phase progressively accumulates higher positive $l_z$, as the kpc-size disk grows through time via condensation.\\

\item {\it Adiabatic rotating flow stirred by turbulence.}\\
From the perspective of the accretion rate and $\rho$, $T$ radial profiles, the stirred hot flow is analogous to the unperturbed adiabatic evolution, with increased variability (factor of 2). The similar $\sim\,$1/3 suppression of $\dot M_\bullet$ is because subsonic turbulence generates local, though not global, vorticity
(which is further enhanced by the baroclinic instability in a stratified medium). As ${\rm Ta_t}\equiv v_{\rm rot}/\sigma_v<1$, the $l_z$ distribution is reshaped and substantially broadened via turbulent diffusion, generating both prograde and retrograde eddies. If ${\rm Ta_t}>1$, the initial PDF is only slightly modified: the flow is again driven and suppressed by coherent rotation ($l_z > 0$) and not by turbulent eddies. \\

\item {\it Chaotic cold accretion (cooling, heating, turbulence).}\\ 
As long as ${\rm Ta_t} < 1$, chaotic cold accretion (CCA) drives the dynamics, as found in G13. 
Within several kpc, thermal instability can grow nonlinearly ($t_{\rm cool}/t_{\rm ff} < 10$), extended multi-temperature filaments condense out of the hot phase and rain toward the BH. The collisions ($r< 1$ kpc) between the cold clouds, filaments, and central clumpy torus enable to efficiently cancel angular momentum, boosting the accretion rate, with impulsive peaks up to the cooling rate or $100\times$ the Bondi rate.
Using $\dot M_\bullet\sim\dot M_{\rm cool}$ is thus a realistic (subgrid) model for large-scale simulations and analytic studies. Without heating, the CCA evolution is analogous, albeit the cold phase properties (cold mass, collisions, and PDF) are magnified by an order of magnitude. 

The condensed cold phase retains the imprint of the stirred hot phase, emerging out of the broadened $l_z$ distribution.
The presence of both prograde and retrograde motions permits the cancellation of angular momentum.
After major collisions the PDF narrows ($\dot M_\bullet$ peaks), while condensation broadens it again. 
Transient incomplete cancellation ($\dot M_\bullet$ valleys) creates a clumpy, highly variable torus, later favoring the interactions with incoming clouds.
In the regime ${\rm Ta_t} > 1$, turbulent diffusion becomes weaker than advection due to rotation, reducing the relative PDF broadening and the efficiency of collisions. The accretion rate thus decreases as ${\rm Ta_t}^{-1}$ until the cold disk drives the evolution again. 
This is aggravated by the increased difficulty of TI and fragmented clouds to form under coherent rotation, as the thermal instability threshold is lowered because of the reduced influence of radial compression.

The synthetic H$\alpha$ maps trace the morphology of the condensed multiphase gas, 
reproducing the main features of observations (e.g., SOAR and {\it Magellan}), such as line fluxes, kpc-scale filaments, or the central rotating disk. 
\\
\end{itemize}

The present work, together with G13, emphasizes the central role of chaotic cold accretion in the evolution
of (supermassive) black holes, even in the presence of rotation. The high and variable accretion rates, 
$\dot M_\bullet\sim\dot M_{\rm cool}$, can trigger the required level and self-regulation of the AGN feedback
to quench cooling flows, star formation and to shape the observed thermodynamic properties of massive galaxies, groups, and clusters (\citealt{Gaspari:2011a,Gaspari:2011b,Gaspari:2012a,Gaspari:2012b}; \S\ref{s:intro}). 

The results obtained in recent years
corroborate the following cosmic accretion and feedback cycle shaping gaseous halos.
As gas cooling starts to overcome heating ($t_{\rm cool}/t_{\rm ff}<10$ or central entropy $K_0 \lta 20$ keV cm$^2$), CCA is triggered\footnote{A nonzero level of subsonic turbulence is always present in real systems (\S\ref{s:init2}), e.g., because of galaxy motions, mergers, stellar evolution. The initial AGN outburst generates further turbulence.},
boosting the accretion rate ($\dot M_\bullet \gg \dot M_{\rm Bondi}$) and consequently the feedback injection via AGN outflows and/or jets (${\rm Ta_t}< 1$). 
This phase can be observationally probed via the extended H$\alpha$ filaments  
(e.g., NGC 5044; \citealt{McDonald:2010,McDonald:2011a,Werner:2013})
or inner flat temperature and $r^{-1}$ density profiles in X-ray, 
as recently discovered in NGC 3115 (\citealt{Wong:2014}) and M 87 (\citealt{Russell:2015}).
As the core entropy rises and turbulence diminishes (${\rm Ta_t} > 1$), nonlinear TI weakens and the rotating disk is left as the sole cold structure
(e.g., NGC 6868 or NGC 7049 in \citealt{Werner:2014}; see also \citealt{Mathews:2003,Young:2011,Alatalo:2013}).
The transition is associated with a gradual decrease in the accretion rate $\propto {\rm Ta_t}^{-1}$ 
(feedback can still be active; e.g., NGC 4261).
Waiting longer times, as the disk is consumed (e.g., via accretion and star formation) and the halo has been overheated by feedback, the hot gas is the only resource available to poorly fuel the SMBH ($t_{\rm cool}/t_{\rm ff} \gg 10$). The main diagnostic of this hot-mode regime is the cuspy X-ray temperature profile, typically associated with more quiescent systems (e.g., NGC 4649, NGC 1332; \citealt{Humphrey:2008,Humphrey:2009}).
Transitioning from the fully cold mode to hot mode, the accretion rate experiences a strong suppression, from $100\times$ to a small fraction of the Bondi rate. 
Feedback heating becomes negligible, entropy starts to decrease, and the gaseous halo is allowed to cool again, restarting a new cycle defined by CCA and boosted feedback to the rotating disk to the hot mode and suppressed feedback, and so on.
The simulated accretion rates cover the realistic range based on the cavity power observed in massive ellipticals (\citealt{Allen:2006}), $\dot M_\bullet = P_{\rm cav}/(\varepsilon\,c^2)\approx 2\times10^{-2}$\,-\,$2\ \msun\,{\rm yr^{-1}}$.

In the subsequent series of works, we will continue to investigate the role of additional physics 
(see Sec.~9.1 in G13 for a discussion)
to further understand CCA and to better interpret new data.
For instance, a rapidly varying potential (e.g., mergers) could facilitate TI and chaotic collisions, thereby promoting
CCA via tidal torques (\S1). Although star formation is observed to be inefficient in elliptical galaxies and in turbulent
molecular clouds ($\lta$\,1\% efficiency; \citealt{Federrath:2015}), it can partially reduce the fueling of the SMBH.
Stellar feedback, on the other hand, may reheat the cold gas, promoting additional turbulence and TI.
Forthcoming observations of cold gas, which combine several wavelengths, e.g., via {\it Herschel}, HST, SOAR, CARMA, SKA, and the newly expanding ALMA observatory, will be instrumental in probing chaotic cold accretion and the related predictions with high accuracy. Besides improving the sample size, the new data should be able to shed light on key astrophysical questions, such as the amount of cold gas `raining' on to the black hole, the kinematics of the cold phase (filaments, clouds, and rotating disk), and its tight coupling with AGN feedback.

\section*{Acknowledgments}
\noindent
The FLASH code was in part developed by the DOE NNSA-ASC OASCR Flash center at the University of Chicago. 
M.G. is grateful for the financial support provided by the Max Planck Fellowship. 
F.B. is in part supported by the Prin MIUR grant 2010LY5N2T.
HPC resources were provided by the NASA/Ames HEC Program (SMD-13-4373, SMD-13-4377, SMD-14-4819; {\it Pleiades}) and CLS center.
The post-processing analysis was in part performed with YT (\citealt{Turk:2011}).
We thank R.~Sunyaev, N.~Werner, M.~Anderson, R.~Khatri, P.~Girichidis, and A.~Gatto for helpful discussions.
We are grateful to N.~Werner, who allowed us to reproduce the H$\alpha$ SOAR images.
We thank the anonymous referee who helped to improve the manuscript.

\bibliographystyle{aa}
\bibliography{biblio}

\label{lastpage}

\end{document}